%% file: sldannrev.tex
\documentclass{ar2e}
\usepackage{epsfig}
\begin{document}
%
\input sldsym.tex  
\jname{Annu. Rev. Nucl. Part. Sci.}
\jyear{2001}
\jvol{51}

\titlepage

\title{
Highlights of the SLD Physics Program \\at the SLAC Linear
Collider\footnote{Work supported in part by Department of Energy contract DE-AC03-76SF00515.}
}


\markboth{ROWSON, SU \& WILLOCQ}
{SLD PHYSICS}

\author{
  P.C.\ Rowson and Dong Su
  \affiliation{Stanford Linear Accelerator Center, Stanford University,
               Stanford, California 94309; ~~~~~~~~~~~~~~~~~~~~~~~~~~~~~~~~
  e-mail: rowson@slac.stanford.edu, sudong@slac.stanford.edu}
  St\'ephane Willocq
  \affiliation{University of Massachusetts, Physics Department,
               Amherst, Massachusetts 01003; ~~~~~~~~~~~~~~~~~~~~~~~~~~~~~~~~~~~
  e-mail: willocq@physics.umass.edu}
}


\begin{abstract}
Starting in 1989, and continuing through the 1990s,
high-energy physics witnessed a flowering of precision
measurements in general and tests of the standard model
in particular, led by \ee\ collider experiments operating at the
\Z\ resonance.  Key contributions to this work came from the SLD
collaboration at the SLAC Linear Collider.
By exploiting the unique capabilities of this pioneering
accelerator and the SLD detector,
including a polarized electron beam, exceptionally small beam dimensions,
and a CCD pixel vertex detector,
SLD produced a broad array of electroweak,
heavy-flavor, and QCD measurements.
Many of these
results are one of a kind or represent the world's standard in
precision.
This article reviews the highlights of the SLD physics
program,
with an eye toward associated advances in experimental technique, and
the contribution of these measurements to our dramatically
improved present understanding of the standard model and its possible
extensions.
~~~~~~~~~~~~~~~~~~~~~~~~~~~~~~~~~~~~~~~~~~~~~~~~~~~~~~~~~~~~~~~~~~~~~~~~
\vspace{1.5 cm}
\end{abstract}

\maketitle


\input sec1_intro.tex


\input sec2_slc.tex



\input sec3_sld.tex


\input sec4_analysis.tex


\input sec5_electroweak.tex


\input sec6_bphysics.tex


\input sec7_interp.tex


\input sec8_epilogue.tex


\input references.tex

\end{document}

%% file: sldsym.tex

\let\emi\en
\def\electron   {\ensuremath{e}}
\def\en         {\ensuremath{e^-}}      
\def\ep         {\ensuremath{e^+}}
\def\epm        {\ensuremath{e^\pm}}  
\def\epem       {\ensuremath{e^+e^-}}
\def\ee         {\ensuremath{e^+e^-}}
\def\emem       {\ensuremath{e^-e^-}}
\def\mmu        {\ensuremath{\mu}}
\def\mup        {\ensuremath{\mu^+}}
\def\mun        {\ensuremath{\mu^-}}    
\def\mumu       {\ensuremath{\mu^+\mu^-}}
\def\mtau       {\ensuremath{\tau}}
\def\taup       {\ensuremath{\tau^+}}
\def\taum       {\ensuremath{\tau^-}}
\def\tautau     {\ensuremath{\tau^+\tau^-}}
\def\ellm       {\ensuremath{\ell^-}}
\def\ellp       {\ensuremath{\ell^+}}
\def\ellell     {\ensuremath{\ell^+ \ell^-}}


\def\g     {\ensuremath{\gamma}}
\def\gaga  {\ensuremath{\gamma\gamma}}  
\def\ggstar{\ensuremath{\gamma\gamma^*}}
\def\Z     {\ensuremath{Z^0}} 


\def\ega   {\ensuremath{e\gamma}}
\def\game  {\ensuremath{\gamma e^-}}
\def\epemg  {\ensuremath{e^+e^-\gamma}}


\def\q   {\ensuremath{q}}
\def\qbar  {\ensuremath{\overline q}}
\def\qqbar {\ensuremath{q\overline q}}
\def\fbar  {\ensuremath{\overline f}}
\def\ffbar {\ensuremath{f\overline f}}
\def\QQbar {\ensuremath{Q\overline Q}}
\def\u  {\ensuremath{u}}
\def\ubar  {\ensuremath{\overline u}}
\def\uubar {\ensuremath{u\overline u}}
\def\d  {\ensuremath{d}}
\def\dbar  {\ensuremath{\overline d}}
\def\ddbar {\ensuremath{d\overline d}}
\def\s  {\ensuremath{s}}
\def\sbar  {\ensuremath{\overline s}}
\def\ssbar {\ensuremath{s\overline s}}
\def\c  {\ensuremath{c}}
\def\cbar  {\ensuremath{\overline c}}
\def\ccbar {\ensuremath{c\overline c}}
\def\b  {\ensuremath{b}}
\def\bbar  {\ensuremath{\overline b}}
\def\bbbar {\ensuremath{b\overline b}}
\def\t  {\ensuremath{t}}
\def\tbar  {\ensuremath{\overline t}}
\def\tbar  {\ensuremath{\overline t}}
\def\ttbar {\ensuremath{t\overline t}}


\def\piz   {\ensuremath{\pi^0}}
\def\pizs  {\ensuremath{\pi^0\mbox\,\rm{s}}}
\def\ppz   {\ensuremath{\pi^0\pi^0}}
\def\pip   {\ensuremath{\pi^+}}
\def\pim   {\ensuremath{\pi^-}}
\def\pipi  {\ensuremath{\pi^+\pi^-}}
\def\pipm   {\ensuremath{\pi^\pm}}
\def\pimp   {\ensuremath{\pi^\mp}}
\def\kaon  {\ensuremath{K}}
\def\Kbar  {\kern 0.2em\overline{\kern -0.2em K}{}}
\def\Kb    {\ensuremath{\Kbar}}
\def\Kpm   {\ensuremath{K^\pm}}
\def\Kp    {\ensuremath{K^+}}
\def\Km    {\ensuremath{K^-}}
\def\KS    {\ensuremath{K^0_{\scriptscriptstyle S}}} 
\def\KL    {\ensuremath{K^0_{\scriptscriptstyle L}}} 
\def\Kstarz  {\ensuremath{K^{*0}}}
\def\Kstarzb  {\ensuremath{\Kbar^{*0}}}
\def\Kstar   {\ensuremath{K^*}}
\def\Kstarb  {\ensuremath{\Kbar^*}}
\def\Kstarp   {\ensuremath{K^{*+}}}
\def\Kstarm   {\ensuremath{K^{*-}}}
\def\Kstarpm   {\ensuremath{K^{*\pm}}}
\def\Kzb   {\ensuremath{\Kbar^0}}
\def\KzKzb {\ensuremath{K^0 \kern -0.16em \Kzb}}
\def\Dz    {\ensuremath{D^0}}
\def\Xpart {\ensuremath{X}}
\def\Xbar{\kern 0.2em\overline{\kern -0.2em X}{}}
%
%
\def\Dbar  {\kern 0.2em\overline{\kern -0.2em D}{}}
\def\Db    {\ensuremath{\Dbar}}
\def\Dzb   {\ensuremath{\Dbar^0}}
\def\DzDzb {\ensuremath{D^0 {\kern -0.16em \Dzb}}}
\def\Dstar   {\ensuremath{D^*}}
\def\Dstarb   {\ensuremath{\Dbar^*}}
\def\Dstarz  {\ensuremath{D^{*0}}}
\def\Dstarzb  {\ensuremath{\Dbar^{*0}}}
\def\Ds    {\ensuremath{D^+_s}}
\def\Dsb   {\ensuremath{\Dbar^+_s}}
\def\Dss   {\ensuremath{D^*_s}}
\def\Bz    {\ensuremath{B^0}}
\def\B     {\ensuremath{B}}
\def\Bbar  {\kern 0.18em\overline{\kern -0.18em B}{}}
\def\Bb    {\ensuremath{\Bbar}}
\def\Bzb   {\ensuremath{\Bbar^0}}
\def\Bu    {\ensuremath{B^+}}
\def\Bub   {\ensuremath{B^-}}
\def\Bpm   {\ensuremath{B^\pm}}
\def\Bs    {\ensuremath{B^0_s}}
\def\Bsb   {\ensuremath{\Bbar^0_s}}
\def\Bd    {\ensuremath{B^0_d}}
\def\Bdb   {\ensuremath{\Bbar^0_d}}
\def\Bq    {\ensuremath{B^0_q}}
\def\Bqb   {\ensuremath{\Bbar^0_q}}
\def\BB    {\ensuremath{B\Bbar}} 
\def\BzBzb {\ensuremath{B^0 {\kern -0.16em \Bzb}}}

%
%
\def\jpsi  {\ensuremath{{J\mskip -3mu/\mskip -2mu\psi\mskip 2mu}}} 
\def\psitwos {\ensuremath{\psi{(2S)}}}
\mathchardef\Upsilon="7107
\def\Y#1S{\ensuremath{\Upsilon{(#1S)}}}
\def\FourS {\Y4S}
%
%
\def\proton     {\ensuremath{p}}
\def\antiproton {\ensuremath{\overline p}}
\def\neutron    {\ensuremath{n}}
\def\antineutron  {\ensuremath{\overline n}}

\mathchardef\Xi="7104
\mathchardef\Lambda="7103
\def\Lc {\ensuremath{\Lambda_c}} 
\def\Lb {\ensuremath{\Lambda_b}} 
\def\Lbar {\kern 0.2em\overline{\kern -0.2em\Lambda\kern 0.05em}\kern-0.05em{}}
\def\Xibar{\kern 0.2em\overline{\kern -0.2em \Xi}{}}
\def\Nbar{\kern 0.2em\overline{\kern -0.2em N}{}}
\def\Lcb {\ensuremath{\Lbar_c}}
\def\Lbb {\ensuremath{\Lbar_b}}


\def\upsbb {\ensuremath{\Upsilon{\rm( 4S)}\to B\Bbar}} 
\def\upsbzbz {\ensuremath{\Upsilon{\rm( 4S)}\to\BzBzb}} 
\def\upsbpbm {\ensuremath{\Upsilon{\rm( 4S)}\to\Bu\Bub}} 

\def\btol       {\ensuremath{b\to\ell}}
\def\btocl      {\ensuremath{b\to c\to\ell}}
\def\ctol       {\ensuremath{c\to\ell}}
\def\bsg        {\ensuremath{b\to s g}}
\def\bmix       {\ensuremath{B^0 \mbox{--} {\Bbar^0}}}
\def\BDDX       {\ensuremath{B\to D\overline{D}X}}
\def\bdmix      {\ensuremath{B_d^0 \mbox{--} {\Bbar_d^0}}}
\def\bqmix      {\ensuremath{B_q^0 \mbox{--} {\Bbar_q^0}}}
\def\bsmix      {\ensuremath{B_s^0 \mbox{--} {\Bbar_s^0}}}

\def\Zto        {\ensuremath{Z^0\to}} 
\def\Ztoe       {\ensuremath{Z^0\to\epem}}
\def\Ztomu      {\ensuremath{Z^0\to\mumu}}
\def\Ztotau     {\ensuremath{Z^0\to\tautau}}
\def\Ztoll      {\ensuremath{Z^0\to\ell^+\ell^-}}
\def\Ztobb      {\ensuremath{Z^0\to\bbbar}}
\def\Ztocc      {\ensuremath{Z^0\to\ccbar}}
\def\Ztoss      {\ensuremath{Z^0\to\ssbar}}
\def\Ztoqq      {\ensuremath{Z^0\to\qqbar}}
\def\ZtoQQ      {\ensuremath{Z^0\to\QQbar}}
\def\Ztoff      {\ensuremath{Z^0\to\ffbar}}
\def\Ztouds     {\ensuremath{Z^0\to\uubar,\ddbar,\ssbar}}
\def\Ztohad     {\ensuremath{Z^0\to{\rm hadrons}}}

\def\Zee        {\ensuremath{Zee}}
\def\Zmumu      {\ensuremath{Z\mu\mu}}
\def\Ztautau    {\ensuremath{Z\tau\tau}}
\def\Zll        {\ensuremath{Z\ell\ell}}
\def\Zbb        {\ensuremath{Z\bbbar}}
\def\Zcc        {\ensuremath{Z\ccbar}}
\def\Zss        {\ensuremath{Z\ssbar}}
\def\Zqq        {\ensuremath{Z\qqbar}}
\def\ZQQ        {\ensuremath{Z\QQbar}}
\def\Zff        {\ensuremath{Z\ffbar}}


\def\ptot       {\mbox{$p$}}
\def\pxy        {\mbox{$p_T$}}
\def\pt         {\mbox{$p_T$}}

\def\mphi       {\mbox{$\phi$}}
\def\mtheta     {\mbox{$\theta$}}
\def\ctheta     {\mbox{$\cos \theta$}}

%
\def\ev   {\ensuremath{\rm \,e\kern -0.08em V}}
\def\kev  {\ensuremath{\rm \,ke\kern -0.08em V}} 
\def\mev  {\ensuremath{\rm \,Me\kern -0.08em V}} 
\def\gev  {\ensuremath{\rm \,Ge\kern -0.08em V}} 
\def\gevc {\ensuremath{{\rm \,Ge\kern -0.08em V\!/}c}} 
\def\tev  {\ensuremath{\rm \,Te\kern -0.08em V}}
\def\mevc {\ensuremath{{\rm \,Me\kern -0.08em V\!/}c}} 
\def\mevcc  {\ensuremath{\rm \,Me\kern -0.08em V}} 
\def\gevcc  {\ensuremath{\rm \,Ge\kern -0.08em V}} 
\def\syin {\ensuremath{^{\prime\prime}}}
\def\in   {\ensuremath{\rm \,in}}
\def\ft   {\ensuremath{\rm \,ft}}
\def\km   {\ensuremath{\rm \,km}}
\def\m    {\ensuremath{\rm \,m}}
\def\cm   {\ensuremath{\rm \,cm}}
\def\cma  {\ensuremath{{\rm \,cm}^2}}
\def\mm   {\ensuremath{\rm \,mm}}
\def\mma  {\ensuremath{{\rm \,mm}^2}}
\def\mum  {\ensuremath{\,\mu\rm m}} 
\def\muma       {\ensuremath{\,\mu\rm m^2}}
\def\nm   {\ensuremath{\rm \,nm}}
\def\fm   {\ensuremath{\rm \,fm}}
\def\barn{\ensuremath{\rm \,b}}
\def\barnhyph{\ensuremath{\rm -b}}
\def\mbarn{\ensuremath{\rm \,mb}}
\def\mbarnhyph{\ensuremath{\rm -mb}}
\def\pb {\ensuremath{\rm \,pb}}
\def\invpb {\ensuremath{\mbox{\,pb}^{-1}}}
\def\fb   {\ensuremath{\mbox{\,fb}}}
\def\invfb   {\ensuremath{\mbox{\,fb}^{-1}}}
\def\mus  {\ensuremath{\rm \,\mus}}
\def\ns   {\ensuremath{\rm \,ns}}
\def\ps   {\ensuremath{\rm \,ps}}
\def\invps {\ensuremath{\mbox{\,ps}^{-1}}}
\def\fs   {\ensuremath{\rm \,fs}}
\def\gm   {\ensuremath{\rm \,g}}
\def\Xrad {\ensuremath{X_0}}
\def\NIL{\ensuremath{\lambda_{int}}}
\let\dgr\degrees

%
%
\def\cms         {\ensuremath{{\rm \,cm}^{-2} {\rm s}^{-1}}}  
\def\nm         {\ensuremath{\rm \,nm}}    
\def\nb         {\ensuremath{\rm \,nb}}
\def\sec{\ensuremath{\rm {\,s}}}       
\def\ms         {\ensuremath{{\rm \,ms}}}      
\def\mus        {\ensuremath{\,\mu{\rm s}}}    
\def\ns         {\ensuremath{{\rm \,ns}}}      
\def\ps         {\ensuremath{{\rm \,ps}}}   
%
%
\def\mhz  {\ensuremath{{\rm \,MHz}}}   
\def\mic  {\ensuremath{\,\mu{\rm C}}}
\def\krad {\ensuremath{\rm \,krad}}
\def\cmc  {\ensuremath{{\rm \,cm}^3}}
\def\yr   {\ensuremath{\rm \,yr}}
\def\hr   {\ensuremath{\rm \,hr}}
\def\degc {\ensuremath{^\circ}{C}}
\def\degk {\ensuremath {\rm K}}
\def\degrees{\ensuremath{^{\circ}}}
\def\mrad{\ensuremath{\rm \,mr}}                
\def\mradhyph{\ensuremath{\rm -mr}}
\def\sx    {\ensuremath{\sigma_x}}     
\def\sy    {\ensuremath{\sigma_y}}    
\def\sz    {\ensuremath{\sigma_z}}     


\def\sinthw{\ensuremath{\sin^2 \theta _W }}
\def\sinsqth{\ensuremath{\sin^2 \theta _W ^{\rm eff}}}
\def\sinsqtl{\ensuremath{\sin^2 \theta _{\rm eff}^{lept}}}
\def\mZ{\ensuremath{m_Z}}
\def\mW{\ensuremath{m_W}}
\def\GZ{\ensuremath{\Gamma_Z}}
\def\Alr{\ensuremath{A_{\rm LR}}}
\def\Af{\ensuremath{A_f}}
\def\Aq{\ensuremath{A_q}}
\def\AQ{\ensuremath{A_Q}}
\def\Al{\ensuremath{A_\ell}}
\def\Ae{\ensuremath{A_e}}
\def\Amu{\ensuremath{A_\mu}}
\def\Atau{\ensuremath{A_\tau}}
\def\Ab{\ensuremath{A_b}}
\def\Ac{\ensuremath{A_c}}
\def\As{\ensuremath{A_s}}
\def\Auds{\ensuremath{A_{uds}}}
\def\Afbb{\ensuremath{A^b_{\rm FB}}}
\def\Afbc{\ensuremath{A^c_{\rm FB}}}
\def\Afbl{\ensuremath{A^\ell_{\rm FB}}}
\def\AfbtQ{\ensuremath{\tilde{A}^Q_{\rm FB}}}
\def\Afbtf{\ensuremath{\tilde{A}^f_{\rm FB}}}
\def\Afbtb{\ensuremath{\tilde{A}^b_{\rm FB}}}
\def\Afbtc{\ensuremath{\tilde{A}^c_{\rm FB}}}
\def\Afbtl{\ensuremath{\tilde{A}^\ell_{\rm FB}}}
\def\AfbQ{\ensuremath{A^Q_{\rm FB}}}
\def\Afbq{\ensuremath{A^q_{\rm FB}}}
\def\Afbf{\ensuremath{A^f_{\rm FB}}}
\def\aeff{\ensuremath{A_\varepsilon}}
\def\apol{\ensuremath{A_{\cal P}}}
\def\aengy{\ensuremath{A_E}}
\def\aback{\ensuremath{A_{bkg}}}
\def\Rf{\ensuremath{R_f}}
\def\Rb{\ensuremath{R_b}}
\def\Rc{\ensuremath{R_c}}
\def\dmd{\ensuremath{\Delta m_d}}
\def\dmq{\ensuremath{\Delta m_q}}
\def\dms{\ensuremath{\Delta m_s}}
\def\blife{\ensuremath{\tau_b}}
\def\tauB{\ensuremath{\tau_B}}
\def\tauBp{\ensuremath{\tau_{B^+}}}
\def\tauBu{\ensuremath{\tau_{B_u}}}
\def\tauBz{\ensuremath{\tau_{B^0}}}
\def\tauBd{\ensuremath{\tau_{B_d}}}
\def\tauBs{\ensuremath{\tau_{B_s}}}
\def\tauLb{\ensuremath{\tau_{\Lb}}}
\def\chib{\ensuremath{\overline{\chi}}}
\def\chid{\ensuremath{\chi_d}}
\def\chis{\ensuremath{\chi_s}}

\def\alphas{\ensuremath{\alpha_s}}
\def\asmz{\ensuremath{\alpha_s(M^2_Z)}}
\def\avxe{\ensuremath{\langle x_E \rangle}}

\def\order{{\ensuremath{\cal O}}}
\def\L{{\ensuremath{\cal L}}}
\def\calL{{\ensuremath{\cal L}}}
\def\calS{{\ensuremath{\cal S}}}
\def\calA{{\ensuremath{\cal A}}}
\def\BR{{\ensuremath{\cal B}}}
\def\calP{{\ensuremath{\cal P}}}
\def\Pe{{\ensuremath{{\cal P}_e}}}
\def\avPe{\ensuremath{\langle{\cal P}_e\rangle}}
\def\eps{\varepsilon}
\def\gsim{{~\raise.15em\hbox{$>$}\kern-.85em
          \lower.35em\hbox{$\sim$}~}}
\def\lsim{{~\raise.15em\hbox{$<$}\kern-.85em
          \lower.35em\hbox{$\sim$}~}}
\def\qsq                {\ensuremath{q^2}}
\def\ra                 {\ensuremath{\rightarrow}}
\def\to                 {\ensuremath{\rightarrow}}

\def\sx    {\ensuremath{\sigma_x}}     
\def\sy    {\ensuremath{\sigma_y}}    
\def\sz    {\ensuremath{\sigma_z}}  

\newcommand{\inverse}{\ensuremath{^{-1}}}
\newcommand{\dedx}{\ensuremath{\mathrm{d}\hspace{-0.1em}E/\mathrm{d}x}}
\newcommand{\chisq}{\ensuremath{\chi^2}}
 
\newcommand{\lum} {\ensuremath{\mathcal{L}}}

\def\fluka      {\mbox{\tt Fluka}}
\def\fortran    {\mbox{\tt Fortran}}
\def\geant      {\mbox{\tt Geant}}
\def\gheisha    {\mbox{\tt Gheisha}}
\def\jetset74   {\mbox{\tt Jetset \hspace{-0.5em}7.\hspace{-0.2em}4}}
\def\jetset     {\mbox{\tt Jetset}}
\def\herwig     {\mbox{\tt Herwig}}
\def\koralz     {\mbox{\tt KoralZ}}
\def\bhlumi     {\mbox{\tt Bhlumi}}
\def\minuit     {\mbox{\tt Minuit}}
\def\squaw      {\mbox{\tt Squaw}}
\def\zvtop      {\mbox{\tt ZVTOP}}

\newcommand{\ace}       [3]  {{{\it Adv.\ Cry.\ Eng.}\ {#1}:{#2} ({#3})}}
\newcommand{\arnps}     [3]  {{{\it Annu.\ Rev.\ Nucl.\ Part.\ Sci.}\ {#1}:{#2} ({#3})}}
\newcommand{\arns}      [3]  {{{\it Annu.\ Rev.\ Nucl.\ Sci.}\ {#1}:{#2} ({#3})}}
\newcommand{\appopt}    [3]  {{{\it Appl.\ Opt.}\ {#1}:{#2} ({#3})}}
\newcommand{\seis}      [3]  {{{\it Bull.\ Seismological Soc.\ of Am.}\ {#1}:{#2} ({#3})}}
\newcommand{\epjc}      [3]  {{{\it Eur.\ Phys.\ J.}\ C{#1}:{#2} ({#3})}}
\newcommand{\ited}      [3]  {{{\it IEEE Trans.\ Electron.\ Devices}~{#1}:{#2} ({#3})}}
\newcommand{\itns}      [3]  {{{\it IEEE Trans.\ Nucl.\ Sci.}\ {#1}:{#2} ({#3})}}
\newcommand{\ijqe}      [3]  {{{\it IEEE J.\ Quantum Electron.}\ {#1}:{#2} ({#3})}}
\newcommand{\ijmp}      [3]  {{{\it Int.\ Jour.\ Mod.\ Phys.}\ {#1}:{#2} ({#3})}}
\newcommand{\ijmpa}     [3]  {{{\it Int.\ J.\ Mod.\ Phys.}\ {A{#1}:{#2} ({#3})}}}
\newcommand{\jl}        [3]  {{{\it JETP Lett.}\ {#1}:{#2} ({#3})}}
\newcommand{\jetp}      [3]  {{{\it JETP}~{#1}:{#2} ({#3})}}
\newcommand{\jpg}       [3]  {{{\it J.\ Phys.}\ {G{#1}:{#2} ({#3})}}}
\newcommand{\jap}       [3]  {{{\it J.\ Appl.\ Phys.}\ {#1}:{#2} ({#3})}}
\newcommand{\jmp}       [3]  {{{\it J.\ Math.\ Phys.}\ {#1}:{#2} ({#3})}}
\newcommand{\jmes}      [3]  {{{\it J.\ Micro.\ Elec.\ Sys.}\ {#1}:{#2} ({#3})}}
\providecommand{\josa}      [3]  {{{\it J.\ Opt.\ Soc.\ Am.}\ {#1}:{#2} ({#3})}}
\newcommand{\nim}       [3]  {{{\it Nucl.\ Instr.\ and Methods}~{#1}:{#2} ({#3})}}
\newcommand{\nima}      [3]  {{{\it Nucl.\ Instr.\ Methods A} {{#1}:{#2} ({#3})}}}
\newcommand{\np}        [3]  {{{\it Nucl.\ Phys.}\ {#1}:{#2} ({#3})}}
\newcommand{\npb}       [3]  {{{\it Nucl.\ Phys.}\ B{#1}:{#2} ({#3})}}
\newcommand{\npps}      [3]  {{{\it Nucl.\ Phys.\ Proc.\ Suppl.}\ {#1}:{#2} ({#3})}}
\newcommand{\npbps}     [3]  {{{\it Nucl.\ Phys.\ B Proc.\ Suppl.}\ {#1}:{#2} ({#3})}}
\newcommand{\ncim}      [3]  {{{\it Nuo.\ Cim.}\ {#1}:{#2} ({#3})}}
\newcommand{\optl}      [3]  {{{\it Opt.\ Lett.}\ {#1}:{#2} ({#3})}}
\newcommand{\optcom}    [3]  {{{\it Opt.\ Commun.}\ {#1}:{#2} ({#3})}}
\newcommand{\partacc}   [3]  {{{\it Particle Acclerators}~{#1}:{#2} ({#3})}}
\newcommand{\pflu}      [3]  {{{\it Physics of Fluids}~{#1}:{#2} ({#3})}}
\newcommand{\ptoday}    [3]  {{{\it Physics Today}~{#1}:{#2} ({#3})}}
\providecommand{\pl}        [3]  {{{\it Phys.\ Lett.}\ {#1}:{#2} ({#3})}}
\providecommand{\pla}       [3]  {{{\it Phys.\ Lett.}\ A{#1}:{#2} ({#3})}}
\providecommand{\plb}       [3]  {{{\it Phys.\ Lett.}\ B{#1}:{#2} ({#3})}}
\providecommand{\prep}      [3]  {{{\it Phys.\ Rep.}\ {#1}:{#2} ({#3})}}
\providecommand{\prl}       [3]  {{{\it Phys.\ Rev.\ Lett.}\ {#1}:{#2} ({#3})}}
\providecommand{\pr}        [3]  {{{\it Phys.\ Rev.}\ {#1}:{#2} ({#3})}}
\providecommand{\pra}       [3]  {{{\it Phys.\ Rev. A}\ {#1}:{#2} ({#3})}}
\providecommand{\prb}       [3]  {{{\it Phys.\ Rev. B}\ {#1}:{#2} ({#3})}}
\providecommand{\prd}       [3]  {{{\it Phys.\ Rev. D}\ {#1}:{#2} ({#3})}}
\providecommand{\pre}       [3]  {{{\it Phys.\ Rev. E}\ {#1}:{#2} ({#3})}}
\providecommand{\progtp}    [3]  {{{\it Prog.\ Theor.\ Phys.}\ {#1}:{#2} ({#3})}}
\providecommand{\rmp}       [3]  {{{\it Rev.\ Mod.\ Phys.}\ {#1}:{#2} ({#3})}}
\newcommand{\rsi}       [3]  {{{\it Rev.\ Sci.\ Instr.}\ {#1}:{#2} ({#3})}}
\newcommand{\sci}       [3]  {{{\it Science}~{#1}:{#2} ({#3})}}
\newcommand{\sjnp}      [3]  {{{\it Sov.\ J.\ Nucl.\ Phys.}\ {#1}:{#2} ({#3})}}
\newcommand{\spu}       [3]  {{{\it Sov.\ Phys.\ Usp.}\ {#1}:{#2} ({#3})}}
\newcommand{\tmf}       [3]  {{{\it Teor.\ Mat.\ Fiz.}\ {#1}:{#2} ({#3})}}
\newcommand{\yf}        [3]  {{{\it Yad.\ Fiz.}\ {#1}:{#2} ({#3})}}
\newcommand{\zpc}       [3]  {{{\it Z.\ Phys. C}\ {#1}:{#2} ({#3})}}
\newcommand{\zp}        [3]  {{{\it Z.\ Phys.}\ {#1}:{#2} ({#3})}}
\newcommand{\zpr}       [3]  {{{\it ZhETF Pis.\ Red.}\ {#1}:{#2} ({#3})}}
\newcommand{\cpc}   [3]  {{{\it Comput.\ Phys.\ Commun.}\ {#1}:{#2} ({#3})}}
\newcommand{\mpl}     [3]  {{{\it Mod.\ Phys.\ Lett.}\ {#1}:{#2} ({#3})}}
\newcommand{\fp}       [3]   {{{\it Fortschr.\ Phys.}\ {#1}:{#2} ({#3})}}
\newcommand{\cnpp}  [3]   {{{\it Comm.\ Nucl.\ Part.\ Phys.}\ {#1}:{#2} ({#3})}}
\newcommand{\rpp}   [3]    {{{\it Rep.\ Prog.\ Phys.}\ {#1}:{#2} ({#3})}}
\newcommand{\app}  [3]    {{{\it Acta Phys.\ Polon.}\ {#1}:{#2} ({#3})}}
\newcommand{\fizika}[3]   {{{\it Fizika}~{#1}:{#2} ({#3})}}
\newcommand{\spd}  [3]    {{{\it Sov.\ Phys.\ Dokl.}\ {#1}:{#2} ({#3})}}
\newcommand{\baps} [3]  {{{\it Bull.\ Am.\ Phys.\ Soc.}\ {#1}:{#2} ({#3})}}
\newcommand{\araa} [3]   {{{\it Ann.\ Rev.\ Astr.\ Ap.}\ {#1}:{#2} ({#3})}}
\providecommand{\apj} [3]     {{{\it Astro.\ Phys.\ J.}\ {#1}:{#2} ({#3})}}
\newcommand{\annp}  [3]  {{{\it Ann.\ Phys.}\ {#1}:{#2} ({#3})}}
\newcommand{\ppnp}  [3]  {{{\it Prog.\ Part.\ Nucl.\ Phys.}\ {#1}:{#2} ({#3})}}
\newcommand{\prsl}  [3]    {{{\it Proc.\ Roy.\ Soc.\ Lond.}\ {#1}:{#2} ({#3})}}
\newcommand{\apas}  [3]  {{{\it Acta Phys.\ Austr.\ Suppl.}\ {#1}:{#2} ({#3})}}
\newcommand{\anp}  [3]  {{{\it Adv.\ Nucl.\ Phys.}\ {#1}:{#2} ({#3})}}
\newcommand{\pan}  [3]  {{{\it Phys.\ Atom.\ Nuclei}~{#1}:{#2} ({#3})}}
\newcommand{\cmp} [3]  {{{\it Commun.\ Math.\ Phys.}\ {#1}:{#2} ({#3})}}
\newcommand{\npaps} [3]  {{{\it Nucl.\ Phys.\ A (Proc.\ Suppl.)}~{#1}:{#2} ({#3})}}

\newcommand{\revise} [2] {#2}

%% file: sec1_intro.tex
\section{INTRODUCTION}

During the 1990s, experimental particle physics underwent
a quiet revolution. Tests of  
electroweak physics in the standard model~\cite{ref:SMrefs}
to unprecedented precision, in some cases at the part-per-mil level or better, 
\revise{raised its status to}{established the standard model as} 
one of the most successful physical theories 
ever devised.  At the center of this activity, starting in 1989, 
were the experimental programs at the \Z\ boson resonance in \ee\
colliders at the Stanford Linear Accelerator Center, SLAC [using
the SLAC Linear Collider (SLC)],
and at the European particle physics laboratory, CERN [home to the
Large Electron Positron (LEP) storage ring].  We \revise{will}{}
review highlights of the physics program at the SLC, where a
longitudinally polarized electron beam proved to 
be a particularly powerful tool for \Z\ physics and was critical   
to an incisive experimental program.

The weak neutral current, and hence the existence of the \Z\ boson,
is perhaps the most significant prediction of the electroweak
standard model.  Maximal parity violation in charged-current
weak interactions, implying \revise{}{that} the $W$ boson couples
only to left-handed fermions, was
well established and was incorporated by the architects of the standard model. 
But unlike the charged $W$ boson, the neutral \Z\ boson couples with both
left- and right-handed fermions (with different strengths),
and the experimental determination  
of these couplings reveals much about the details of the theory.  
In addition, left-handed
and right-handed fermions carry different values of a new
quantum number, weak isospin, with value $I_W = 1/2$ for left-handed fermions
and $I_W = 0$ for right-handed fermions.
The left-handed fermions are grouped into weak-isospin doublets, where the
upper members (up, charm, and top quarks, and the three neutrinos in the lepton
case) carry a weak isospin third component $(I_W)_3 = +1/2$, and the lower
members (down, strange, and bottom quarks, and the charged leptons)
\revise{a}{carry the} value $(I_W)_3 = -1/2$.
All right-handed fermions are weak-isospin singlets.
A consequence of the model is that left- and
right-handed fermions behave as different particles, 
underlining the pivotal role of helicity.
The relative strengths of the left- and right-handed couplings to the \Z\
depend \revise{upon}{on} the electroweak mixing parameter, $\sin^2\theta_W$,
as follows:
\begin{equation}
g_L = (I_W)_3 - q \sin^2\theta_W  \ \ \ \ \ \ {\rm and}\ \ \ \ \ \
   g_R = q \sin^2\theta_W ,
\end{equation}
where $q$ is the electric charge. It is useful to define
the linear combinations given by
\begin{equation}
 g_V = g_L - g_R \ \ \ \ \ \ {\rm and } \ \ \ \ \ \ g_A = g_L + g_R ,
\end{equation}
where $g_V$ and $g_A$ then appear as the vector and axial-vector
couplings, respectively, in the neutral-current Lagrangian
density,
\begin{equation}
 { \cal L}_{\rm NC} \sim {\overline \psi}
   (g_V - g_A \gamma^5) \gamma^\mu \psi Z_\mu.
\end{equation}
Here $\psi$ and $Z_\mu$ are the fermion and \Z\ boson fields.
Note that the left-handed and right-handed terms are included, 
since $(g_V - g_A \gamma^5) = 
(g_V + g_A){1 \over 2}(1 - \gamma^5) + (g_V - g_A){1 \over 2}(1 + \gamma^5)$,
explicitly exhibiting the helicity projection operators.
The utility of a longitudinally polarized $e^-$ beam 
for the study of the weak neutral current is clear: With a
polarized beam, one can separately generate the two different
``particles'' $e^-_L$ and $e^-_R$, and directly measure their
couplings to the \Z\ boson.

The relatively large peak cross section
(30.5 nb for hadronic final states) at the \Z\ pole, 
the clean and well-controlled \ee\ environment, and the
universality of \Z\ boson interactions
(the \Z\ couples to all known fermions with comparable strength)
combine to make an ideal environment for a broad experimental program. 
As we will demonstrate, when beam polarization is available as
it was at the SLC,
one has a perfect laboratory for testing the standard model.
 
At the time of the 1988 International Conference on High Energy
Physics~\cite{ref:Munich88-Langacker},
direct and indirect data led to an uncertainty on
the \Z\ mass of 0.7 GeV, the limit on the number of light Dirac neutrinos was  
$N_\nu < 6.7$ at a 90\% confidence level (CL), the error on the electroweak
mixing parameter of the standard model was $\pm 0.0048$ 
from a combination of all available data,
the allowed range for the top quark
mass was about 50 to 200 GeV, and there was essentially no constraint at all
on the standard-model Higgs boson mass
($\sim 5$~GeV~$< m_{H} <$ ~$\sim 1$~TeV).

Contrast this situation with the present state
of affairs~\cite{ref:LEPEWWG2001}:
thanks to LEP
precision measurements of the \Z\ resonance, the \Z\ mass is known
to 2.1 MeV and the \Z\ ``invisible'' width determines
$N_\nu = 2.9841\pm0.0083$; the precise
measurement of \sinsqth, from the SLC Large Detector (SLD) 
experiment at SLAC and from the LEP experiments at CERN,
comes with an error of $\pm0.00017$.
Electroweak data can now be used to predict the top quark mass with remarkable
precision  ($m_{t} = 181^{+11}_{-9}$~GeV). 
In addition, the discovery of the top quark at Fermilab in 1994 and the
subsequent determination that $m_{t} = 174.3\pm5.1$~GeV~\cite{ref:FNALtop},
in excellent agreement with the current
standard-model prediction,
has greatly improved the predictive power of
the theory. The standard-model Higgs boson mass is now 
constrained by the worldwide
electroweak data to be $<$ 195 GeV 
at a 95\% CL, where \Z-pole weak mixing angle results are the leading 
contributors.
Direct searches performed at the higher-energy LEP-II machine constrain this
mass to be $>$ 114.1 GeV~\cite{ref:HiggsWG}---hence, 
the allowed window for the
standard-model Higgs has shrunk dramatically.
\clearpage
  
The \ee\ \Z-pole programs also contributed to significant advances in
heavy-flavor (bottom and charm) physics, in particular at the SLC.
The unique features of SLC operation led to 
superlative decay-vertex reconstruction in the SLD detector and
to highly efficient identification of the short-lived bottom and charm
hadrons. Thanks to the polarized electron beam, 
the SLC program provides the only direct measurements
of parity violation in the bottom and charm neutral-current 
couplings.  Heavy-flavor data from LEP, along with current data on the
leptonic neutral-current couplings (coming from SLC and LEP), can be combined
with the SLD results.  The resulting 
uncertainties are 2.4\% and 1.0\% for the vector and
axial-vector bottom couplings, and
3.7\% and 1.1\% for the vector and axial-vector charm couplings.
The corresponding results from the 1988 International
Conference were one to two orders of 
magnitude less precise; for example, about 260\% and 13\%
for bottom vector and axial-vector couplings, respectively.

The most impressive marriage of all of the special features of \Z\ physics,
linear collider
operation, sophisticated detector instrumentation, and beam polarization 
is evident in the study of \bsmix\ mixing at the SLC.  
These measurements are sensitive to the details of quark mixing, which are
parameterized
by the Cabibbo-Kobayashi-Maskawa
(CKM) mixing matrix~\cite{ref:CKM} in the standard model. These data
significantly constrain one of the least-known elements
of the CKM matrix and thereby
contribute to our understanding of $CP$ violation.
The \bsmix\ mixing measurements were also performed at LEP with
comparable precision owing to the much larger event totals
of these experiments. The 1988 International 
Conference predated these
first successful \bsmix\ mixing results, but it is safe to say that
the quality of the initial- and final-state heavy-flavor identification
and the decay vertex resolutions achieved at the \Z, particularly at the SLC,
were not anticipated. 
  
The SLD collaboration collected data at the SLC 
during the period 1992--1998.
Table \ref{tab:data-samples} shows a breakdown of the data samples,
along with the average 
electron beam polarization at the interaction point.
\begin{table}[tbp]
\caption{SLD data samples for the various data-taking periods}
\vspace{0.3 cm}
\label{tab:data-samples}
\begin{center}
\begin{tabular}{@{}cccccc@{}}%
\toprule
 Year           &  1992 & 1993  & 1994--1995 & 1996  & 1997--1998  \\
\colrule
\Ztohad\ events &  10K  &  49K  &    94K     &  52K  &    332K    \\
Polarization    &  22\% &  63\% &   77\%     &  76\% &    73\%    \\
\botrule
\end{tabular}
\end{center}
\end{table}
Since 1996, data were recorded with the upgraded vertex detector
VXD3 (see Section~\ref{sec:sld-tracking}).
In what follows, we summarize
the highlights of the physics program
in precision electroweak measurements and heavy-flavor physics.
The reader is referred to published work for studies
of $\tau$~\cite{ref:TAUatSLD} and QCD~\cite{ref:QCDatSLD} physics at SLD.

%% file: sec2_slc.tex
\section{THE SLAC LINEAR COLLIDER}
\label{sec:slc}

The SLC is, so far, the only
$\ee$ linear collider. Consequently,
the SLC/SLD program was both a particle physics experiment and an
accelerator R\&D project, and key features of these parallel
activities were closely intertwined.  
A linear collider is a ``single-pass'' device---the \ep\ and \en\ bunches are
generated anew for every single collision---and hence the operation of
the SLC introduced technical challenges 
in the control of bunch trajectory, intensity, energy, and focusing 
not common to the more forgiving storage ring design.  
Commissioning of the machine began in 1987 and lasted
about two years.  The SLC luminosity
was initially far below expectations, although the brief 
Mark II physics program (1989--1990)~\cite{ref:markII} produced some
important early results before being eclipsed by the much higher-luminosity
LEP $\ee$ storage ring.  A breakthrough for the SLC
came in 1992, when a longitudinally polarized ($\sim 25\%$) electron beam 
first became available.  Because of advances in photocathode
technology, the electron beam polarization was increased to 
$\sim 75\%$ by 1994. Over the years, the luminosity of the machine 
improved steadily, most dramatically during the final run (1997--1998),
when a peak of about $3 \times 10^{30} {\rm ~cm}^{-2}{\rm s}^{-1}$
was reached and roughly 70\% of the entire SLD dataset was accumulated.

The cornerstone of the SLC program at the \Z\ pole was 
the polarized electron beam, which proved essential not only
for precision electroweak measurements but for most heavy-flavor
physics as well.
The basic operation of the SLC is described below, where we 
highlight features most relevant to the SLD physics,
particularly polarized beam production and transport.
More complete descriptions are
available in an earlier review~\cite{ref:ARNPPseeman} and in
a more recent technical report~\cite{ref:phinney}. 

\subsection{SLC Operation with Polarized Electron Beams}

\begin{figure}[tbp]
\begin{center}
  \epsfxsize16cm
  \epsfbox{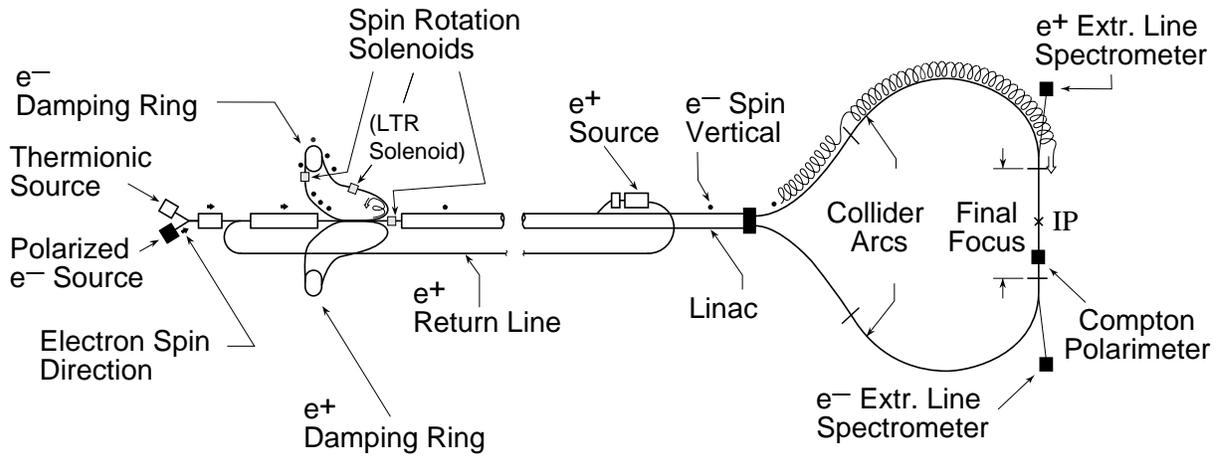}
\end{center}
\vspace{-0.5cm}
\caption{Layout of the SLC.}
\label{fig:slc}
\end{figure}

Figure~\ref{fig:slc} illustrates the basic layout of the SLC.
The machine produced $\ee$ collisions at a repetition rate of 120 Hz.
Longitudinally polarized electrons were produced by illuminating a GaAs
photocathode with a circularly polarized Ti-Sapphire laser~\cite{ref:PLS}.
Since the advent of high-polarization ``strained lattice'' GaAs
photocathodes in 1994~\cite{ref:sld-strain}, where mechanical strain
induced in a thin (0.1 \mum) GaAs layer lifts an angular-momentum degeneracy
in the valence band of the material,
the average electron polarization at the $\ee$ interaction point
(slightly lower than the value produced at the source)
was in the range 73\% to 77\%.
Corresponding values were about 22\% in 1992 using an unstrained
``bulk'' GaAs cathode, and 63\% in 1993 using a thicker (0.3 \mum) strained
layer cathode design. 
The electron helicity
was chosen randomly pulse to pulse at the machine repetition rate
of 120 Hz by controlling the circular polarization of the source laser.
This pulse randomization minimized unwanted correlations between
the pulse helicity and other periodic phenomena in the accelerator, 
as well as in the SLD and polarimeter data.  
On each SLC cycle, two closely spaced (61 ns) electron bunches were
produced, the second of which was used to produce the positrons.

The electron spin orientation 
was longitudinal at the source and remained longitudinal until the
transport to the damping ring (DR).
In the linac-to-ring (LTR) transport line, the electron spins precessed
in the dipole magnets, where
the spin precession angle is given in terms of the anomalous
magnetic moment $g$:
\begin{equation}
 \theta_{\rm precession} = \left({g-2 \over {2} }\right)
 {E \over m}\: \theta_{\rm bend}.
\end{equation}
By design, the bend angle $\theta_{\rm bend}$ resulted in 
transverse spin orientation at the entrance to the LTR spin
rotator magnet.  This superconducting
solenoid magnet was used to rotate the polarization 
about the beam direction
into the vertical orientation for storage in the DR.
This was necessary because any horizontal spin components
precessed rapidly and net horizontal polarization
completely dissipated during the 8.3 ms (1/120 s)
storage time owing to energy spread in the bunch.
In addition, the
polarized electron bunches could be stored in one of two possible
configurations by reversing the LTR spin
rotator solenoid magnet.
These reversals, typically done every three months, were useful for
identifying and minimizing the small [${\cal O}(10^{-4})$]
polarization asymmetries produced at the source. 

The product of the horizontal and vertical
emittances of the electron (positron) bunches, a measure
of the transverse phase-space of the beam, was
reduced in the DR by a factor of roughly 100 (1000).
One of two stored positron bunches was
extracted from the positron DR and led 
the two electron bunches down the linac,
the bunches spaced by 61 ns.  
The second electron ``scavenger'' bunch was diverted
to the positron target after 30 
GeV of acceleration. Positrons produced at the target were accelerated
to 200 MeV for transport via the positron return 
line to the positron DR, where 
they were stored for 16.6 ms.    
The remaining $e^-$ and $e^+$
bunches continued down the linac and reached $\sim 46.5$ GeV.
They were then transported through the arcs 
(energy loss of about 1 GeV occured in the arcs owing to synchrotron
radiation) and final focus to the interaction point (IP).  
Following the IP, the bunches entered the extraction lines
where their energies were measured by precision spectrometers,
after which they were transported to beam dumps.  

After leaving the DR, the electrons were accelerated in the linac
with their polarization oriented vertically,
and brought into the SLC north arc.
The electron spin orientation was manipulated in the
north arc (see Figure~\ref{fig:slc}) using controllable betatron
oscillations (known as
``spin bumps'') to achieve longitudinal polarization at the
IP.  This was \revise{made possible by the
fact that}{possible because} the betatron phase advance closely
matched the spin precession
(1080$^\circ$ and 1085$^\circ$, respectively)
in each of the 23 bending-magnet assemblies (``acromats'') used in the arc 
---and hence the north
arc operated close to a spin-tune resonance.  As a result,
excepting 1992 running,
the two additional SLC spin rotator solenoids 
were not necessary for
spin orientation and were used only in
a series of specialized polarization experiments.
This unanticipated simplification had an additional benefit. In order
to achieve higher luminosity, starting in 1993 the SLC was operated with 
``flat beams'' ($\sigma_x > \sigma_y$), a setup that would have been precluded
by the spin rotators \revise{due to}{because} $x$-$y$ coupling in the 
ring-to-linac spin rotator would
have spatially rotated the electron bunches. 

\subsection{Energy Spectrometers}
\label{sec:sld-espec}

The SLC employed a pair of energy spectrometers~\cite{ref:sld-espec}
located in the
electron and positron extraction lines (see Figure~\ref{fig:slc}).  
Beam deflection, and therefore beam energy, 
was measured from the spatial separation of synchrotron radiation
emitted by the beam in deflector magnets located before
and after a precision dipole magnet.  These two spectrometers
together nominally
measured the $\ee$ collision energy with a precision of 20 MeV.
The importance of the energy measurement and some details 
concerning spectrometer precision are
discussed in Section~\ref{sec:phys-ewlepton}.

\subsection{Luminosity Improvements}

The luminosity $\cal L$ of a linear collider is given by
\begin{equation}
 {\cal L} = {{N^+ N^- f} \over{4 \pi \sigma_x \sigma_y}} H_d,
 \label{eq:lumi}
\end{equation}
where $N^\pm$ are the number \revise{}{of} positrons and electrons
in each bunch at the IP, $f$ is the collision repetition rate, and
$\sigma_{x,y}$ are
the average horizontal and vertical beam sizes. The additional
parameter $H_d$ is a disruption
enhancement factor that is a function of $N^\pm$ and the transverse
and longitudinal beam dimensions, and for moderate beam intensities $H_d$
is equal to one.
 
The route to high luminosity at the SLC was somewhat different than
originally conceived, with a greater emphasis on reduced beam size 
(design performance was exceeded by about a factor of three) 
because the maximum bunch
size was limited to about $4 \times 10^{10}$ due to wakefield effects and 
DR instabilities ($7 \times 10^{10}$ was indicated in the initial plans).
In the final focus, the $\ee$ transverse bunch dimensions were
demagnified to about 1.5 $\mu$m in $x$ and 0.7 $\mu$m in $y$
for collision at the IP. 
\begin{figure}[tbp]
\begin{center}
  \epsfxsize16cm
  \epsfbox{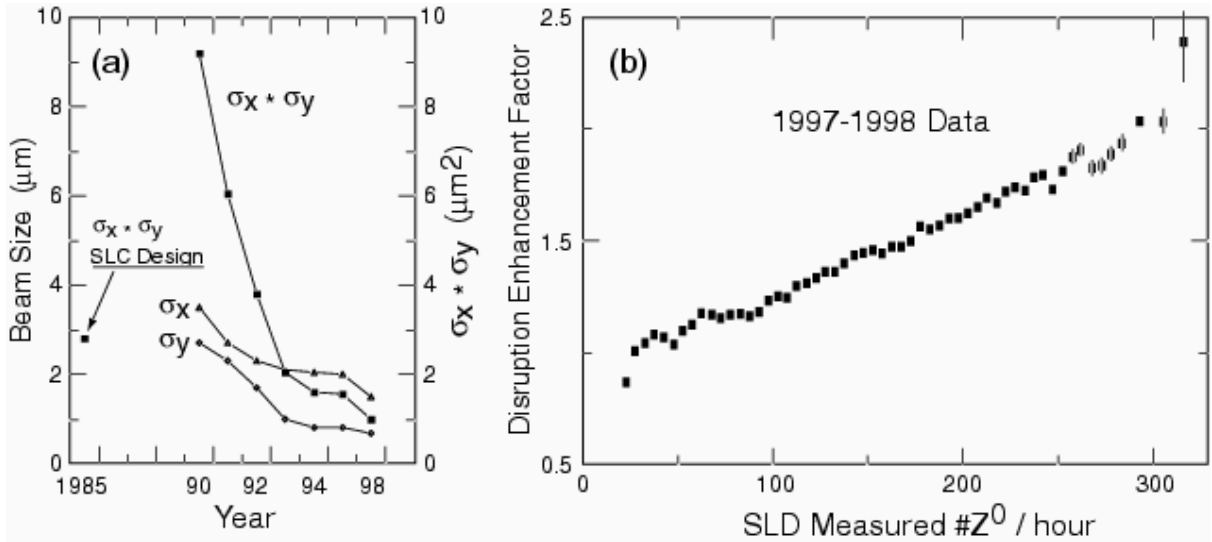}
\end{center}
\vspace{-0.5cm}
\caption{($a$) Horizontal ($\sigma_x$) and vertical ($\sigma_y$)
 beam spot sizes, as well as the cross-sectional area
 ($\sigma_x\cdot\sigma_y$) of the beam as a function
 of time. The original design value for the area is also shown.
 ($b$) Luminosity enhancement due to beam-beam disruption.}
\label{fig:spotsize}
\end{figure}
Figure~\ref{fig:spotsize}$a$ shows the historical trend in the size of the
SLC beams.  A consequence of the very small SLC beam sizes at the IP
was that a significant luminosity enhancement due to beam-beam disruption
($H_d > 1$) was observed at the highest luminosities.
From a comparison between the luminosity measured with the SLD luminosity
monitor and that computed from Equation~\ref{eq:lumi}
with $H_d~=~1$,
enhancements of up to a factor of two
were seen, as shown in Figure~\ref{fig:spotsize}$b$. 

  Though not directly relevant to our discussion of
the SLD physics
program, one important feature of SLC operation deserves mention.
Producing and maintaining the small beam emittances required
for these very small spot sizes, and sustaining their collisions, all
on a pulse-to-pulse basis, was challenging. An important lesson
was that a linear collider is an inherently unstable machine 
and extensive feedback systems are indispensable.  
In particular, the importance of fast feedback (at the 120 Hz
machine repetition rate) became apparent. 
By the end of the SLD
program in 1998, the SLC featured over 50 feedback systems controlling
250 machine parameters.  These features included intensity feedback
at the source,
energy feedback, linac orbit feedback, and an elaborate collision
optimization system that modulated five final focus corrections for
each beam while monitoring a direct measure of luminosity (beamstrahlung).

\subsection{Impact on the SLD Physics Program}

Three features of the SLC contributed to the unique capabilities
of the SLD physics program.  Most important was the polarized electron
beam.  The SLC achieved a precise determination of the weak mixing
angle with more than 30 times the sensitivity per \Z\ event of any competing
method at LEP
and with smaller systematic error. The beam also
enabled a number of the world's
best measurements in heavy-quark electroweak 
\revise{physics due in part to}{physics, partly through} the
greatly enhanced statistical power and simplicity
afforded by polarization. \revise{Secondly,}{Second,}
the small size of the luminous region at the collision point, smaller
in the transverse ($x$,$y$) and longitudinal ($z$) dimensions than 
at LEP by factors of typically 100, 4, and 10, respectively, allowed for
excellent localization of the primary vertex in reconstructed \Z\ decays.
Finally, the low repetition rate of the SLC (120 Hz compared with 45 kHz
at LEP), a liability insofar as luminosity is concerned, permitted the
use of charge-coupled device (CCD) sensors with inherently long readout times
in the SLD vertex detector, a device that subsequently set
the world standard for precision in decay-length reconstruction.
All these issues are addressed 
in more detail in our discussion of
physics results.

%% file: sec3_sld.tex
\section{THE SLD DETECTOR}
\label{sec:slddet}

\subsection{Overall Description}
\label{sec:sld-gen}
\input sec3_gen.tex 

\subsection{The Tracking and Vertexing System}
\label{sec:sld-tracking}
\input sec3_tracking.tex

\subsection{The Cherenkov Ring Imaging Detector}
\label{sec:sld-crid}
\input sec3_crid.tex


\subsection{Polarimetry}
\label{sec:sld-pol}
\input sec3_pol.tex


%% file: sec3_gen.tex

\begin{figure}[tbp]
\begin{center}
  \epsfxsize12cm
  \epsfbox{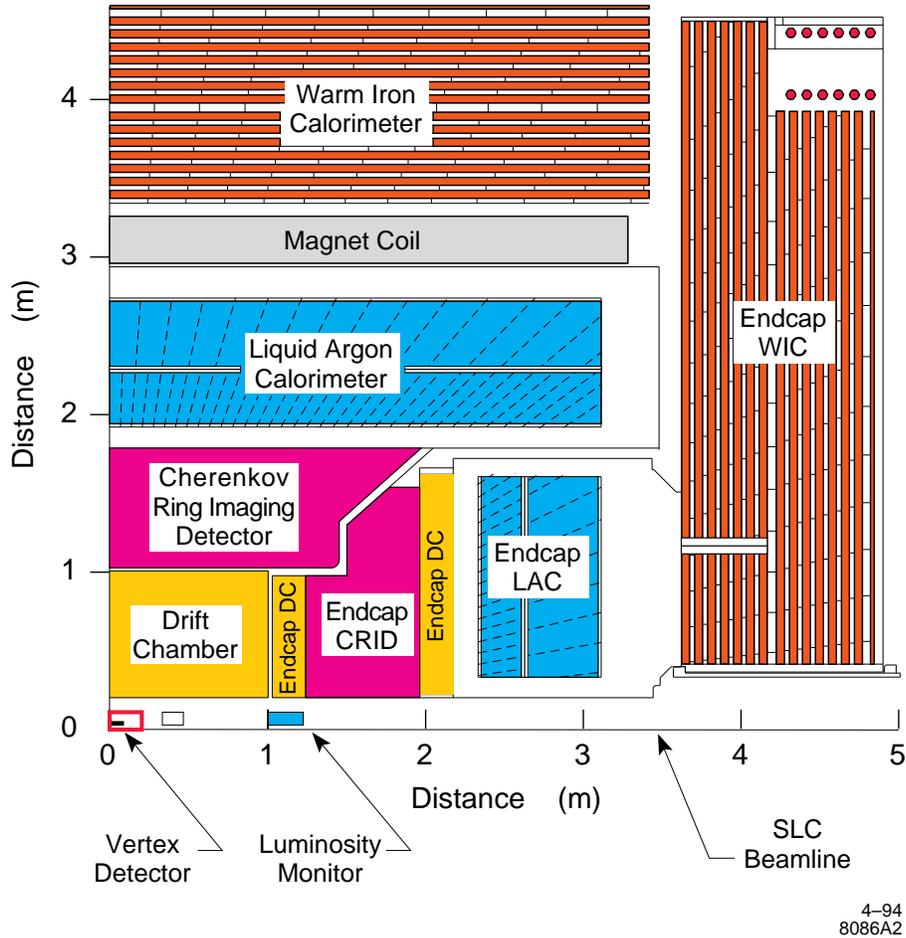}
\end{center}
\vspace{-0.8cm}
\caption{Side view of one quadrant of the SLD detector.}
\label{fig_SLDquad}
\end{figure}

The SLD was a general-purpose device~\cite{SLDDR}
covering most of $4\pi$ sterradians (sr).
It was optimized for physics at the \Z\ resonance,
i.e., it was designed for the reconstruction and identification of particles
in a momentum range \revise{between}{of} several hundred \mevc\ up to nearly 50 \gevc.
The detector consisted of multiple concentric layers
(see Figure~\ref{fig_SLDquad}).
In the following, we refer to a right-handed coordinate system
with a $z$-axis along the \ep\ beam direction and
a $y$-axis along the vertical direction.

Precision charged-particle tracking was performed with
the CCD pixel Vertex Detector (VXD) surrounded by the
Central Drift Chamber (CDC).
Outside the tracking systems,
the Cherenkov Ring Imaging Detector (CRID) identified the
charged hadrons $\pi$, $K$, and $p$,
and the Liquid Argon Calorimeter (LAC) provided electromagnetic
and hadronic calorimetry as well as electron identification
(and muon identification to a lesser extent).
These systems were contained inside a large solenoid producing
a 0.6~T axial magnetic field.
Finally, the Warm Iron Calorimeter (WIC) served as the
magnet flux return yoke and identified muons.
In the following, we describe the calorimetry systems and devote separate
sections to the tracking, particle identification,
and polarimetry systems.

The main calorimeter was the lead--liquid~argon 
sampling calorimeter \cite{LAC},
which satisfied the design requirements for hermiticity (it covered 98\% of
$4\pi$~sr), good energy and position resolution, and
good uniformity of response to electromagnetic and hadronic showers.
To provide information on longitudinal shower development,
the LAC was divided radially into two separate electromagnetic sections
totaling 21 radiation lengths
and two hadronic sections totaling 2.8 interaction lengths.
Energy resolution was measured to be
$\sigma / E =$ $0.15 / \sqrt{E ({\rm GeV})}$ for electromagnetic showers
and estimated to be
$\sigma / E =$ $0.60 / \sqrt{E ({\rm GeV})}$ for hadronic showers.
The LAC was placed inside the magnet coil to minimize the amount of
material in front of the calorimeter, thus achieving good resolution
down to low energies.
Finally, the LAC was used to trigger the data acquisition system and,
in conjunction with the drift chamber, to identify electrons.

Outside the magnet coil stood the WIC~\cite{WIC},
a coarsely sampled iron-gas calorimeter operated in limited streamer mode,
which was designed to catch the remaining $\sim 15\%$
of hadronic energy leaking out from the LAC.
Its main use, however, was as a muon tracking and identification device.
The material in the WIC amounted to four interaction lengths,
which, in addition to the 2.8 (0.7) interaction lengths from
the LAC (magnet coil), gave a total of over seven interaction lengths
for particles traversing the whole detector from the IP.

A third calorimeter system~\cite{LUM}, made of tungsten and silicon pads,
detected electromagnetic showers at small angle with respect to
the beam line: $28 < \theta < 68$ mrad. The purpose of this system
was to monitor the SLC luminosity.

%% file: sec3_tracking.tex

\subsubsection{The Central Drift Chamber}

The CDC~\cite{ref:CDC} had a cylindrical geometry with a length of 
1.8\m,
extending radially from 0.2\m\ to 1.0\m. It consisted of ten
superlayers, four coaxial to the beam,
interleaved with pairs of superlayers 
having stereo angles of $\pm$41 mrad. 
Each superlayer consisted of cells $50\mm$ along the radius ($r$) and 
$\sim 59\mm$ wide in azimuth ($\phi$) at the midpoint. Each of the cells 
contained eight sense wire layers. An average spatial resolution 
of 82\mum\ was 
obtained with a gas of CO$_2$~:~argon~:~isobutane 
in the ratio of 
\mbox{75\%~:~21\%~:~4\%}, respectively. The azimuthal and polar angle
resolutions on the track direction at the inner radius of the CDC were 
0.45 mrad and 3.7 mrad, respectively, for high-momentum tracks.
The momentum resolution of combined CDC$+$VXD tracks in the central 
region ($|\ctheta|<$0.7) was
$\frac{\sigma_{\pxy}}{\pxy}=0.010\oplus 0.0024\:\pxy$,
where $\pxy$ is the momentum transverse to the beam 
axis, measured in \gevc.

\subsubsection{The CCD Vertex Detector}

The detection of short-lived particles 
using vertex detectors has played a vital role in the advance of our 
understanding of flavor physics. In the LEP/SLC era in particular, 
the usage of vertex detectors extended beyond the traditional 
field of heavy-flavor decay physics and became an important tool 
for electroweak and QCD physics, as well as for new particle searches.      
The low beam-crossing rate of the SLC compared \revise{to}{with} the 
circular colliders offered an opportunity for using
\revise{Charge-Coupled Devices (CCD)}{CCDs} as the basic vertex detector elements.
Some of the main advantages over silicon-strip vertex detectors 
employed at LEP are as follows: 
\begin{itemize}
\item The SLD vertex detector was a factor of two closer to the beam line, 
      which was crucial for reducing track errors due to extrapolation 
      and multiple scattering. 
\item The genuine three-dimensional nature of CCD pixel hits
      led to fewer tracking hit misassignments.  In contrast,
      double-sided silicon-strip detectors yield uncorrelated $\phi$ and $z$ 
      hits, which result in a quadratic increase in hit 
      ambiguity.
\item The spatial resolution provided by the SLD CCD pixel segmentation was 
      better by a factor of two. 
\item The active silicon layer of the CCDs was typically 15 times 
      thinner than that in silicon-strip detectors.
      This feature largely prevented the 
      drastic $z$-resolution degradation at shallow polar angles that
      is inevitable for silicon strips. It also allowed for
      thinner detectors to reduce multiple scattering.     
\end{itemize}     
      
\begin{figure}[tbp]
\begin{center}
  \epsfxsize16cm
  \epsfbox{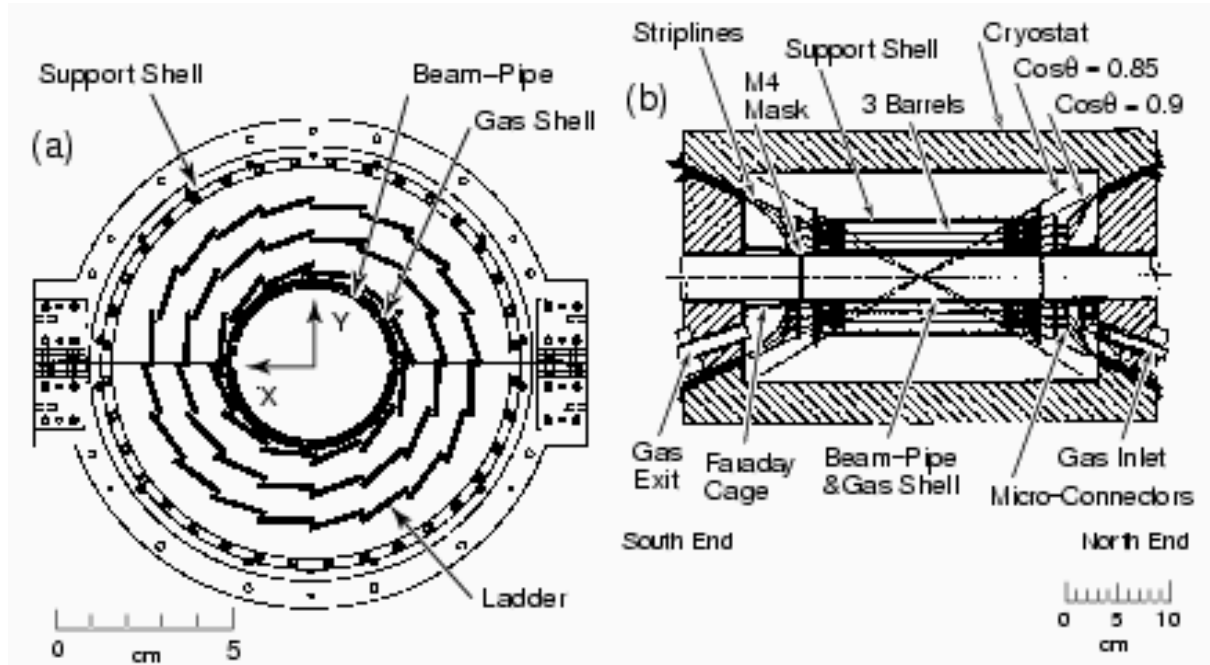}
\end{center}
\vspace{-0.5cm} 
\caption{Layout of VXD3 and mechanical support structure showing 
         ($a$) cross-section view; ($b$)~longitudinal section view.}
\label{fig:VXDgeom}
\end{figure}

The first version of the SLD CCD vertex detector (VXD2)~\cite{ref:VXD2}
was installed in late 1992. Its reliable operation and 
impressive performance inspired the design of an upgraded vertex 
detector (VXD3)~\cite{ref:VXD3}, with a more optimized
geometry, improved coverage, and reduced material. 
This detector was operational starting in 1996 and hence for the 
majority of the SLD data sample. VXD3 consisted of
48 ``ladders,'' each containing two \mbox{8$\times$1.6 cm} custom-made 
CCDs. The ladders were arranged in \revise{3}{three} cylindrical layers (see 
Figure~\ref{fig:VXDgeom}) at radii of 2.7, 3.8, and 4.8\cm\, respectively,
covering $|\ctheta|<$0.85 with three layers and $|\ctheta|<$0.90
with two layers. 
The entire detector was supported by a 
beryllium structure and operated at 190~K using nitrogen gas cooling.    
The CCDs were thinned to 200\mum\ and 
supported by a beryllium motherboard so that the material per layer
for normal incidence was 0.40\% of a radiation length. The 
beryllium beam pipe and inner gas shell added 0.52\% radiation length
of material inside the VXD.         
 
Each VXD3 CCD contained 800$\times$4000 pixels that were 20\mum$\times$20\mum\ 
in size, and the active silicon depth was also 20\mum. VXD3 was 
continuously read out at a rate of 5\mhz\ and the entire detector 
took 200\ms\ (25 SLC beam crossings) to read out. The electronic 
noise per pixel was 55 electrons, compared with the most probable
energy loss of 1200 electrons for minimum ionizing particles,  
typically spread over 4--5 pixels. The cluster centroid positions 
were calculated using the measured ionization charge accumulated 
in each pixel. The average single-hit efficiency was over 99.5\% for 
the data taken since 1997; the remaining inefficiency mainly 
came from a small number of faulty electronic channels during 
limited time periods. The average efficiency
for the 1996 data sample was $\sim$2\% worse because of
early electronics problems and radiation damage effects, which were
cured in 1997--1998 by operating at a lower temperature. 

The VXD3 geometry was designed with careful consideration for 
track-based alignment, with CCD overlaps in $z$ on the same ladder 
and in $\phi$ between adjacent ladders in the same layer. The CCD
shape distortions were first corrected based on optical survey data. The 
CCD positions, fully described by three translations and three rotations, 
were subsequently determined using a track-based internal alignment.  
This alignment also included long-range constraints from 
$\Ztoe$ and $\Ztomu$ events. The average 
single-hit spatial resolution was 3.5\mum\ in the $r$-$\phi$ plane for 
all polar angles and 3.7\mum\ (5.7\mum) in the $r$-$z$ plane for 
tracks with $|\ctheta|<0.6$ (0.6$<|\ctheta|<$0.85). The resolution 
degradation in $z$ at high $|\ctheta|$ was partly due to some
residual errors in the CCD shape distortion corrections.

The central beam pipe and VXD assembly was mechanically 
supported from the CDC. Its mechanical coupling to the rest of the 
beam line was reduced by a factor of ten with a set of bellows separating
out the central section. This loose coupling provided a very stable relative
position of the VXD with respect to the CDC, at the level of a few 
microns, during long periods of operation. Global alignment between 
VXD and CDC was necessary only after major detector accesses. 

\subsubsection{Tracking Performance}

The tracking algorithm first reconstructed three-dimensional track 
segments in the VXD with three or more hits and 
$\pxy>$ 80\mevc. These VXD track segments were then extrapolated
to the CDC to search for CDC hits to form tracks. 
On the second pass, the tracks found with the remaining CDC hits
not used by the \revise{first pass tracks,}{first-pass tracks} 
were extrapolated into the VXD to search for \revise{hits,
demanding at least 2 VXD hits.}{hits; at least two VXD hits were required.} 
All pattern-recognized tracks were subject to a Kalman fit 
\cite{ref:Kalman} to determine the track parameters, taking into account 
multiple scattering effects. 
The SLC beam caused a VXD background hit density of 2--7 
hits/\cm$^2$ depending on the CCD layer and $\phi$ position. This 
occupancy turned out to be quite manageable thanks to the extremely
fine granularity of the CCD pixels.
For ``prompt'' tracks in hadronic events, the CDC$+$VXD full tracking 
efficiency was 96\%. Prompt tracks are defined as 
tracks that originate from the IP or from heavy hadron decays
with $\pxy>400$ \mevc\ and $|\cos\theta|< 0.85$, and neither decayed nor 
interacted before exiting the CDC.
 
Among the prompt tracks with full CDC$+$VXD information, 97.8\% 
had three or more VXD hits,
and 99.8\% had all VXD hits correctly assigned.
The VXD track segments without a CDC link were used to improve 
vertex charge determination. Despite the rather short radial 
span ($\sim 2$ cm), the charge was correctly determined for 90\%
of the VXD track segments up to 1.5\gevc\ (3.5\gevc) without (with)
a vertex constraint.
 
\begin{figure}[tbp]
\vspace{-1.2 cm}
\begin{center}
  \epsfxsize15cm
  \epsfbox{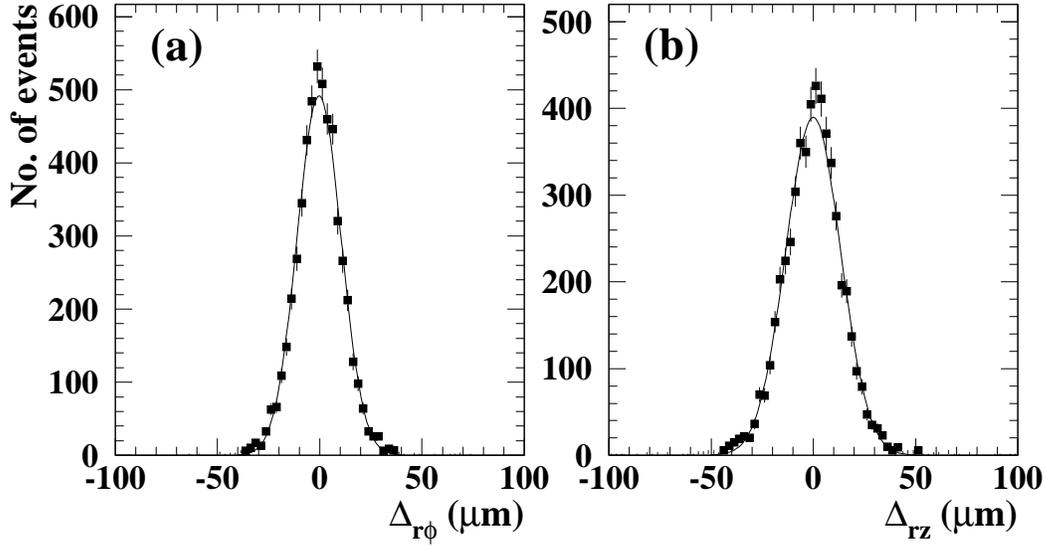}
\end{center}
\vspace{-0.8cm} 
\caption{Two-track miss-distance distributions for \Ztomu\ events:
         ($a$) $r$-$\phi$ miss-distance; ($b$) $r$-$z$ miss-distance.}
\label{fig:2prong-miss}
\end{figure}

The track impact parameter resolution can be roughly 
parameterized as 
$\sigma_{r\phi}=7.7\oplus\frac{33}{p \sin^{3/2}\theta}$~\mum\
and  
$\sigma_{rz}=9.6\oplus\frac{33}{p \sin^{3/2}\theta}$~\mum,
where $p$ is the track momentum, measured in \gevc.
The constant terms were determined from a study
of two-track miss-distance distributions near the IP
in \Ztomu\ events, see Figure~\ref{fig:2prong-miss}.
This resolution performance is a factor of two better 
than the best LEP detectors in both the constant and multiple
scattering terms.      

\subsubsection{The SLC Interaction Point}

To achieve the best identification of secondary vertices from
heavy-flavor decays, the primary vertex resolution is crucial. 
To enable the micron-sized SLC beams 
to collide, it was essential to maintain a stable
IP. For this reason, the average beam position 
in the $x$-$y$ plane is a very good estimate of the event primary vertex.
In the \epem\ storage ring experiments,
the beam spot in $y$ is also fairly narrow (a few to 10\mum),
but the beam size in $x$ is typically 100\mum\ or more,
so that event-by-event reconstruction of the primary vertex
$x$ position was required.  
This typically resulted in a primary vertex $x$ resolution
$> 20$ \mum\ for LEP experiments.
The average SLC beam position in the $x$-$y$ plane was determined 
from $\sim$30 consecutive hadronic events by fitting all tracks 
coming from a common origin. The resolution of the beam 
position determination was found to be $\sigma_{xy} =3.5$~\mum\,
using the impact parameter distribution of tracks in
\Ztomu\ events.
This resolution was the result of the combined effect 
of beam jitter and tracking resolution. 
  
The event primary vertex in $z$ was measured
event by event owing to the large luminous region along the beam axis
($\sim$700\mum). The $z$ position of the primary vertex
was determined from the median of the $z$-coordinate distribution
for all tracks consistent with the IP in the $x$-$y$ plane.
This simple approach was found to be particularly
stable against track impact parameter resolution tails.
The event primary vertex $z$ resolution for ($uds,c,b$) events was 
estimated to be (10,11,17)\mum.

%% file: sec3_crid.tex

  The CRID covered the region $|\cos\theta| < 0.68$.
To perform $\pi/K/p$ separation over the full
momentum range of interest ($\sim$0.2 to 45 \gevc),
two different radiators were used~\cite{CRID}.
The liquid $C_6 F_{14}$ radiator identified particles in the
lower portion of the momentum range, whereas the gas
$C_5 F_{12} + N_2$ radiator covered the higher end of the range.
Thresholds for emission of Cherenkov radiation in the liquid (gas) were
0.17 (2.4) \gevc\ for $\pi^\pm$,
0.62 (8.4) \gevc\ for $K^\pm$,
and 1.17 (16.0) \gevc\ for protons.
Cherenkov photons emitted in a cone around the charged-particle trajectory
were focused onto an array of 40 time projection chambers (TPCs)
filled with ethane and $\sim 0.1\%$ tetrakis dimethylamino ethylene
(TMAE) which served as the photosensitive agent.
The three-dimensional location of each photoelectron conversion
was recorded in the TPCs and
the production angle relative to a given charged track was reconstructed.
For typical $\beta \simeq 1$ particles passing through the
liquid (gas) radiator, an average of 12.8 (9.2) photons
were detected per full ring and the Cherenkov angle resolution was
measured to be 16 (4.5) mrad.

%% file: sec3_pol.tex

In Compton scattering of longitudinally polarized electrons
from circularly polarized photons,   
the differential cross section in terms
of the normalized scattered photon energy fraction $x$ is given
by
\begin{equation}
 {{d \sigma}\over{dx}} = {{d \sigma_0}\over{dx}}
   [1 - {\cal P}_\gamma {\cal P}_e A(x)],
\end{equation}
where ${d \sigma_0}\over{dx}$ is the unpolarized differential cross section,
${\cal P}_\gamma$ and ${\cal P}_e$ are the photon and electron polarizations,
and
$A(x)$ is the Compton asymmetry function.  The asymmetry arises
from the difference between cross sections for parallel and anti-parallel
spins ($\sigma_{j = 3/2} > \sigma_{j = 1/2}$), and is defined in the usual way
as the cross-section difference divided by the sum of the two cross sections. 
In a polarimeter, the Compton-scattered photons or electrons are
detected, and the requisite instrumental effects are incorporated into
a detector response function.  The normalized convolution of 
$A(x)$ with ${d \sigma_0}\over{dx}$ and the
response function (all functions of $x$) is the ``analyzing power.'' 
When ${\cal P}_\gamma$ and the analyzing power are
known, the experimental determination of the $j = 3/2$ to
$j = 1/2$ scattering asymmetry determines ${\cal P}_e$ and 
hence the utility of this
elementary QED process to electron polarimetry.

The SLD precision Compton polarimeter detected beam electrons
that had been scattered by photons from a circularly polarized laser.
The choice of a Compton-scattering polarimeter was dictated by the requirements
that the device be operated continually while beams were in collision
and that uncertainties in the physics of the
scattering process not be a limiting factor in the systematic error---both are
troublesome issues for M\o ller scattering instruments because of their
magnetic foil targets.  \revise{In addition,
the}{The} pulse-to-pulse controllability of the laser polarization, 
as well as the high polarization (99.9\%), are additional advantages
over other options.  

\begin{figure}[tbp]
\begin{center}
  \epsfxsize15cm
  \epsfbox{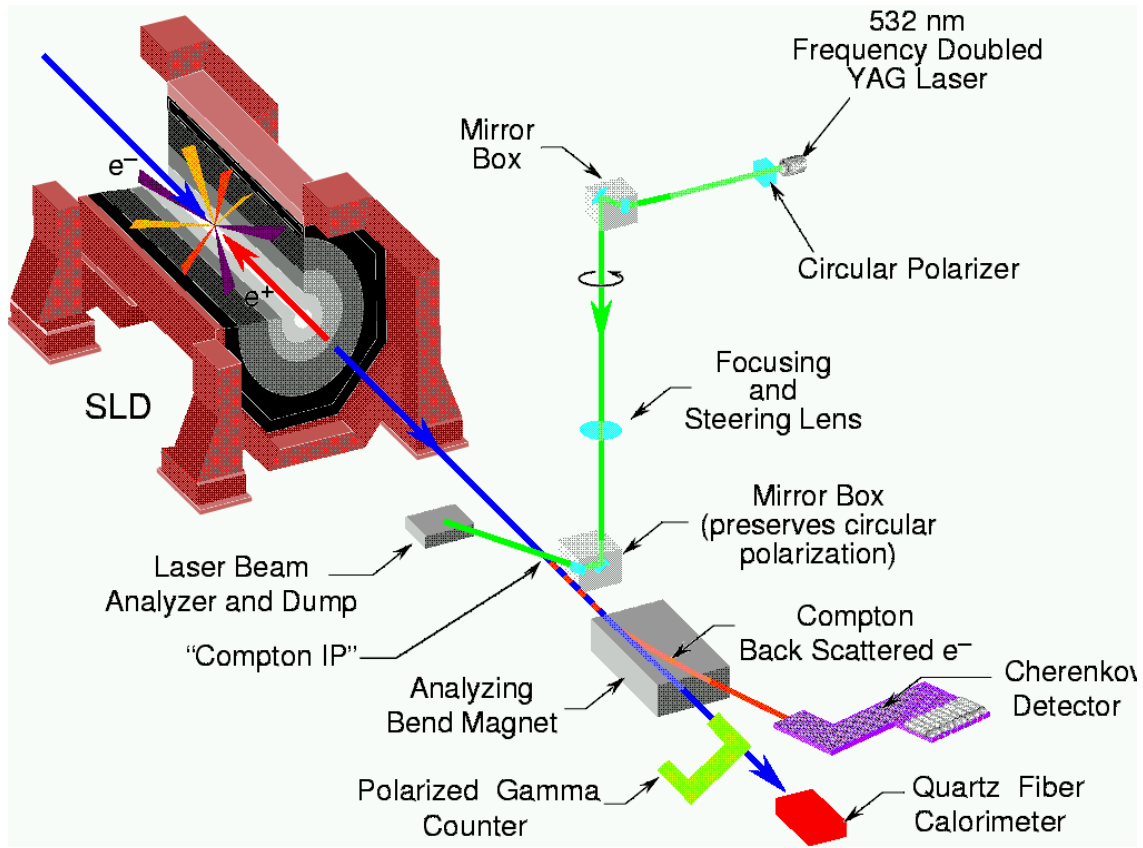}
 \vspace{-0.2 cm}
\caption[The SLC Compton polarimeter setup]
{A conceptual diagram of the SLD Compton polarimeter.}
\label{fig:compton}
\end{center}
\end{figure}

Figure~\ref{fig:compton} illustrates the essential features of the
polarimeter: Frequency-doubled (532 nm) Nd:YAG laser pulses of
8 ns duration and peak power of
typically 25 MW were produced at 17 Hz, circularly polarized by 
\revise{}{a} linear polarizer and a Pockels cell
pair.  The laser beam was transported to the SLC beam line by four sets
of phase-compensating mirror pairs and into the vacuum chamber
through a reduced-strain quartz window.  About 30 \revise{meters}{m} 
downstream from the
IP, the laser beam was brought into
collision with the outgoing electron beam (10 mrad crossing angle) at the
Compton \revise{Interaction Point}{interaction point} (CIP) and then left
the beam pipe
through a second window to an analysis station.
The pair of Pockels cells on the optical bench allowed for full
control of elliptical polarization and was 
used to automatically scan the laser
beam polarization at regular intervals in order to monitor 
and maximize laser polarization at the CIP.
Downstream from the CIP, a pair of 
bend magnets swept out the off-energy
Compton-scattered electrons 
(typically of order 1000 per laser pulse) \revise{which}{that} passed
through a thin window out of the beam-line vacuum and into a nine-channel
(1 cm per channel) transversely
segmented gas Cherenkov detector.  
By detecting scattered electrons with a threshold Cherenkov device, the
large beamstrahlung backgrounds in the SLC environment were 
dramatically reduced.

\begin{figure}[tbp]
\begin{center}
  \epsfxsize14cm
  \epsfbox{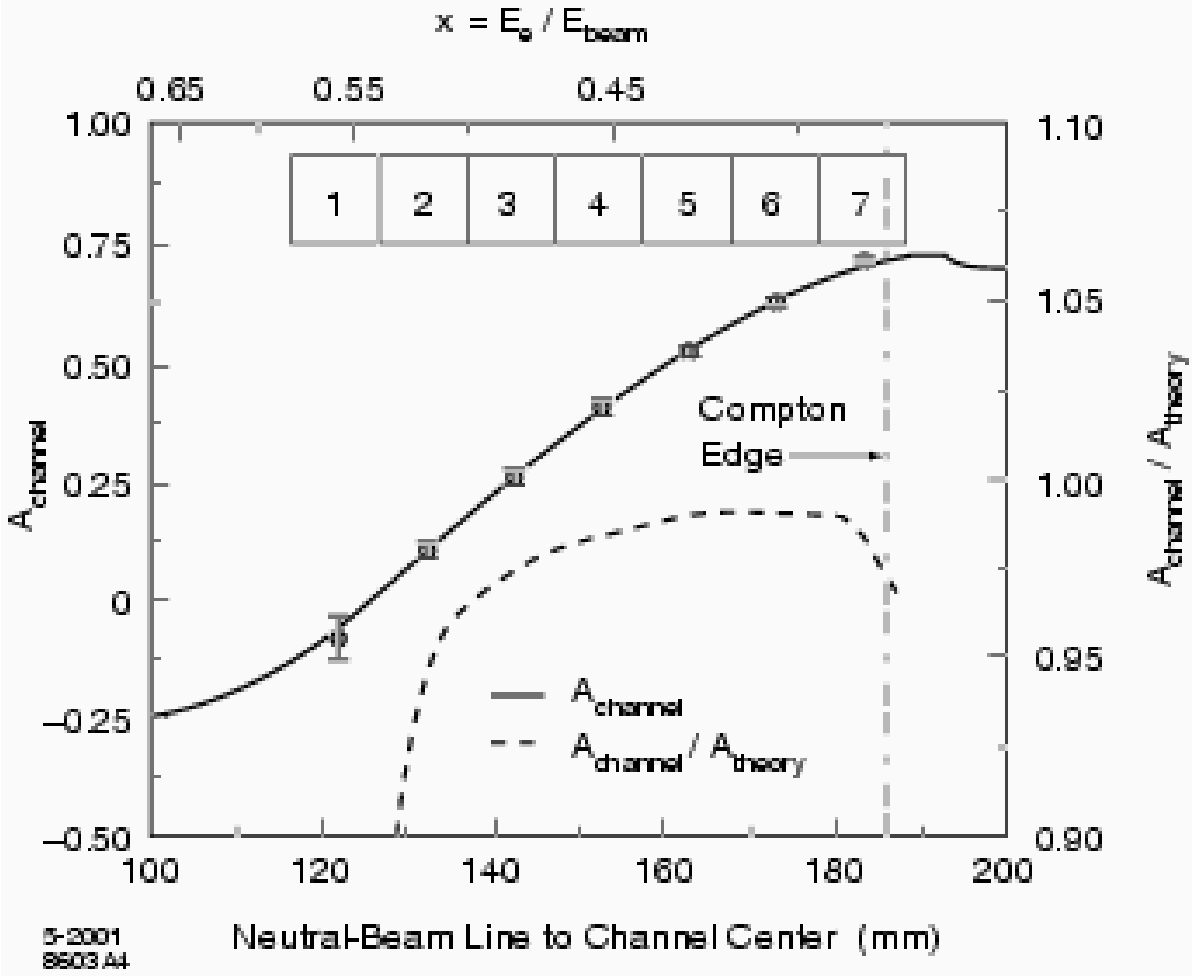}
 \vspace{-0.2 cm}
\caption{Compton scattering asymmetry as a function of channel position.
The horizontal axis gives the distance in mm from the center of the
detector channels (1 cm wide each) to the path of the hypothetical
undeflected electron beam (neutral-beam line). 
The inset shows the seven inner detector channels, sized to match
the horizontal scale.
The per channel data are plotted as open circles, and the 
corrected asymmetry function is the solid curve.
The size of the correction from the theoretical QED calculation is
indicated by the dashed curve.}
\label{fig:comptondata}
\end{center}
\end{figure}

The minimum-energy 17.4 GeV electrons,
corresponding to
full backscattering, generally fell into the seventh detector
channel.  At this point in the
electron spectrum, known as the Compton edge, 
the energy-dependent Compton asymmetry function 
reached its maximum
value of 0.748.  Small deviations from the theoretical 
asymmetry function (of order 1\% in channel 7)
were modeled using an electromagnetic
shower simulation~\cite{ref:sld-EGS4}
in a detailed model of the detector geometry
and the magnetic spectrometer~\cite{ref:sld-torrence}.
The detector was mounted on a movable platform and the
Compton edge was scanned
across several channels at regular intervals in order to 
monitor its location and to experimentally 
constrain the detector channel 
response functions.  This procedure was essential for the
precise determination of the analyzing power of the important outer channels.
Figure~\ref{fig:comptondata} presents
data showing the corrected Compton asymmetry, as
well as the magnitude of the correction,
as a function of position and scattered electron energy.
There is good agreement between
the corrected asymmetry and the data in each channel.

Starting in 1996,
two additional polarimeters~\cite{ref:sld-pgc-qfc},
\revise{that}{which} detected the Compton-scattered
photons and which were operated in the absence of positron beam
due to their sensitivity to beamstrahlung backgrounds,
were used to verify the precision polarimeter calibration.
These two devices were of different design (one employed a gas Cherenkov
detector and the other a quartz-fiber calorimeter)
with different systematic errors, 
and they had in common with the primary electron polarimeter only  
the instrumental errors due to the polarized
laser.  The cross-check
provided by these photon detectors was used to establish a calibration
uncertainty of 0.4\%.
Table~\ref{tab:polsys} summarizes
the systematic uncertainties in the polarization measurement.
During the period 1992--1998,
the total fractional systematic error decreased from 2.7\% to its final
value of 0.50\%,
with the most significant reductions coming from greatly improved understanding
of the laser
polarization and the Cherenkov detector nonlinearity. 
The dominant error is now
due to the
analyzing power calibration discussed above.

\begin{table}[tbp]
\begin{center}
\caption[Compton polarimeter systematic errors.]
{Compton polarimeter systematic errors on the beam polarization
 and, in italics, the total accelerator-related systematic error}
\vspace{0.3 cm}
\label{tab:polsys}
\begin{tabular}{lc}
\hline
Uncertainty (\%) & $\delta\Pe/\Pe$ \\
\hline
Laser polarization & 0.10 \\
Detector linearity & 0.20 \\
Analyzing power calibration & 0.40 \\
Electronic noise & 0.20 \\ \hline
Total polarimeter uncertainty  & 0.50 \\
{\it Chromaticity and \revise{IP}{interaction point} corrections}  & 0.15  \\
\hline
\end{tabular}
\end{center}
\end{table}

The polarimeter result was corrected for higher order QED and
accelerator-related effects (a total of
$-0.22\pm0.15$\% for 1997--1998 data). The higher-order QED offset
was very small and well-determined ($-0.1\%$)~\cite{ref:sld-morrisQED}.
The primary accelerator-related
effect arose from energy-polarization correlations
that caused the average beam polarization
measured by the Compton polarimeter to
differ slightly from the luminosity-weighted average beam polarization
at the IP.  In 1994--1998,
a number of measures in the operation of the SLC 
and in monitoring procedures (smaller and better-determined 
beam energy spread and polarization energy dependence)
reduced the size of this chromaticity
correction and its associated error
from its value of $1.1\pm1.7 \%$ when
it was first observed in 1993 to below $0.2\%$.
An effect of comparable magnitude
arose from the small
precession of the electron spin in the final focusing
elements between the SLC IP and the CIP.
The contribution of depolarization during collision was determined
to be negligible, as expected, by comparing polarimeter data
taken with and without beams in collision.  All effects combined
yielded a correction with the uncertainty given in Table~\ref{tab:polsys}.
\vspace{0.9 cm}

Table~\ref{tab:polresults} gives
the fully corrected, luminosity-weighted, average polarizations
corresponding to each of the SLD runs.
Improvements in GaAs photocathode performance are evident in 
the 1993 run (first use
of a strained-lattice material) and 
the 1994--1995 run
(in which the active layer was three times thinner).
Changes in the achieved polarization
in later years mainly reflect
variation in photocathode manufacture.

\begin{table}[tbp]
\caption{Luminosity-weighted average polarization values for all SLD data}
\vspace{0.3 cm}
\label{tab:polresults}
\begin{center}
\begin{tabular}{ccccc}
\hline
 1992 & 1993 & 1994--1995 & 1996 & 1997--1998 \\
\hline
  $0.224$    & $0.630$    & $0.7723$
  & $0.7616$    & $0.7292$    \\
  $\pm\:0.006$ & $\pm\:0.011$ &
$\pm\:0.0052$
 & $\pm\:0.0040$ & $\pm\:0.0038$ \\
\hline
\end{tabular}
\end{center}
\end{table}


A number of experiments and redundant systems were used to
verify the high-precision polarimeter.  Most important were the following:

\begin{itemize}

\item {\bf Moderate precision M\o ller and Mott polarimeters confirmed
the high-precision Compton polarimeter result to $\sim 3\%$ (1993--1995),
and gamma polarimeters were operated in parallel with the
electron detector polarimeter (1996--1998).}
M\o ller polarimeters located at the end of the SLAC linac
and in the SLC
electron extraction line were used to cross-check
the Compton polarimeter.  The perils of using 
a less reliable method to
test a precision device were apparent when large
corrections for atomic electron momentum effects in the M\o ller target
were discovered~\cite{ref:sld-levchuk} and accounted for, 
after which good agreement was obtained.
In addition, a less direct comparison was provided by Mott polarimeter
bench tests of the GaAs photocathodes~\cite{ref:sld-Mott}.
As described above,
during the last years of the program, two photon detectors
were used to cross-check the primary electron polarimeter and
to experimentally establish calibration uncertainties.

\item {\bf SLC arc spin transport was extensively studied (1993--1998)
and was frequently monitored and adjusted.}
A series of experiments studied the beam polarization
reported by the Compton polarimeter as a function of beam energy, beam
energy
spread, and beam trajectory in the SLC arcs~\cite{ref:sld-arctests}.
Two spin rotators (in the \revise{Linac,}{linac} and in the ring-to-linac
return line) were scanned in order to determine the IP polarization
maximum.
An important result of these experiments was the discovery that the SLC
arcs operate near a spin tune resonance, leading to
the advent of spin manipulation via 
``spin bumps" in the SLC arcs mentioned
above.  This procedure eliminated the need for
these spin rotators 
and allowed one to minimize the spin chromaticity ($d\Pe/dE$),
which reduced the resulting polarization correction from $>1\%$ in
1993 to $<0.2\%$ by 1995.  
\end{itemize}

%% file: sec4_analysis.tex
\section{THE PHYSICS ANALYSIS TOOLS}


\subsection{Monte Carlo Simulation}
\label{sec:analysis-mc}
\input sec4_mc.tex



\subsection{Vertexing Algorithm}
\label{sec:analysis-vertexing}
\input sec4_vertexing.tex


\subsection{$b$ and $c$ Tagging}
\label{sec:analysis-bctag}
\input sec4_bctag.tex

\subsection{Hadron Identification}
\label{sec:analysis-hadronid}
\input sec4_hadronid.tex


\subsection{Lepton Identification}
\label{sec:analysis-leptonid}
\input sec4_leptonid.tex

%% file: sec4_mc.tex

The Monte Carlo simulation for \Ztohad\ events employed the
\jetset\ 7.4 generator~\cite{ref:JETSET}.  
For the $B$ meson decay simulation, the CLEO QQ generator was adopted, with 
adjustments to improve the agreement with inclusive particle
production measurements from ARGUS and CLEO. A detailed  
simulation of the detector was performed with \geant\ 3.21~\cite{ref:GEANT}. 
Simulated \Z\ decays were overlaid with signals from events taken
on random beam crossings in close time-proximity to each recorded 
\Z\ and then processed using the standard data reconstruction.
The Monte Carlo \Ztohad\ events were produced with 
statistics exceeding data by factors of 4, 15, and 22
for $uds$, \ccbar, and \bbbar\ events, respectively. 

For many of the measurements described in this article,
the systematic errors 
associated with tracking efficiency and resolution uncertainties 
are important \revise{issues which}{and} require a consistent evaluation. 
The uncertainty in the tracking efficiency was evaluated from a
comparison between data and \revise{MC}{Monte Carlo} for the fraction of good CDC tracks 
extrapolating close to the IP \revise{which}{that} did not have associated 
VXD hits. These studies 
indicated that the \revise{MC}{Monte Carlo simulation} overestimated tracking efficiency
by $\sim$1.5\% on average. A procedure for the random removal of tracks
in bins of $\pt$, $\phi$, and $\theta$ was used to 
correct the \revise{MC}{Monte Carlo} for this difference.
Most analyses quoted the effect of the entire 
correction as a systematic error. The effect of residual detector 
misalignment on the tracking was estimated from the observed shifts
in the track impact parameter distributions as a function of $\phi$
and $\theta$ in both $r$-$\phi$ and $r$-$z$ planes.
The typical impact parameter biases observed 
in the data were $\sim2.5\mum$ ($\sim5\mum$) in the $r$-$\phi$ ($r$-$z$) plane.
A correction procedure was applied so that the \revise{MC}{Monte Carlo} 
tracks match the mean bias values of the data 
in various $(\phi,\theta)$ regions. 
This was a more realistic evaluation of alignment bias effects, 
where tracks passing the same detector region were biased in
a correlated manner. The full effect of this correction was
taken as the systematic error.

%% file: sec4_vertexing.tex

Conventional identification of heavy-flavor decays with vertex 
detectors typically relies on either counting tracks with 
large impact parameter significance or vertexing kinematically 
selected decay tracks. SLD took advantage of its CCD vertex 
detector and the small and stable SLC IP to pioneer inclusive 
topological vertexing methods that provided additional 
information on decay kinematics, charge, and flavor.   

The topological vertexing algorithm \zvtop\ \cite{ref:ZVTOP}
aimed to identify all separated vertices and all secondary decay 
tracks in the event. Well-measured tracks with VXD hits were 
used to search for secondary vertices by looking for areas of high 
track overlap density in \revise{3-D}{three-dimensional} 
coordinate space, taking into account
the individual track resolution functions.   
Because of the excellent decay
length resolution available at SLD, it was possible in a significant 
fraction of decays to distinguish 
secondary ($b$-hadron decay vertex) tracks from tertiary (cascade 
charm decay vertex) tracks, instead of lumping all secondary tracks 
into a single vertex,
as was typically done by other experiments. 
The vertexing algorithm was 
therefore designed to allow the search for multiple vertices 
associated with one $b$-hadron decay.  

The vertexing was done in each hemisphere or jet separately.
The $b$~hadrons produced in \Z\ decays have an average momentum
of $\sim$30\gevc, corresponding to $\gamma\beta\sim$6.
This large boost distributes the tertiary vertices along the 
$b$-hadron flight direction. The original algorithm, \revise{as described 
in}{as described by Jackson} \cite{ref:ZVTOP}, 
simply used a 50\mum-radius cylindrical volume 
around the jet or hemisphere axis as a preferred secondary vertex 
volume to suppress fake vertices. An improved version of 
\zvtop, used by some of the analyses in this
article, deployed a ``ghost'' track to best represent the $b$-hadron
flight direction. This was achieved by finding the direction 
that minimized the vertex $\chi^2$ sum for the vertices between the 
ghost track and all other tracks in the jet. 
It should be noted that once the ghost track was removed, one could 
be left with single-track ``vertices,'' which is in fact a fairly common 
decay topology (e.g., one-prong $D^{\pm}$ decays). This algorithm
found at least one secondary vertex in 73\% (29\%) of the hemispheres
in \bbbar\ (\ccbar) events. Among the $b$ hemispheres \revise{which}{that} had 
at least one secondary vertex, 
two or more secondary vertices were found in 30\% of them.

\begin{figure}[tbp]
\begin{center}
  \epsfxsize7cm
  \epsfbox{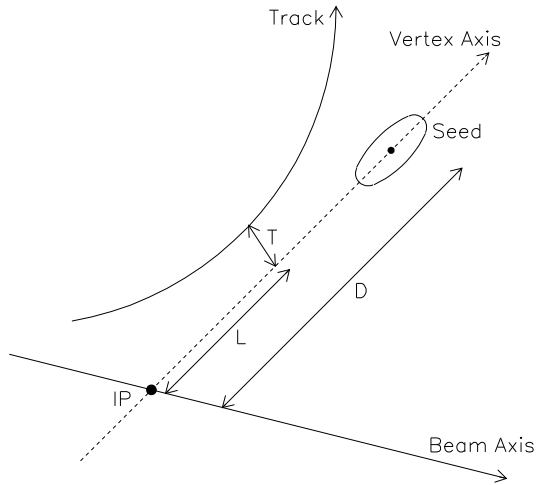}
\end{center}
\caption{Variables used in \zvtop\ track-vertex assignment.}
\label{fig:zvattach}
\end{figure}

To obtain the best description of a $b$ hadron, it is crucial 
to identify all its decay products. Since the basic decay 
vertex topology was established with well-measured tracks only, 
an attempt was made to incorporate lower-quality tracks or even 
unlinked VXD track segments. To do so, a vertex axis 
was formed by joining the IP to a ``seed'' vertex combining 
both secondary and tertiary tracks, and the distance $D$ between 
the IP and this vertex was computed. For each track, 
the three-dimensional
distance of closest approach $T$ to the vertex axis
and the longitudinal 
distance $L$ between the IP and the point of closest approach 
on the vertex axis were calculated (see Figure~\ref{fig:zvattach}).
Typical requirements for attaching a track to the seed vertex
were $T<1$~\mm, $L>250$~\mum, and $L/D>0.25$. This track attachment 
procedure was further improved by means of a neural-network algorithm
combining the information on $T$, $L$, $L/D$, and track angle with 
respect to the vertex axis. There were on average 3.9 quality 
secondary tracks in the seed vertex, and 0.9 additional tracks were 
attached per $b$ hemisphere. The VXD track segments were used only
for vertex charge determination, with an average of 0.2 such
segments attached per $b$-hadron decay. The sum of the identified 
secondary tracks and VXD track segments corresponds to an average of 82\% 
of all prompt $b$-hadron decay tracks, with a track assignment 
purity of 96.8\%.

%% file: sec4_bctag.tex

Prior to the advent of vertex detectors, events with $b$ or $c$ quarks 
were typically identified by the presence of high-$p_T$ leptons or
fully reconstructed $D^\ast$ decays. However, these tagging methods 
suffered from relatively low branching ratios as well as limited
purity.
The development of vertex detectors opened a new avenue in 
heavy-flavor tagging by exploiting the surprisingly long $b$-hadron lifetime. 
Limited success in $b$-tagging had already been achieved with vertex drift 
chambers at PEP/PETRA \cite{ref:TASSO-btag}, but $b$-tagging
developed into a powerful tool in the LEP/SLC era only with the advent 
of silicon vertex detectors. 
SLD contributed pioneering vertexing techniques to fully 
explore the high resolution offered by the CCD pixel vertex detector
and elevated inclusive $b$- and $c$-tagging to a new level.

There are various ways of \revise{utilizing}{using} the basic tagging tools
(tracks with large impact parameters and secondary vertices),
which are all quite efficient for $b$-tagging. However, 
it is difficult to achieve $b$-tag purities $>95\%$ without drastic 
loss of \revise{efficiency, as}{efficiency because} 
the main remaining background consists \revise{}{of}
irreducible \ccbar\ events with genuine long decay length. 
An important improvement in $b$-tagging was the 
utilization of the secondary vertex mass as a powerful discriminator
to separate $b$ and $c$ quarks.

\begin{figure}[tbp]
\begin{center}
  \epsfxsize16cm
  \epsfbox{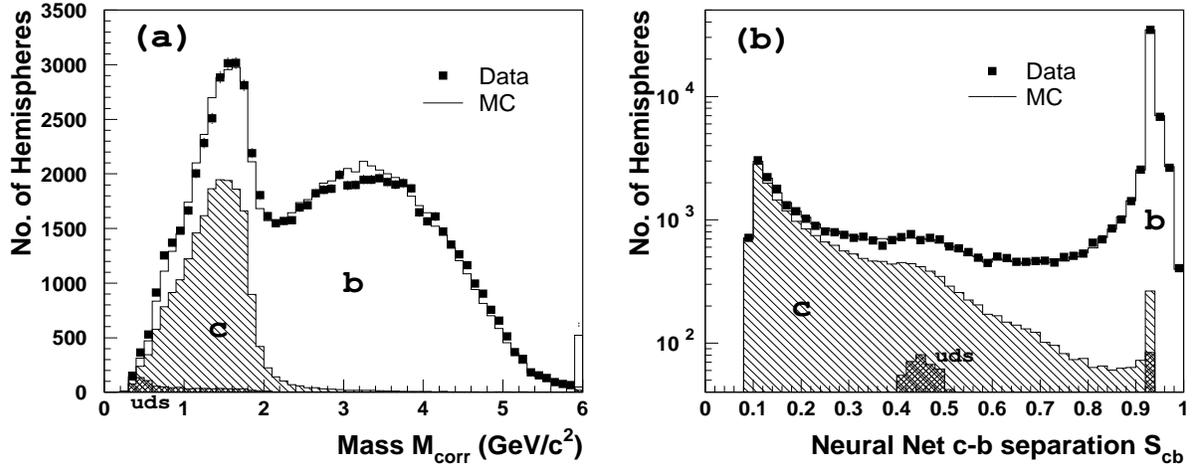}
\end{center}
\vspace{-0.5cm} 
\caption{Distributions of ($a$) the $p_t$-corrected mass $M_{\rm corr}$ 
and ($b$) the neural-net $c-b$ separation variable $S_{cb}$ comparing
data and Monte Carlo.}
\label{fig:bmasscb}
\end{figure}

The SLD implementation of the vertex mass tag included additional 
kinematical information to improve the efficiency while still maintaining 
high purity. The tracks identified as secondary by the procedure
described in the previous section were used to calculate the raw 
vertex invariant mass ($M_{\rm raw}$) for each hemisphere. For vertices 
with $M_{\rm raw}$ close to the charm cutoff of 2\gevcc, 
the vertex momentum is typically collinear with the vertex axis for 
vertices in \ccbar\ events whereas vertices in \bbbar\ events at this 
low mass typically have large 
missing transverse momentum ($p_t$) with respect to the vertex axis 
\revise{due to}{because of} missing neutral particles. 
This information was utilized by
forming a ``$p_t$-corrected mass'':
\begin{equation}
 M_{\rm corr} = \sqrt{M_{\rm raw}^{2}+p_t^{2}} + |p_t|.
\end{equation}
This minimum-$p_t$ correction can be simply derived by boosting the 
system of identified secondary tracks along the vertex axis until
their longitudinal momentum sum is zero, and assuming that a massless 
neutral system is recoiling with the missing $p_t$ in that frame.    
To prevent $udsc$ events from gaining an artificially large $p_t$ due to 
detector resolution effects, the $p_t$ was actually the minimum 
$p_t$ allowed by the uncertainties in the primary and secondary 
vertex positions.  $M_{\rm corr}$ was further restricted to be 
$\leq 2 \times M_{\rm raw}$ to reduce the contamination from fake vertices 
in light-quark events. The effectiveness of $M_{\rm corr}$ as a $b$-tagging
variable can be clearly seen in Figure~\ref{fig:bmasscb}$a$. Many 
analyses using only a basic secondary track classification and 
requiring a mass tag of $M_{\rm corr}>2$\gevcc\ 
benefited from a hemisphere $b$-tagging efficiency of 53\% and a 
$b$ purity of 98\%. In this simple scheme, a cut 
on the raw mass of 
$M_{\rm raw}>2$\gevcc\ would give a similar purity, but the efficiency 
would be only 35\%. 
The partial correction for missing neutral particles 
depends critically on the ability to precisely measure the primary
and secondary vertex positions to ensure that the vertex flight
direction was well determined. The fact that
so far only SLD has effectively applied this underscores
the crucial 
role of the small and stable SLC IP and the high-precision 
vertexing achieved with the CCD vertex detector.  

Although the requirement $M_{\rm corr}>2$\gevcc\ cleanly selected the 
high-mass $b$ hemispheres, further kinematical information can be used 
to achieve good separation between $b$ and $c$ in the mass region
below 2\gevcc. A neural-network algorithm was constructed
to process information 
on the secondary vertex momentum, track multiplicity, and decay length,  
in addition to the $p_t$-corrected mass, leading to a single 
$c-b$ separation variable $S_{cb}$ as shown in 
Figure~\ref{fig:bmasscb}$b$. The $b$~tags were given by large values of 
$S_{cb}$, whereas a clean $c$~tag also emerged at small $S_{cb}$. 
A typical $b$~tag requiring $S_{cb}>0.75$ gave a hemisphere $b$-tagging 
efficiency of $\epsilon_b$=62\% and purity of $\Pi_b$=98.3\%.

We not only enhanced the $b$-tag performance but also simultaneously 
obtained a $c$~tag with a typical cut of $S_{cb}<0.30$, 
giving $\epsilon_c$=18\% and $\Pi_c$=84\%. 
Given the shorter lifetime and much less distinctive 
kinematical properties of \ccbar\ events, this is also a very
impressive performance, which is approaching the
earlier era of $b$-tagging performance. It should be noted that the 
$uds$ events amounted to only a very small fraction of the events 
containing a secondary vertex, and they were concentrated near
$S_{cb} = 0.5$. Furthermore, a large fraction of the $uds$ background 
was actually due to $g\ra\bbbar$ and $g\ra\ccbar$ production in $uds$
events. The success of this inclusive charm tag is unique to 
SLD.

%% file: sec4_hadronid.tex

SLD was one of two \Z\ resonance experiments with
a Cherenkov ring imaging system for hadron
identification. The primary goal of the CRID was to identify charged
pions, kaons, and protons, but the system proved valuable
in lepton identification also (see next section).
Hadron identification at SLD relied exclusively on the CRID,
i.e., no $dE/dx$ information from the CDC was used.
Tracks were extrapolated outward from the end of the drift chamber
through the liquid and gas radiators.
The TPC hit positions were then transformed into Cherenkov angles
with respect to the track trajectory for each of the radiators.
Finally, a global likelihood approach was used to identify the
charged particles.
A likelihood $\calL_i$ was computed for each particle hypothesis
($i=$ $e$, $\mu$, $\pi$, $K$, or $p$).
The likelihood made use of the expected number of photons and
Cherenkov angle, which both depend on the charged-particle momentum and
the mass hypothesis chosen,
the number and angles of reconstructed photons,
and a background term.
The latter incorporated information about background from random hits
as well as the contribution from genuine Cherenkov photons originating
from other charged particles in the event.
The identification proceeded by cutting on differences between the
logarithms of these likelihoods (the log-likelihoods from the two radiators
were simply added to form an overall likelihood).

Kaon identification is very important in heavy-flavor physics,
and it is preferable to keep as high an efficiency as possible 
by focusing mostly on pion rejection.
Proton rejection was difficult in the momentum range
$3 \lsim p \lsim 9$ \gevc\ because both kaons and protons were
below Cherenkov threshold in the gas radiator and were thus primarily identified
by the absence of Cherenkov radiation.
Pion rejection was achieved by requiring
$\log \calL_K - \log \calL_\pi > 3$ ($ > 5$) 
for $p < 2.5$ ($> 2.5$) GeV/$c$, corresponding to the momentum
region below (above)
threshold for charged pion radiation in the gas radiator.
Loose proton rejection was obtained with
$\log \calL_K - \log \calL_{\rm p} > -1$. 
Other cuts were also applied to guarantee that the detector was fully
operational and that the quality of the track extrapolation was high.
For kaons with  $p > 0.8$ \gevc\ and $|\cos\theta| < 0.68$, the
selection had an efficiency between 40 and 55\%,
depending on the momentum of the track.
Using clean pion samples from $\KS \to \pi^+ \pi^-$ decays,
the $\pi \to K$ misidentification rate was measured
to be nearly constant at 2.5\% for $0.8 < p < 2.5$ \gevc,
rising to about 10\% between 2.5 and 4 \gevc\ and remaining
$\sim$10\% for $p > 4$~\gevc.

%% file: sec4_leptonid.tex

  Like hadron identification, lepton identification 
required that charged tracks be
extrapolated outward from the drift chamber to the LAC and WIC subsystems.
The extrapolation transported the track parameters as well as the error matrix,
taking multiple scattering into account.

Electron identification was performed in the range $|\cos\theta| < 0.72$
for tracks with $p > 1$ \gevc\ and relied primarily on information from
the LAC. A special feature of the identification
was the use of the CRID to remove charged hadron contamination.
A neural network, tuned to the Monte Carlo simulation and
cross-checked with data,
optimized the electron selection based on the LAC and CRID variables.
The selection efficiency was estimated using the
Monte Carlo simulation to be 64\%, whereas the purity
was 64\% for $p > 2$ \gevc.
A pion misidentification rate of approximately $1.0\%$ was measured with
a clean sample of pion tracks from $\KS \to \pi^+ \pi^-$ decays,
in good agreement with the simulation.
Electrons from photon conversion were identified and rejected with
a 73\% efficiency.

Muon identification was performed in the range $|\cos\theta| < 0.70$.
Extrapolated charged tracks with $p > 2$ \gevc\
were matched with hit patterns in the WIC.
Cuts on the quality of the hit pattern and CDC/WIC match,
as well as the penetration depth, provided a sample of muons
with 81\% efficiency and 68\% purity,
as determined in the simulation.
This muon selection also took advantage of the CRID to reject approximately
half of the charged kaon and proton contamination in most of the momentum
range, and roughly a third of all charged pions with $p < 6$ \gevc\
(with an efficiency loss of only 5\%).
Information from the pattern of energy deposition in the LAC was also used
in the identification and was especially useful at larger $|\cos\theta|$.
Pion misidentification rates of about 0.3\% were estimated
from $\KS \to \pi^+ \pi^-$ and $\tau^\pm$ decays in the data,
in reasonable agreement with the simulation.

%% file: sec5_electroweak.tex
\section{ELECTROWEAK PHYSICS: \Z\ BOSON COUPLINGS}

\subsection{Introduction}
\label{sec:phys-ewintro}
\input sec5_ewintro.tex

\subsection{Lepton Couplings}
\label{sec:phys-ewlepton}
\input sec5_ewlepton.tex

\subsection{Quark Couplings}
\label{sec:phys-ewquark}
\input sec5_ewquark.tex


%% file: sec5_ewintro.tex

In Section 1, we mentioned that in the
standard model the $\Zff$ couplings depend
on the weak-isospin of the fermions and on a
single parameter, \sinthw, but we did not discuss 
the origin of this parameter.
Recall that in the standard model, the weak isotriplet $\vec{A}_\mu$
and the isosinglet $B_\mu$ gauge fields, with gauge couplings $g$
and $g^\prime$,
are mixed by electroweak symmetry breaking due to the finite
vacuum expectation value of the Higgs scalar field. The 
charged fields $A^1_\mu$ and $A^2_\mu$
and neutral fields $B_\mu$ and $A^3_\mu$ combine linearly into the physical
charged $W^+$ and $W^-$ and the neutral photon and \Z\ gauge bosons.
This change of basis is parameterized by a 
\revise{{\it weak mixing angle}}{weak mixing angle} $\theta_W$, 
given by
$g^\prime = g \tan(\theta_W)$, and the masses of the physical gauge bosons
(and the fermions) 
emerge as \revise{a by-product}{by-products} of electroweak symmetry breaking
(the ``Higgs mechanism'').

Electroweak tests of the standard model reached an important turning point once
the \Z\ boson mass was determined at LEP to a precision of two parts
in $10^5$.   
The measurement of $M_Z$ provides a third precision
constant, which together with the Fermi constant $G_F$ (constrained by muon
decay) and the fine structure constant $\alpha$ (evaluated at $Q^2 = M^2_Z$) 
is sufficient to determine the three universal parameters of the
electroweak standard model: the $SU(2)_{L} \times U(1)$
couplings $g$ and $g^\prime$, and the vacuum expectation value 
of the Higgs field. 
The couplings of fermions to the \Z\ boson are, by virtue of weak mixing,
a function of $\theta_W$---hence,
their determination provides a fundamental
test of electroweak symmetry breaking and, if sufficiently precise, of 
higher-order corrections.
The electroweak measurements made by the SLD collaboration can generally
be described as measurements of the fermion-to-\Z\ couplings.

The differential cross section for
$e^+ e^- \rightarrow \Z \rightarrow f \bar{f}$
is expressed as
\begin{equation}
  \frac{d\sigma}{d\cos\theta} = (1-\Pe A_e)
  (1 + \cos^2\theta) + 2\cos\theta(A_e - \Pe) A_f ,
  \label{Equ_cross}
\end{equation}
where $\cos\theta$ is the cosine of the angle between the final-state
fermion $f$ and the incident electron directions, $\Pe$ is the electron
beam longitudinal polarization, and $A_e$ and $A_f$ are the asymmetry
parameters for the initial- and final-state fermions, respectively.
The asymmetry parameter for a given fermion represents
the extent of parity violation at the \Ztoff\ vertex and is defined as
\begin{equation}
  A_f = \frac{2 g_V g_A}{g^2_V + g^2_A} = \frac{g^2_L - g^2_R}{g^2_L + g^2_R}.
\end{equation}
The parameter \Af\ can be isolated by the measurement of various cross-section
asymmetries, which is experimentally attractive
because systematic effects are minimized.  In addition, $g^2_V + g^2_A$
(or equivalently, $g^2_L + g^2_R$) is determined from
the \Z-decay partial widths, 
which are normalized by the hadronic decay partial width
to control systematic effects.

The simplest of the asymmetries  
is the left-right cross-section asymmetry
\begin{equation}
  A_{LR}^0 = \frac{\sigma_L - \sigma_R}{\sigma_L + \sigma_R} = A_e,
\end{equation}
for which all angular dependence and all dependence
on the final state
cancel\footnote{The dependence of $\Alr$ on
the final state couplings and polar (and azimuthal) angle
completely vanishes,
provided that the efficiency for detecting a fermion at
some polar angle (with respect to the electron direction)
is equal to the efficiency for detecting
an anti-fermion at the same polar angle.
This condition is satisfied by the SLD detector.}.
As a result, \Alr\ is
a particularly robust quantity, with smaller systematic
effects than any other asymmetry.
This asymmetry provides
a direct measurement of the coupling between the \Z\
and the $e^+ e^-$ initial state, and \revise{}{provides,} as we shall see,
the best sensitivity to \sinthw.

Asymmetries that retain angular information are
sensitive to the final-state couplings.
For \Ztoff\ decays, 
the forward-backward asymmetry, measured at LEP
for lepton and heavy quark final states, can be
expressed in terms of $z=\cos\theta$ as 
\begin{equation}
 \Afbf(z) = \frac{\sigma^f(z)-\sigma^f(-z)}{\sigma^f(z)+\sigma^f(-z)}
            = \Ae \Af \frac{2z}{1+z^2}.
\label{eqn:afbf}
\end{equation}  
The asymmetry $\Afbf$ for fermions is a composite observable 
sensitive to both the initial-state \Ae\ and the final-state \Af. 
For example, from 
the magnitudes of the parity violation parameters \Af\ as 
listed in Table \ref{tab:zffcoupling}, it can be seen in the 
case of the $b$ quark that the large value of \Ab\ makes \Afbb\ 
particularly sensitive to \Ae\ (and hence to \sinthw). In general,
the final-state 
\Zff\ asymmetry parameter \Af\ can be deduced by 
taking the lepton coupling \Ae's from other 
measurements such as the $\tau$ polarization, the lepton pair 
forward-backward \revise{asymmetries and}{asymmetries, and} 
\Alr.  Note that \Afbf\ is
subject to systematic errors in the determination of
detector acceptance and efficiency.

\begin{table}[tbp]
\caption{The approximate magnitude of the various fermion-to-\Z\ 
         coupling parameters, for $\sinsqth = 0.23$}
\label{tab:zffcoupling}
\begin{center}
\begin{tabular}{@{}crrccc@{}}%
\toprule
   & $g^f_L$ & $g^f_R$ & 
    $\Rf=\frac{\Gamma(\Ztoff)}{\Gamma(Z^0\ra hadrons)}$ 
   & \Af & $\frac{\delta\Af}{\delta\sin^2\theta_w}$ \\
\colrule 
  $e,\mu,\tau$  & $-$0.27  & $-$0.23  &  0.05  &   0.15  & $-$7.9 \\
  $u,c$         & ~0.35    & ~0.15    &  0.17  &   0.67  & $-$3.5 \\
  $d,s,b$       & $-$0.42  & $-$0.08  &  0.22  &   0.94  & $-$0.6 \\
\botrule
\end{tabular}
\end{center}
\end{table}

At the SLC, 
the polarized forward-backward asymmetry can be measured.
Due to the factor $(\Ae-\Pe)$ in Equation~\ref{Equ_cross},
manipulation of the helicity of the $e^-$ beam 
($\Pe < 0$ for left-handed electrons) distinguishes two different 
forward-backward asymmetries. 
In particular, for a highly polarized beam, the forward-backward 
asymmetry is not only of the opposite sign for left-handed beam, 
the magnitude is also expected to be larger compared with
the asymmetry for right-handed beam. 
Separate measurements of the left and right cross sections  
can be combined into the left-right forward-backward asymmetry 
\begin{equation}
 \Afbtf(z) = 
 \frac{[\sigma^f_L(z)-\sigma^f_L(-z)]-[\sigma^f_R(z)-\sigma^f_R(-z)]}
      {[\sigma^f_L(z)+\sigma^f_L(-z)]+[\sigma^f_R(z)+\sigma^f_R(-z)]}
               = |\Pe| \Af \frac{2z}{1+z^2}. 
 \label{eqn:afbpol}
\end{equation}
The use of \Afbtf\ eliminates the dependence on \Ae\ and 
measures \Af\ directly. 
Compared with $\Afbf$, $\Afbtf$ benefits from a large gain 
in statistical power for the determination of \Af. 
Given an SLC electron beam polarization 
of $\sim$75\%, this improvement factor is $(\Pe/\Ae)^2\sim25$.
The effects of nonuniformity of the detector acceptance 
and efficiency cancel to first order for 
this double asymmetry.  
The SLD collaboration has measured \Afbtf\
for bottom, charm, and strange \revise{quarks, and}{quarks as well as} 
the charged leptons, 
providing direct measurements of the associated
fermion asymmetry parameters.  With the exceptions of \Ae\ and \Atau,
these direct measurements are unique to SLD.   

Measurements of partial \Z-decay widths provide information
\revise{that is complementary to}{complementing} 
that obtained from the asymmetries.  Because
\begin{equation}
  R_f= \frac{\Gamma(\Z\rightarrow f\bar{f})}
  {\Gamma(\Z\rightarrow\mbox{hadrons})}
      \propto g^2_L + g^2_R~,
\end{equation}
\Rf\ measures the \Ztoff\ coupling strength,
\revise{while}{whereas} \Af\ measures the extent of 
parity violation at the \Ztoff\ vertex.  
From another perspective, consider for example
the sensitivity of \Rb\ and \Ab\ to 
the left- and right-handed \Zbb\ couplings:
\begin{eqnarray}   
{\delta R_b}/{R_b} &\sim&-3.57\,\delta g^b_L-0.65\,\delta g^b_R 
 \hspace{4mm}{\rm and} \nonumber\\
{\delta A_b}/{A_b} &\sim&-0.31\,\delta g^b_L+1.72\,\delta g^b_R.
\end{eqnarray}
We see that
\Rb\ is more sensitive to the possible deviations 
in the left-handed \Zbb\ coupling, and \Ab\ is more sensitive to 
deviations in the right-handed coupling. 
The SLD collaboration has measured both \Rb\ and \Rc.

\begin{figure}[tbp]
\begin{center}
  \epsfxsize9cm
  \epsfbox{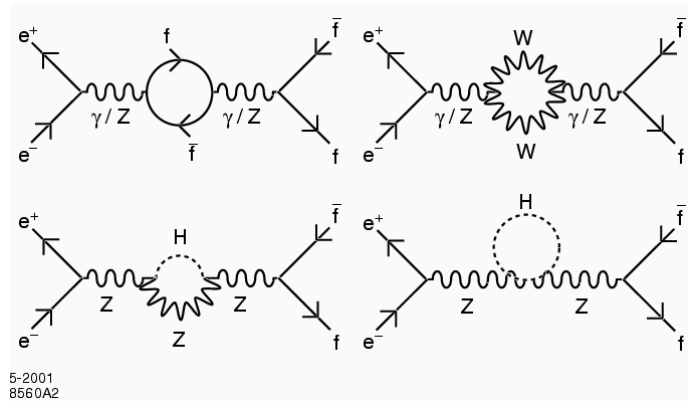}
\end{center}
\vspace{-5mm}
\caption{Feynman diagrams showing radiative corrections affecting the
  $Z^0$ propagator, including contributions from fermions, gauge bosons
  and the Higgs boson.}
\label{Fig_radcor1}
\end{figure}

We have discussed how the well-determined constants
$M_Z$, $G_F$, and $\alpha(M^2_Z)$ constrain the standard model.
However, these data are a sufficient set only at tree level, and to 
test the standard model, including loop effects, at least one additional 
precise measurement is required.  
In the standard model, aside from the effects of real and virtual photons, 
radiative corrections to the fermion-to-gauge boson couplings are
dominated by corrections to the boson propagators.  
These effects are known as \revise{``oblique''
corrections, as}{oblique corrections because} the virtual
loop is not directly coupled to the initial- or final-state fermions
(see Figure~\ref{Fig_radcor1}).
Loop corrections to the boson propagators
are dominated by those due to the heaviest fermion, 
the top quark, and are proportional
to $(m^2_t - m^2_b)/M^2_Z$ (note that this factor depends on 
weak-isospin symmetry breaking in the top-bottom doublet).  
This arises \revise{due to}{because of} the proportionally large Yukawa 
couplings of the heaviest fermions to the
Higgs boson.  The sensitivity of
the radiative corrections to $m_t$ means that precise electroweak measurements
can be used to constrain $m_t$.  Now that $m_t$ has been measured directly,
precision measurements have become sensitive to much smaller
effects, such as those due to the Higgs boson, which are only logarithmic
in the Higgs mass, and those due to new physics processes.

\begin{figure}[tbp]
\begin{center}
  \epsfxsize9cm
  \epsfbox{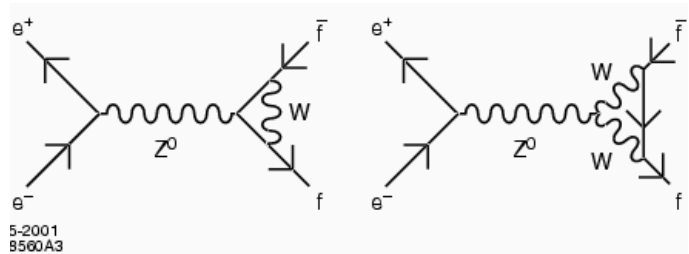}
\end{center}
\vspace{-5mm}
\caption{Feynman diagrams showing radiative corrections affecting the
  $Z^0$ coupling to the final state fermions.}
\label{Fig_radcor2}
\end{figure}

In the standard model, direct vertex corrections 
(see Figure~\ref{Fig_radcor2}) are small, 
but they are expected to be largest for the heaviest
fermions. In particular,   
the \Zbb\ vertex corrections,
which act only on the left-handed coupling, are particularly 
large owing to the large top quark mass and the fact 
that the relevant quark-mixing matrix element ($|V_{tb}|\sim 1$)
is large. 
In general,  
heavy quark partial widths and asymmetry parameters are most
sensitive to vertex corrections
but with different sensitivity to left- and right-handed coupling constants.
In contrast, oblique corrections are best isolated by measurements of 
the lepton asymmetries, and this distinction 
has an important consequence.
In contrast to the \sinthw\ sensitive 
lepton asymmetries, the vertex-sensitive
observables are largely independent of the Higgs boson and top quark masses,
so their standard-model predictions are relatively unambiguous.
Hence, the SLD precision electroweak results constitute a 
complementary set of neutral-current coupling measurements. 

%% file: sec5_ewlepton.tex

\subsubsection{The \Alr\ Measurement}

The measurement of the left-right cross-section asymmetry 
$\Alr$ 
\cite{ref:sld-prlalr,ref:sld-alrtheses} at the SLC provides a
determination of the asymmetry parameter \Ae\ and is
presently the most precise single measurement, with the smallest
systematic error, of this quantity.  In addition, \Ae\ is particularly
sensitive to the effective weak mixing angle with
$\delta \Ae \approx 7.85 ( \delta \sinsqth )$
(for the $\Afbl$ measurements, the analogous factor is 1.77).  
Hence, the most precise available determination
of the effective weak mixing angle derives from the \Alr\ measurement.

In principle, the analysis is
\revise{straightforward: one}{straightforward. One}  
counts the numbers of \Z\ bosons produced by left-
and right-handed longitudinally polarized electrons
($N_{\mathrm{L}}$ and $N_{\mathrm{R}}$),
forms an asymmetry, and
then divides by the luminosity-weighted \en\ beam polarization
(the \ep\ beam is not polarized):
\begin{equation}
  \label{eq:ALR}
  \Alr = \frac{N_{\mathrm{L}} - N_{\mathrm{R}}}
              {N_{\mathrm{L}} + N_{\mathrm{R}}}
         \frac{1}{\avPe}.
\end{equation}
The method requires no
detailed final-state event identification (\epem\ final-state events
are removed because of their
nonresonant t-channel contributions, 
as are all other backgrounds not due to $\Z$ decay) and is insensitive to all
acceptance and efficiency effects.
In order to convert the \Alr\ into a determination of the effective
weak mixing angle, the result is converted into a ``\Z-pole'' value
by the application of a small (typically about 2.0\%) correction for
initial-state radiation and $\gamma - \Z$ interference~\cite{ref:sld-ZFITTER}:
\begin{equation}
  \label{eq:ALR0}
  \Alr(E_{\rm cm}) \rightarrow \Alr^0 \equiv \Ae.
\end{equation}
This calculation requires accurate and precise
knowledge of the luminosity-weighted average center-of-mass collision
energy $E_{\rm cm}$.

For the most recent data (1997--1998), the small total systematic error
of a relative 0.65\% is dominated by the 0.50\%
relative systematic error in the determination of the e$^-$ polarization,
with the second largest error (0.39\%) arising from uncertainties in the
determination of the luminosity-weighted average center-of-mass energy.
\revise{A number of very}{Some} 
much smaller contributions to the systematic error \revise{will be}{are} 
discussed below.  The relative statistical error on \Alr\
from all data is about 1.3\%.

Below, we describe some
details of the \Alr\ measurement
and provide some historical context for the \Alr\ program at SLC/SLD
(1992--1998).

\noindent{\bf Systematic Effects and Their Control~~~~}
%
%
The \Alr\ measurement is remarkably resistant to
detector-dependent systematic effects and
Monte Carlo modeling uncertainties that are significant issues for
most other electroweak precision measurements.  By far the dominant 
systematic effects arise from polarimetry and from the determination
of the collision energy, rather than from any details of the
analysis or the operation of SLD.
The simple expression given in Equation~\ref{eq:ALR} applies to the
ideal case in the absence of additional systematic effects, 
and as such it is 
a good approximation to better than 0.2\%. Nevertheless,   
systematic
left-right asymmetries in luminosity, polarization, beam energy, and
acceptance, as well as background and positron polarization
effects, can be incorporated into an extended expression for the
cross-section asymmetry.  One finds that
the measured asymmetry $A_m$ is related to \Alr\ by the following
expression, which incorporates a number of small correction terms in
lowest-order approximation:
\begin{eqnarray}
\label{eq:ALRsys}
\Alr & = & \frac{A_m}{\avPe} +\frac{1}
{\avPe}\biggl[f_{\rm bkg}(A_m-A_{\rm bkg})-A_{\cal L}+A_m^2\apol \nonumber \\
 & & ~~~~~~~~~~~~~~~~~~
-E_{\rm cm}\frac{\sigma^\prime(E_{\rm cm})}{\sigma(E_{\rm cm})}\aengy
-\aeff + \avPe\calP_p \biggr],
\end{eqnarray}
where \avPe\ is the mean luminosity-weighted polarization;
$f_{\rm bkg}$ is the background fraction;
$\sigma(E)$ is the unpolarized \Z\ boson cross section at energy $E$;
$\sigma^\prime(E)$ is the derivative of the cross section with
respect to $E$;
$A_{\rm bkg}$, $A_{\cal L}$, $\apol$, $\aengy$, and
$\aeff$ are the left-right asymmetries\footnote{The left-right asymmetry
for a quantity $Q$ is defined as
$A_Q\equiv(Q_L-Q_R)/(Q_L+Q_R)$ where the subscripts $L$ and $R$ refer to
the left- and right-handed beams, respectively.}
of the residual background,
the integrated luminosity, the beam polarization,
the center-of-mass energy, and
the product of detector acceptance and efficiency, respectively;
and $\calP_p$ is any longitudinal positron 
\revise{polarization which}{polarization, which} is assumed
to have constant helicity. Because the colliding electron and positron
bunches
were produced on different machine cycles and because the
electron helicity of each cycle was chosen randomly, any positron
helicity arising from the polarization of the production electrons
was uncorrelated with the electron helicity at the IP.
The net effect of positron
polarization from this process vanished rigorously.  However,
positron polarization of constant helicity would affect the measurement.

The close ties
between the $\Alr$ measurement and the SLC accelerator complex are highlighted
by numerous accelerator-based
experiments dedicated to the SLD physics program. 
Particularly important among these experiments was
a \Z-peak scan to calibrate the sum of the beam energies.
The SLC energy spectrometers were briefly described in
Section~\ref{sec:sld-espec}.
These devices were first operated in their final configuration in
1989 by the Mark~II experiment, and the calibration of the two precision
spectrometer
magnets was performed in 1988~\cite{ref:sld-espec}.  Their expected
precision was about $\pm 20$ MeV on the measured center-of-mass collision
energy $E_{\rm cm}$.  The importance of these devices to the $\Alr$ measurement
is quantified by the approximate rule of thumb that an 80 MeV uncertainty
in $E_{\rm cm}$ corresponds to a $1\%$ error on the $\Z$-pole asymmetry
$\Alr^0$.
For this reason, a \Z-peak scan was performed in 1998 to calibrate
the spectrometers to the LEP measurement of the \Z\ mass.  The scan used two
optimized off-peak points at $+0.88$ and $-0.93$ GeV and
a luminosity approximately
equivalent to 9000 produced \Z\ bosons to reach a statistical precision on
the peak position of 20 MeV.  The results of a complete
analysis of systematic effects determined an offset
of $-46$ MeV and a total $E_{\rm cm}$ uncertainty of 29 MeV [the 0.39\% 
uncertainty on $\Alr^0$ mentioned \revise{earlier,}{above} and 
given in Table~\ref{tab:alr:alrcorr}~\cite{ref:sld-zpeak}].

Other examples of accelerator-based experiments include the following:

\begin{itemize}
\item {\bf The $e^-$ bunch helicity transmission was verified
by setting up a large current-helicity correlation in the SLC, allowing
for the use of the LAC to verify data synchronization (1992--1993).}
Although the electron bunch polarization state was transmitted via
reliable and redundant paths to the SLD detector/polarimeter complex,
the SLD electroweak group proposed a series of independent tests of the
synchronization of
this information and the SLD event data.  In one such test,
the laser optics at the SLC polarized source
were temporarily modified by the addition of a polarizer and quarter-wave
plate
so that photocathode illumination was nulled for one
of the two circular polarization states.  The positron beam was turned
off,
and the electron beam was delivered to the IP.  Beam-related
background in the LAC was detected, but only for the
non-extinct
pulses.  By this means, the expected correlation between helicity and
the presence of beam, and hence the LAC data stream, was
verified~\cite{ref:sld-extinct}.

\item {\bf Positron polarization was experimentally constrained.}
In 1998, a dedicated experiment was performed in order to directly test
the expectation that accidental polarization of the positron beam was
negligible.  The positron beam was delivered to the SLAC End Station A
(ESA),
where a M\o ller polarimeter was used. Experimental control was assured
by first
delivering the polarized electron beam, and then an unpolarized
electron beam (sourced from SLAC's thermionic electron gun), to the
ESA, confirming polarimeter operation.
In addition, the spin rotator magnet located in the linac was
reversed halfway through the positron beam running, thereby reversing the
sense of polarization at the M\o ller target and reducing systematic
error.  The final result verified that
e$^+$ polarization was consistent with
zero ($-0.02\pm 0.07$\%)~\cite{ref:sld-posipol}.
\end{itemize}

A simple calorimetric event selection in the LAC,
supplemented by track multiplicity
and topology requirements in the CDC, 
was used to select hadronic \Z\ decays. 
For each event candidate,
energy clusters were reconstructed in the LAC.  Selected
events were required to contain at least 22~GeV of energy observed in
the clusters and to have a normalized energy
imbalance\footnote{The energy imbalance is defined
as a normalized vector sum of the energy clusters as follows,
$E_{\rm imb}=|\sum \vec E_{\rm cluster}|/\sum |E_{\rm cluster}|$.} of less
than 0.6.
The left-right asymmetry associated
with final-state $\ee$ events is expected to be diluted by the t-channel
photon exchange subprocess.  Therefore, we excluded $\ee$ final states
by requiring that each event candidate contain at least four
selected CDC tracks,
with at least two tracks in each hemisphere (defined with respect to
the beam axis), or at least four tracks in either hemisphere.
This track topology requirement excluded Bhabha events that contained
a reconstructed gamma conversion.
Small backgrounds in the $\Alr$ data sample were due to
residual $e^+e^-$ final-state events,
and to two-photon interactions, beam-related noise, and cosmic rays.
For the most
recent data (1996--1998), the total background contamination was estimated to
be $<0.05\%$ for a selection efficiency of $(91\pm1)\%$.

The asymmetries in luminosity, polarization, and beam energy
were all continually monitored
using  small-angle Bhabha counters, the Compton polarimeter,
and energy spectrometer data available at the SLC repetition rate of 120
Hz.  These asymmetries
were limited to approximately $10^{-4}$, $10^{-3}$, and $10^{-6}$,
respectively.  The long-term average values of all asymmetries of this
type were reduced by the roughly trimonthly
reversal of the transverse polarization sense
in the electron damping ring (DR)
referred to in Section~\ref{sec:slc}.
The dominant cause of the observed asymmetries was
the small current asymmetry produced at the SLC polarized
source.  This effect arose \revise{due to}{because of}
the source photocathode sensitivity to linear polarized light, together with 
residual linear polarization in the source laser
light that was correlated with the light helicity.
The source current asymmetry was minimized by a polarization control and
intensity feedback system (starting in 1993) and was generally maintained
below $10^{-4}$. 

The value of \Alr\ is unaffected by decay-mode-dependent
variations in detector acceptance and efficiency,
provided that the efficiency for detecting a fermion at
some polar angle (with respect to the electron direction)
is equal to the efficiency for detecting
an anti-fermion at the same polar angle.  This fact, and the
high degree of polar symmetry in the SLD detector, render
$\aeff$ negligible.  Finally, $\calP_p$ was
experimentally demonstrated to be consistent with zero
to a precision of $7 \times 10^{-4}$,
as described above.  Calculations based on polarization buildup
in the positron DR suggested a much smaller number,
$\calP_p < \order(10^{-5})$; hence, no correction for $\calP_p$
was applied to the data.
\begin{table}[tbp]
\caption[]
    {\Z\ event counts and corrections for
     all SLD run periods.
%
%
Also shown are the total
     polarimetry errors (including chromaticity and IP
     effects) and the relative error due to the
     electroweak interference correction needed for the
     conversion of $\Alr$ to $\Alr^0$.
     Note that \revise{due to}{because of} low statistics, a number of
     effects were ignored for the 1992 data.
%
%
}
\label{tab:alr:alrcorr}
\begin{center}
\renewcommand{\arraystretch}{0.9}
\begin{tabular}{cccccc}
\hline
      & 1992 & 1993 & 1994--95 & 1996 & 1997--98 \\
\hline
\hline
$N_{L}$  &   5,226 & 27,225 & 52,179 & 29,016 & 183,335 \\
\hline
$N_{R}$  &   4,998 & 22,167 & 41,465 & 22,857 & 148,259 \\
\hline
$A_{m}$  & $0.0223$ & $0.1024$ & $0.1144$ & $0.1187$ & $0.1058$  \\
         &$\pm 0.0099$ & $\pm 0.0045$ & $\pm 0.0032$ & $\pm 0.0044$ & $\pm 0.0017$ \\
\hline
\hline
$f_{\rm bkg}$ (\%) & $1.4$  & $0.25$ & $0.11$ & $0.029$ & $0.042$     \\
                   & $\pm 1.4$ & $\pm 0.10$ & $\pm 0.08$ & $\pm 0.021$ & $\pm 0.032$ \\
\hline
$A_{\rm bkg}$ & & $0.031$     & $0.055$     & $0.033$     & $0.023$      \\
              & & $\pm 0.010$ & $\pm 0.021$ & $\pm 0.026$ & $\pm 0.022$ \\
\hline
$A_\calL$ $(10^{-4})$ & $1.8$ & $0.38$  & $-1.9$  & $+0.03$ & $-1.3$  \\
                      & $\pm\: 4.2$ & $\pm\: 0.50$ & $\pm\: 0.3$ 
                       & $\pm\: 0.50$ & $\pm\: 0.7$ \\
\hline  
$A_{\cal P} (10^{-4})$ & $-29$ & $-33$ & $+24$  & $+29$ & $+28$      \\
                       & & $\pm\: 1$ & $\pm\: 10$ & $\pm\: 43$ & $\pm\: 69$ \\
\hline
$A_{E}$ $(10^{-4})$ & & $0.0044$ & $0.0092$ & $-0.0001$ & $+0.0028$      \\
                    & & $\pm 0.0001$ & $\pm 0.0002$ 
                                & $\pm 0.0035$ & $\pm 0.0014$ \\
\hline
$E_{\rm cm} \frac{\sigma '(E_{\rm cm})}{\sigma (E_{\rm cm})}$
                  & & $-1.9$ & $0.0$       & $2.0$       & $4.3$  \\
                  & &        & $\pm\: 2.5$ & $\pm\: 3.0$ & $\pm\: 2.9$ \\
\hline
$A_{\varepsilon}$ & 0 & 0 & 0 & 0 & 0 \\
                  & & & & & \\
\hline
$\calP_p$ $(10^{-4})$  & $<0.16$ & $<0.16$ & $<0.16$ & $<0.16$ & $-2$ \\
                       &     &         &         &         & $\pm\: 7$ \\
\hline
\hline
Total correction,   & & $+ \: 0.10$  & $+ \: 0.2$   & $+0.02$ & $+0.16$   \\
$\Delta \Alr / \Alr$, (\%) & & $\pm\: 0.08$ & $\pm\: 0.06$ 
                           & $\pm\: 0.05$ & $\pm\: 0.07$ \\
\hline
\hline
$\delta\Pe/\Pe$ (\%) & 2.7 & 1.7 & 0.67 & 0.52 & 0.52 \\
\hline
\hline
Electroweak interference &   & 0.3 & 0.4 & 0.4 & 0.39 \\
correction [relative (\%)] & &     &     &     &      \\
\hline
\end{tabular}
\end{center}
\end{table}

Table~\ref{tab:alr:alrcorr} summarizes the systematic
effects discussed in this section.
The total uncertainties due to backgrounds and accelerator-induced asymmetries
were much smaller than the leading systematic effects 
due to polarimetry and energy uncertainties,
as can be seen by comparing the last three rows
of Table~\ref{tab:alr:alrcorr}.

Table~\ref{tab:sld-alrresults} shows the run-by-run \Alr\ results.
The previously discussed $E_{\rm cm}$-dependent radiative correction and 
its
uncertainty are evident in the difference between \Alr\ and $\Alr^0$. 
\begin{table}[tbp]
\caption[\Alr\ and \sinsqth\ measurements: summary
of results for all SLD runs.]
    {Summary of \Alr\ and \sinsqth\ measurements for all SLD runs
     (statistical and systematic errors are listed separately)}
\label{tab:sld-alrresults}
\begin{center}
\begin{tabular}{cccc}
\hline
 Data Set & \Alr\ & $\Alr^0$ & \sinsqth \\
\hline
1992 & $0.100$ 
     & $0.097$  
     & $0.2378$ \\
     & $\pm 0.044 \pm 0.004$ 
     & $\pm 0.044 \pm 0.004$  
     & $\pm 0.0056\pm0.0005$ \\
\hline
1993 & $0.1628$ 
     & $0.1656$ 
     & $0.2292$ \\
     & $\pm 0.0071 \pm 0.0028$ 
     & $\pm 0.0071 \pm 0.0028$ 
     & $\pm 0.0009 \pm 0.0004$ \\
\hline
1994--1995 &  $0.1485$ 
        &  $0.1512$ 
        &  $0.23100$ \\
        &  $\pm 0.0042 \pm 0.0010$ 
        &  $\pm 0.0042 \pm 0.0011$ 
        &  $\pm 0.00054 \pm 0.00014$ \\
\hline
1996    &  $0.1559 $ 
        &  $0.1593 $ 
        &  $0.22996$ \\
        &  $ \pm 0.0057 \pm 0.0008$  
        &  $ \pm 0.0057 \pm 0.0010$  
        &  $\pm 0.00073 \pm 0.00013$ \\
\hline
1997--1998 &  $0.1454 $ 
        &  $0.1491 $  
        &  $0.23126$ \\ 
        &  $ \pm 0.0024 \pm 0.0008$  
        &  $ \pm 0.0024 \pm 0.0010$  
        &  $ \pm 0.00030 \pm 0.00012$ \\
\hline
\hline
All & & $0.15138 \pm 0.00216$ & $0.23097 \pm 0.00027$ \\
\hline
\end{tabular}
\end{center}
\end{table}
These five results show a $\chi^2$ of 7.4 for four degrees of freedom,
corresponding to a probability of 11.6\%.
The \sinsqth\ results are derived from
the equivalence $\Alr^0 \equiv \Ae$, which implies
\begin{equation}
\Alr^{0} = \frac{2(1 - 4 \sinsqth)}
{1 + (1 - 4 \sinsqth)^{2}}.
\label{eq:alrsin}
\end{equation}

The average for the complete SLD data sample is
\begin{eqnarray}
 \Alr^{0} & = & 0.15138 \pm\: 0.00216,\nonumber\\
 \sinsqth & = & 0.23097 \pm\: 0.00027.
\end{eqnarray}
Small correlated systematic effects are accounted for in
the calculation of this average.

\subsubsection{Leptonic Asymmetries}

The individual lepton asymmetry parameters were determined
from lepton final-states~\cite{ref:sld-al1997,ref:sld-al2000}.
The \Alr\ measurement determined the initial-state parameter \Ae, but
electron polarization \revise{allows one also
to directly measure}{also allows direct measurement of} the final-state 
asymmetry parameter
\Al\ for lepton $\ell$
using the left-right forward-backward asymmetry
($\Afbtl = \frac{3}{4} \avPe \Al$). 
If lepton universality is assumed, the results for all three lepton flavors
can be combined to yield a determination of \sinsqth, which in turn can
be combined with the more precise result from \Alr.  The event sample used
for \Alr\ was purely hadronic (there was a very small $0.3\pm0.1\%$ admixture
of $\tautau$ events), and hence, the left-right asymmetry of the lepton events
constituted an independent measurement.
\revise{While}{Although} the lepton final-state analysis described
\revise{in what follows}{below} is more sophisticated than a simple \Alr-style
counting measurement,
essentially all the information on \sinsqth\ is obtained from the left-right
asymmetry of these events.  For example in the 1996-1998 data set, 
the inclusion of the distributions in polar angle that are
essential for the extraction of the final-state asymmetries improved the
resulting precision on \sinsqth, but only to $\pm 0.00076$,
compared with about $\pm0.00078$
obtained from a simple left-right event count.

\noindent{\bf The Analysis Method~~~~}
Table~\ref{tab:sld:alevts} summarizes the selection efficiencies,
backgrounds, and numbers of selected candidates
for $e^+e^-$, $\mu^+\mu^-$, and $\tau^+\tau^-$ final states.
Event-by-event maximum likelihood fits, for each lepton flavor, were used
to incorporate the contributions of all the terms in the 
polarized differential cross section and to
include the effect of initial-state radiation.    
Photon exchange terms and, if the final-state leptons were electrons,
t-channel contributions, were taken into account.
All three lepton asymmetry parameters,
\Ae\ and \Amu\ (\Atau), were obtained from $\mu^+\mu^-$
($\tau^+\tau^-$) final states.
These \Ae\ results were combined with \Ae\ obtained from the $e^+e^-$
final state.

\begin{table}[tbp]
\caption{Summary of event selections, efficiency, and purity for
$e^+e^- \to \ell^+\ell^-$ for the 1997--1998 SLD data}
\vspace{0.4cm}
\label{tab:sld:alevts}
\begin{center}
\begin{tabular}{llcl}
\hline
Event  & Background   & Efficiency in   & \# of selected \\
sample & fraction & $|\cos\theta|<0.9$ & events\\ \hline
$e^+e^-\rightarrow e^+e^-$ &
   0.7\% $\tau^+\tau^-$  & 75\%  & 15,675 \\ \hline
$\Z\rightarrow\mu^+\mu^-$   &
   0.2\%  $\tau^+\tau^-$ & 77\% & 11,431 \\ \hline
$\Z\rightarrow\tau^+\tau^-$ &
   $e^+e^-$\,:\,$\mu^+\mu^-$\,:\,2-$\gamma$\,:\,hadrons
 &  70\% & 10,841    \cr
 & 0.9\%\,:\,2.9\%\,:\,0.9\%\,:\,0.6\%                  &       &  \\\hline
\end{tabular}
\end{center}
\end{table}

\begin{figure}[tbp]
\vspace{-0.5 cm}
\begin{center}
  \epsfxsize14cm
  \epsfbox{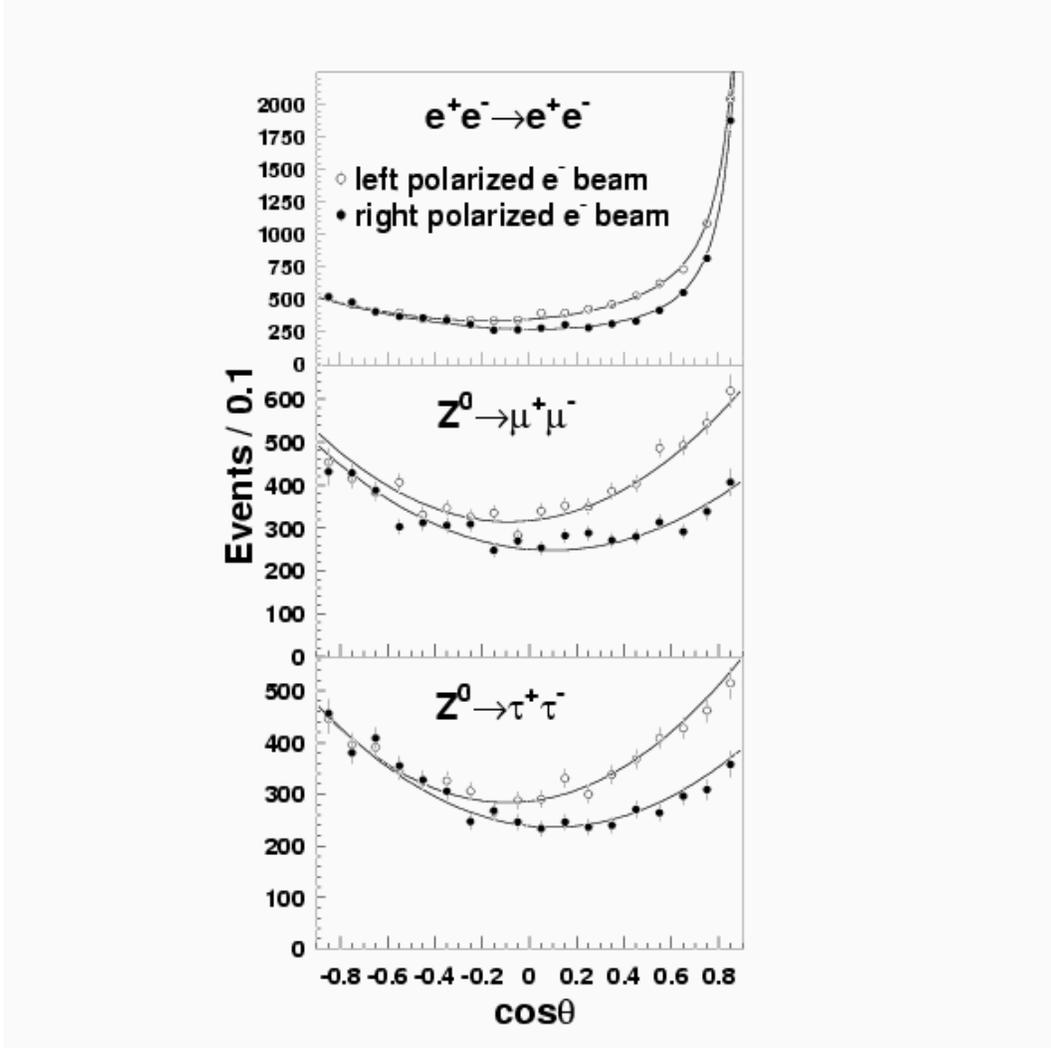}
\end{center}
\vspace{-0.7 cm}
\caption{Distributions of the lepton $\cos\theta$ for $e^+e^-$,
$\mu^+\mu^-$, and $\tau^+\tau^-$ candidates from the 1997--1998 data set.}
\label{fig:alepton}
\end{figure}

Figure~\ref{fig:alepton} shows the $\cos\theta$ distributions for $e^+e^-$,
$\mu^+\mu^-$, and $\tau^+\tau^-$ candidates for the 1997--1998 data.
The pre-1997 results are similar
but have smaller acceptance ($|\cos\theta| \le 0.8$)
(note that the SLD event totals including this older data are 22,254,
16,844, and 16,084
for the $\ee$, $\mumu$ and $\tautau$ final states respectively).

The high-$|\cos\theta|$ region is increasingly
sensitive to the asymmetry parameters, most apparent
in the $\mumu$ and $\tautau$ distributions in Figure~\ref{fig:alepton}.
In 1996, the upgraded vertex detector and a new trigger system
for forward $\mumu$ events were installed.
The improved acceptance of VXD3 allowed for efficient
track finding up to
$|\cos\theta|=0.9$~\cite{ref:sld-al2000}.
The new trigger for $\mu^+\mu^-$ events
covered the angular range up to $|\cos\theta|<0.95$
by requiring two back-to-back tracks that pass through the IP
and reach the endcap WIC used for
muon identification.
\vspace{1.2 cm}

\noindent{\bf Systematic Errors~~~~}
This measurement is statistically limited.
The 1997--1998 dataset, which dominates the sample,
typically has statistical errors between 4 and 10 times larger
than the total systematic error for the various asymmetry parameters. 
The systematic errors arise from polarimetry, backgrounds, radiative
corrections, $\tau^{\pm}$ polarization effects (``V$-$A''),
incorrect charge assignment for tracks at large $\cos\theta$,
and nonuniformities in the detector efficiency and
forward-backward asymmetries.
These errors are given in Table~\ref{tab:sld:alsys}.

\begin{table}[tbp]
\caption{\label{tab:sld:alsys}
Summary of statistical and systematic uncertainties in units of $10^{-4}$
for the 1997--1998 data.
%
%
  The superscript on each asymmetry refers to the
lepton sample from which it was derived (electrons, muons, or taus).
}
\label{Table:systematics}
\begin{center}
\begin{tabular}{lccccc}
\hline 
Source & $\Ae^e$ & $\Ae^{\mu}$ & $\Ae^{\tau}$ & $\Amu^{\mu}$ &
$\Atau^{\tau}$   \\ \hline
Statistics & 110  & 130 & 130 & 180 & 180 \\
Polarization            &  8  &  8 &  8  &  8  &  8 \\
Backgrounds             &  5  & --    & 13 & --     & 14 \\
Radiative correction    & 23 &  2 &  2  &  3  &  2 \\
V-A                     & --     & --    & --     & --     & 18 \\
Charge confusion        & --     & --    & --     &  7 & 11  \\
Detector asymmetry      & --     & --    & --     & --     & 4 \\
Nonuniform efficiency   & 2  & --    & --     & --     & -- \\
\hline 
\end{tabular}
\end{center}
\end{table}

One error deserves some discussion because of its specific
relevance to polarized forward-backward asymmetry measurements.
The dominant systematic error in the $\tau^{\pm}$ analysis resulted from
final-state $\tau^{\pm}$ polarization effects, 
that introduced
a selection bias.
The term V$-$A 
(``vector minus axial vector'') used in Table~\ref{tab:sld:alsys}
arises from the $(1 - \gamma^5)\gamma^\mu$ 
Lorentz structure of the charged-current Lagrangian
that governs $\tau^{\pm}$ decay.
For example, if both $\tau^+$ and $\tau^-$ decay to $\pi\nu$,
helicity conservation requires that both pions generally have
lower momentum for a left-handed $\tau^-$ and right-handed $\tau^+$ and
higher momentum otherwise.
This effect, which biased the reconstructed event mass, was significant
because of the $e^-$ beam polarization, which induced a large and
asymmetric $\tau^{\pm}$ polarization as a function of polar angle.
The value of \Ae\ extracted from $\tau^+\tau^-$ final states was
not affected, since the overall relative efficiencies for
left-handed beam and right-handed beam events were not changed significantly
(only the polar angle dependence of the efficiencies was changed).

\noindent{\bf Results~~~~}
Results for all SLD data sets were combined, taking into
account small effects due to correlations in systematic uncertainties
(polarization and average SLC center-of-mass energy).
From purely leptonic final states, one obtains
$\Ae=0.1544 \pm 0.0060$.
This \Ae\ result was combined with the left-right asymmetry measurement
in the tabulation of leptonic asymmetry results:
\begin{equation}
\begin{array}{cclcll}
\Ae\    & = & 0.1516 &\pm& 0.0021, \  (\mbox{with}\ \Alr^0)& \cr
\Amu\   & = & 0.142  &\pm& 0.015, &  \mbox{and}\cr
\Atau\  & = & 0.136  &\pm& 0.015 .&
\end{array}
\end{equation}
These results are consistent with lepton universality and hence can be
combined to obtain \Al, which in the context of the standard model is simply
related to the weak mixing angle.
The result is discussed in the following section.

\subsubsection{Combined Results for \sinsqth}

Assuming lepton universality, the \Alr\ result and the results on the
leptonic left-right forward-backward asymmetries are combined, where
small correlated systematics are accounted for,
yielding
\begin{equation}
 \Al = 0.15130 \pm 0.00207.
\end{equation}
This measurement is equivalent to the determination
\begin{equation}
 \sinsqth = 0.23098 \pm 0.00026,
\end{equation}
where the total error and corresponding systematic error of
$\pm 0.00010$ are more precise than those obtained
by any other technique.

%% file: sec5_ewquark.tex

It is particularly interesting to measure the couplings of
individual quark flavors separately, and thereby 
probe the radiative corrections to the \Zqq\ vertex.  
The $b$ and $c$ quarks are the heaviest charge 1/3 and charge 2/3 
\revise{quarks respectively}{quarks, respectively, that are} 
accessible at the \Z\ energy. Potential \revise{new physics}{new-physics} 
signatures may preferentially appear in the heavy-quark couplings,
e.g., those \revise{involving}{that involve} 
Yukawa-type couplings favoring fermions
with large masses. \revise{From a more general standpoint,}{More generally,} 
any unexpected
difference in quark coupling of one flavor compared with other flavors 
could be a vital clue 
toward a solution to the puzzle of fermion family 
degeneracy. The relatively similar production rates of all quark flavors in 
\Z\ decays, combined with our ability to tag bottom and charm hadron 
decays, offers the possibility to test the \Z\ coupling to 
the individual quark flavors with high precision.  In addition to the 
measurements with heavy quarks, comparing the measurements for the $s$ quark 
to those of the $b$ quark 
provides a unique test of the universality 
of couplings for charge 1/3 quarks.  

For most of the measurements in this section, the hadronic \Z\ events
were selected by requiring at least seven charged tracks 
and a visible charged
energy of at least 18\gev. The events were typically required to be 
well-contained in the high-quality tracking fiducial volume of 
$|\cos\theta_{\rm thrust}|<0.7$. The flavor-tagging efficiencies referred
to are generally for these selected fiducial events. 
       
\subsubsection{\Rb\ and \Rc\ Measurements}

The formulation of \Rb\ and \Rc\ as ratios of hadronic cross sections
ensures that the propagator (oblique) electroweak radiative corrections 
and QCD radiative corrections that are common to all flavors largely 
cancel, isolating the heavy quark to \Z\ coupling vertex radiative 
corrections. In the standard model, the large top quark mass introduces a $-1.5$\% 
correction on \Rb\ \cite{ref:Rb-topmass}, compared with the tree-level 
prediction. Extensions to the standard model can also produce potential deviations 
in \Rb\ at $\sim$1\%~\cite{ref:Rb-exotic}. 
The effort toward $< 1\%$ precision $R_b$ 
measurements has therefore been one of the primary activities of the 
\Z-pole experiments. The \Zcc\ vertex corrections in the standard model are much
smaller, so that any deviations of the measured \Rc\ from
the standard-model prediction would signal exotic new physics processes. 

At the peak of the \Rb\ and \Rc\ ``crisis'' in early 1996, the world average 
for \Rb\ was over $3 \sigma$ higher 
than the standard model, whereas
that for \Rc\ was over $2 \sigma$ lower than the 
standard model~\cite{ref:Moriond96}.
SLD's crucial contribution was to introduce an improved analysis method
that was eventually adopted by other experiments, resulting in
a significant increase in precision.
      
\noindent
%
%
{\bf \Rb\ Measurement~~~~} 
Recent \Rb\ measurements have generally adopted a double-tag technique 
to reduce modeling uncertainty. Events were divided 
into two hemispheres by the plane perpendicular to the thrust axis,
and a $b$-tagging algorithm was applied to each hemisphere in turn. 
The measured hemisphere tag rate $F_s$ and event double-tag rate $F_d$
allow the extraction of both \Rb\ and the hemisphere $b$-tagging 
efficiency $\epsilon_b$ from the data by solving two simultaneous 
equations:
\begin{eqnarray} 
F_s & = &\epsilon_b R_b + \epsilon_c R_c + \epsilon_{uds}(1-R_b-R_c), 
\nonumber\\
F_d & = & C_b \epsilon_b^2 R_b + C_c \epsilon_c^2 R_c + 
          \epsilon_{uds}^2 (1-R_b-R_c).
\label{eqn:rbdtag}
\end{eqnarray} 
The small background tagging efficiencies for $uds$ and 
charm hemispheres, $\epsilon_{uds}$ and $\epsilon_c$, 
as well as the $b$-tagging hemisphere correlations 
$C_b=\frac{\epsilon^{\rm double}_b}{\epsilon^2_b}$ and 
$C_c=\frac{\epsilon^{\rm double}_c}{\epsilon^2_c}$, were estimated from 
the \revise{MC}{Monte Carlo simulation}. Furthermore, a standard-model value of \Rc\ was assumed.
The $R_b$ statistical error is approximately $\propto 1/\epsilon_b$, 
whereas the $uds$ and charm systematic errors scale as 
$\epsilon_{uds}/\epsilon_b$ and $\epsilon_c/\epsilon_b$,
indicating the need to maintain both high efficiency and high purity 
for the $b$ tag.

\begin{figure}[tbp]
\vspace{-1.0cm} 
\begin{center}
  \epsfxsize16cm
  \epsfbox{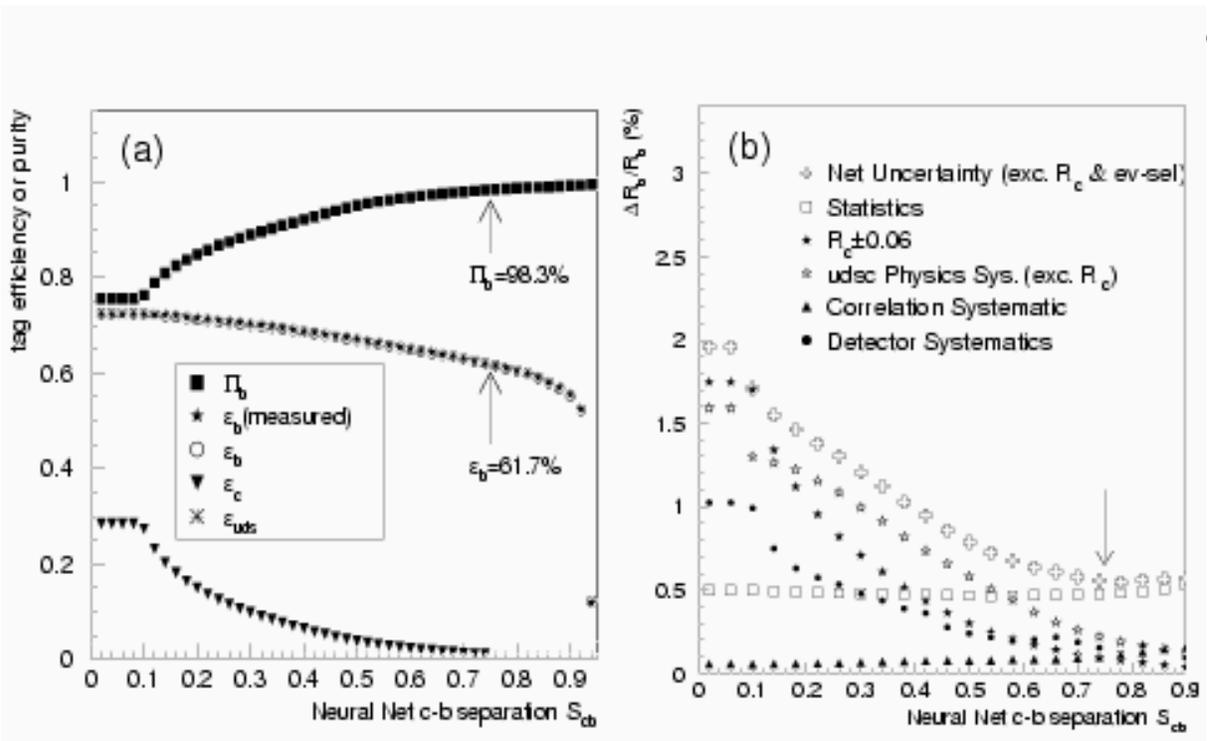}
\end{center}
\vspace{-0.5cm} 
\caption{($a$) $b$-tagging efficiency and purity \revise{vs.}{versus} $S_{cb}$ cut. ($b$) $R_b$
statistical and systematic errors \revise{vs.}{versus} $S_{cb}$ cut.}
\label{fig:rbtagsys}
\end{figure}

For the preliminary analysis of the 1996--1998 data, a cut on the 
neural-net $c-b$ separation variable of $S_{cb}>0.75$ (see Section
\ref{sec:analysis-bctag}) was used as the $b$ tag.
Figure~\ref{fig:rbtagsys}$a$ shows the hemisphere tagging 
efficiencies and $b$ purity for 
the 1997--1998 data as a function of the $S_{cb}$ cut. It can be seen that 
the measured $b$-tagging efficiency is in good agreement 
with the Monte Carlo simulation.
At the nominal cut of $S_{cb}$=0.75, the tagging 
efficiencies for various flavors were $\epsilon_{uds}$=0.07\%,
$\epsilon_c$=1.1\%, and $\epsilon_b^{\rm data}$=61.8\%, which corresponded
to a $b$ purity of $\Pi_b$=98.3\%. The small hemisphere correlation 
($1-C_b=0.00007\pm0.00050$) was due to a combination of several
effects related to the IP, $b$-tagging efficiency dependence on 
$b$-hadron \revise{momentum and}{momentum, and} $\cos\theta$,     
which were all \revise{checked}{verified} to be small. 
Table~\ref{tab:rbrcsys} presents the preliminary \Rb\ measurement result and systematic errors, combined with the published 
1993--1995 result. The choice of the $S_{cb}>0.75$ cut was the
result of an optimization that minimized the total 
error, as seen in figure~\ref{fig:rbtagsys}$b$. 

\noindent
%
%
{\bf \Rc\ Measurement~~~~}
All \Rb\ measurements now rely on the long $b$-hadron lifetime to
exploit the double-tag technique, but the shorter charm lifetime
and limited vertex detector resolution prevent the effective use
of that technique for \Rc\ measurements in most experiments.
This led to more complicated paths for the \Rc\ 
measurements at \revise{LEP with variety of techniques mostly}{LEP, 
mostly involving  techniques that}
relied on fully 
or partially reconstructed charm hadrons \cite{ref:LEPEWsum}. These
methods were 
not only limited statistically, owing to low efficiency, but also limited
systematically at the $\pm0.010$ level, owing to the lack of clean 
self-calibration of tagging efficiencies from data.
In 1997, SLD introduced an efficient double-tag analysis, which led to,
by far, the most precise \Rc\ measurement.  
          
The current preliminary SLD measurement of \Rc\ is based on 
the 1996--1998 data only. 
Since the $b$ tag separated out the majority of 
$b$ hemispheres with high purity, 
the \Rb\ analysis used a simple double-tag technique.
The \Rc\ measurement, on the other hand, benefited significantly 
from the additional statistics of \ccbar\ hemispheres in a wider 
range of $S_{cb}$ values. Although the charm purity was 
moderate compared with the $b$-tagging purity, the remaining impurities
were mainly $b$ hemispheres
\revise{for which their fractions}{whose fractions} could be precisely
measured from data. The \Rc\ 
measurement therefore adopted a multi-tag scheme, which was
also used in some LEP \Rb\ measurements. 
Hemispheres were classified into five categories, corresponding 
to four exclusive categories of $b$ and $c$ tags defined by various 
$S_{cb}$ ranges and \revise{the remainder}{a fifth} category with no tag. 
Table~\ref{tab:rctags} 
lists the efficiencies and purities of the four different $b$ and $c$
tags for the 1997--1998 data analysis.   
\begin{table}[tbp]
\caption{Hemisphere tagging efficiencies and purities for tags 
         used in the \Rc\ analysis, based on the $c-b$ separation 
         variable $S_{cb}$ (boxed efficiency values were 
         measured values from data)}
\vspace{0.4 cm}
\label{tab:rctags}
\begin{center}
\begin{tabular}{@{}crrrr@{}}%
\toprule
  Tag Name & ``$c$-pure'' & ``$c$-like'' & ``$b$-like'' & ``$b$-pure''  \\
  Tag Cuts & $S_{cb}<0.3$ & $0.3<S_{cb}<0.5$ & $0.5<S_{cb}<0.75$ 
           & $S_{cb}>0.75$ \\ 
\colrule 
  $\epsilon_b$ (\%)     & \framebox{2.53}  & \framebox{2.96} 
                        & \framebox{5.10}  & \framebox{62.02} \\   
  $\epsilon_c$ (\%)     & \framebox{17.94} & \framebox{5.04}
                        & \framebox{2.29}  &  1.12 \\
  $\epsilon_{uds}$ (\%) &    0.05          &  0.10            
                        &    0.12          &  0.07 \\
\colrule                   
   $b$ purity (\%)      &  15.0  &  40.9 & 70.4  & 98.3  \\
   $c$ purity (\%)      &  84.2  &  55.3 & 25.1  &  1.4  \\
   $uds$ purity (\%)    &   0.9  &   3.8 &  4.5  &  0.3  \\
\botrule
\end{tabular}
\end{center} 
\end{table}
The measured event rates, $N_{ij}/N_{\rm tot}$ for two hemispheres  
(categories $i$ versus $j$), were used in a $\chi^2$ fit for all 
tag combinations:
\begin{equation} 
\chi^2 =  \sum_{ij} \frac{ N_{ij}/N_{\rm tot} 
          - \Rb\epsilon_i^b\epsilon_j^b C_{ij}^b   
          - \Rc\epsilon_i^c\epsilon_j^c C_{ij}^c   
          - (1-\Rb-\Rc)\epsilon_i^{uds}\epsilon_j^{uds} C_{ij}^{uds} } 
            {\sigma(N_{ij}/N_{\rm tot})}. 
\end{equation} 
Besides \Rb\ and \Rc, all $b$-tagging 
and $c$-tagging efficiencies (except the $c$-tagging efficiency
for the $b$-pure tag) 
were also measured from data. The small $uds$ 
efficiencies and the hemisphere correlations ($C_{ij}$) were estimated  
from the Monte Carlo simulation. The hemisphere correlation factors 
were found to be very close to 1 for all tags. The essence of this 
multi-tag scheme is that by looking at the tags opposite the high-purity 
$b$-pure and $c$-pure tags,
the flavor composition in the 
less pure tags was measured from data. 
However, this efficiency 
cross calibration is reliable only when the $uds$ fractions are kept 
as low as they were in this analysis. 

Table~\ref{tab:rbrcsys} summarizes the preliminary \Rc\ result
and its systematic errors.
The value of the simultaneously fitted \Rb\ was also very consistent with the
simple double-tag \Rb\ result. The various measured $b$- and $c$-tagging 
efficiencies were found to be close to the Monte Carlo simulation.

\noindent
%
%
{\bf \Rb\ and \Rc\ Summary~~~~}
The SLD measurements of \Rb\ and \Rc\ are summarized in 
Table~\ref{tab:rbrcsys}, where the results were corrected for
$\gamma$-exchange by $+0.0003$ for \Rb\ and $-0.0003$ for \Rc. 
\begin{table}[tbp]
\caption{Preliminary results and systematic summary for 
the \Rb\ and \Rc\ measurements}
\vspace{0.3 cm}
\label{tab:rbrcsys}
\begin{center}
\begin{tabular}{@{}lrr@{}}%
\toprule
  Observable                  &  \Rb\        & \Rc\       \\
\colrule 
  Data sample                 & 1993--1998   & 1996--1998  \\
  References                  & \cite{ref:Moriond01,ref:Rb935}
                              & \cite{ref:Moriond01,ref:RcSLD} \\
\colrule                     
  Measurement value           &  0.21641     &  0.17382  \\
  Statistical error           &  0.00092     &  0.00308  \\ 
\colrule 
  Monte Carlo statistics      &  0.00024     &  0.00095  \\
  Event selection bias        &  0.00026     &  0.00027  \\
  $uds$ and charm physics     &  0.00042     &  0.00142  \\
  $b$-hemisphere correlation  &  0.00023     &  0.00025  \\
  $g\ra\ccbar, g\ra\bbbar$    & $-$0.00023   & $-$0.00082  \\
  Detector effects            &  0.00043     &  0.00079  \\
  $R_c$ ($\pm$0.006)          & $-$0.00020   &    ---~~~ \\
  $R_b$ ($\pm$0.0015)         &    ---~~~    & $-$0.00020  \\
\colrule 
  Total systematic error      &  0.00080     &  0.00209  \\
\botrule
\end{tabular}
\end{center}
\end{table}

\begin{figure}[tbp]
\vspace{-0.4 cm}
\begin{center}
  \epsfxsize16cm
  \epsfbox{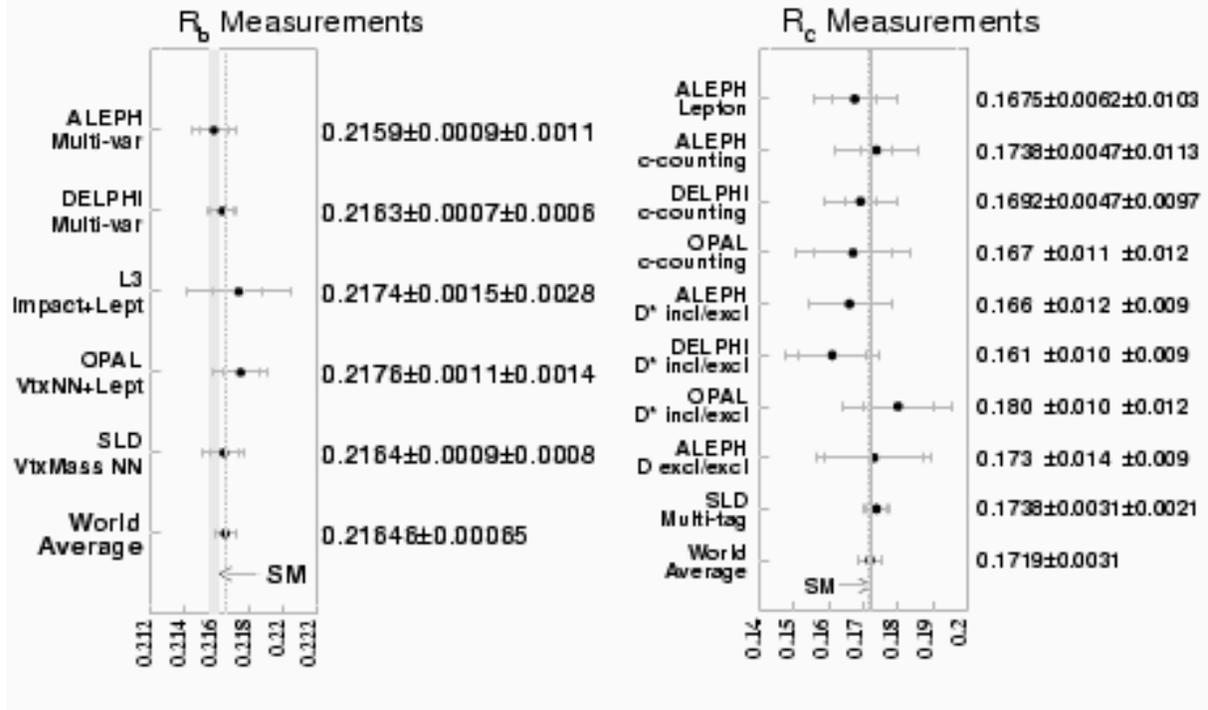}
\end{center}
\vspace{-1.0cm} 
\caption{Summary of \Rb\ and \Rc\ measurements from LEP and SLD.
         The shaded band for the \Rb\ standard model (SM) prediction
         reflects the uncertainty in the top quark mass.}
\label{fig:rbrcmeas}
\end{figure}

These results are in good agreement with the 
LEP measurements \cite{ref:LEPEWsum} and the standard model (see 
Figure~\ref{fig:rbrcmeas}). Despite the disadvantage of 
event statistics, the SLD \Rb\ measurement is competitive in 
precision, and the SLD \Rc\ measurement
dominates the world average with 
a systematic error three times smaller than that of 
other measurements.
The \Rb\ crisis is now believed to have been caused by
underestimated detector resolution effects that resulted in an excess 
of light-flavor tags in one of the early LEP measurements.
The \Rc\ deviation from the standard model was related to
incorrect assumptions on charm production and decay branching ratios,
which were later measured directly from the data.
Recent measurements are less prone to these problems
as improved analysis methods have largely reduced the sensitivity
to physics and detector modeling uncertainties.

\subsubsection{\Ab\ and \Ac\ Measurements}

The direct measurements of \Ab\ and \Ac\ at SLD are mostly sensitive 
to deviations in the right-handed \Zbb\ and \Zcc\ couplings, which 
have unambiguous standard-model predictions with small radiative corrections.
Most proposed extensions to the standard 
model  predict changes
in the left-handed \Zbb\ coupling, with little or no effect on the
right-handed coupling. Any deviations from the standard-model prediction 
for \Ab\ and \Ac\ will therefore be sensitive to exotic new physics processes.
   
The heavy-quark asymmetry analyses typically involved quark flavor 
tags to enrich the fraction of the desired heavy-quark species. 
The primary $\QQbar$ production axis was typically approximated
by the event thrust axis.
Various techniques were then used to determine the direction of  
the quark as opposed to the antiquark. We refer \revise{}{to} this process as
the quark charge assignment. All analyses adopted maximum likelihood 
fits to the polarized differential cross section 
(see Equation~\ref{Equ_cross}) as a function of $\cos\theta$, 
and events were weighted appropriately based on the correct
quark charge assignment probability, event quark-flavor fractions,
and electron beam helicity. The information drawn from this procedure 
is equivalent to that obtained from the left-right 
forward-backward asymmetry $\AfbtQ$ at first order.
The measurements were corrected for QCD and QED radiative effects.
The dependence of radiative corrections on center-of-mass energy 
was negligible.      

In the following, we first describe the analysis procedure,
emphasizing features that are special to SLD. The results are 
presented in Table~\ref{tab:abacsys}.

\noindent
%
%
{\bf \Ab\ and \Ac\ with Leptons~~~~} 
The use of leptons to identify heavy-hadron decays has been a 
traditional technique for the $b$- and $c$-asymmetry measurements. 
The lepton tags not only enrich the $b$- and $c$-quark events but
also provide the crucial separation between quark and antiquark.
In this SLD measurement, the estimation of correct charge 
assignment probability for this $Q/\overline{Q}$ separation was 
based on the probabilities from the Monte Carlo for the candidate 
lepton to be from various sources, e.g., \btol, \btocl, \ctol\, or
lepton misidentification, etc. This analysis benefited 
from the well-understood kinematics of semileptonic decays of 
heavy hadrons and the generally well-measured decay branching ratios 
for the various lepton sources. Besides the conventional lepton 
total momentum and transverse momentum with respect to 
the jet axis, the vertex mass $b$ tag and other vertexing 
observables were also used to improve the lepton source 
classification. A special variable assisting this separation was 
the lepton longitudinal position along the vertex axis, $L/D$ (see 
Section~\ref{sec:analysis-vertexing}), which had significant 
resolving power between direct secondary \btol\ and cascade 
tertiary \btocl\ leptons. The major sources of physics systematic 
uncertainty were \revise{due to}{} the various semileptonic branching ratios 
and $B$ mixing rates, which were taken from the LEP combined
lepton fit results \cite{ref:LEPEWsum}.  
      
\noindent{\bf \Ab\ with Jet Charge~~~~}
%
%
The use of momentum-weighted track charge (more commonly 
refered to as ``jet charge'') to sign the $b$-quark direction is now 
a standard technique for $b$-asymmetry measurements. 
The method is based on the correlation between the primary quark 
charge and the net charge of high-momentum tracks in the jet.
In this analysis, \bbbar\ events were selected by requiring at 
least one hemisphere with a vertex mass tag of $M_{\rm corr}>2$\gevcc. 
The momentum-weighted track charge was calculated from  
\begin{equation}
 Q = \sum_{\rm tracks}q_i \cdot \mbox{sign}(\vec{p}_i \cdot 
  {\hat T}) |(\vec{p}_i \cdot {\hat T} )|^{\kappa},
\end{equation}
where $q_i$ and $\vec{p}_i$ are the charge and momentum vector of track 
$i$, $\hat{T}$ is the thrust axis direction and $\kappa$ was chosen to be 
0.5 to optimize the charge determination. 

The correct charge assignment probability was calibrated from  
data. The approximately Gaussian shape of the jet charge 
distribution allowed a simple parameterization of the correct 
charge assignment probability as $p_b(|Q|)=1/(1+e^{-\alpha_b |Q|})$. 
The widths of the hemisphere charge sum and difference 
distributions were used to calculate the single parameter 
$\alpha_b$.
The correct charge assignment probability was evaluated for
each event and used in the likelihood fit.
On average, this probability was $\sim$69\%.
The small $udsc$ background was subtracted 
based on the Monte Carlo 
simulation, but the $b$ purity was measured from data using the 
double-tag technique, as in the case of the \Rb\ measurement. 
There was a small hemisphere charge correlation, analogous 
to the hemisphere tag correlation in the \Rb\ 
analysis, which was estimated from the Monte Carlo. 
\vspace{1.0 cm}

\noindent 
%
%
{\bf \Ac\ with $D^{(\ast)}$ and inclusive $D^{\ast +}\ra D^0\pi^+$~~~~}
A traditional way of tagging charm is simply to reconstruct 
$D^{(\ast)}$ decays. Two measurements of \Ac\ were 
performed. The first analysis used fully or partially reconstructed
$D^{\ast+}$, $D^+$, and $D^0$ 
mesons;\footnote{Charge conjugate modes are implied
throughout this paper.}
the second used inclusively reconstructed  
$D^{\ast+}\ra D^0\pi^+$ decays. These methods not
only selected the \ccbar\ events, the $D^{(\ast)}$ 
($\overline{D}^{(\ast)}$) mesons naturally tagged $c$ (\cbar) quarks
with high purity. The behavior of the random combinatorial background 
(RCBG) events can be conveniently studied with the events in the 
mass sidebands. A simple $b$ tag with mass requirement of 
$M_{\rm corr}>2$\gevcc\ was used to veto real $D^{(\ast)}$ background
from \bbbar\ events. 

The effective rejection 
of RCBG using precision vertexing
allowed the inclusion of many decay modes in this measurement.
The exclusive $D^{\ast+}\ra D^0\pi^+$ analysis utilized four $D^0$ 
decay modes: $K^-\pi^+$, $K^-\pi^+\pi^0$, 
$K^-\pi^+\pi^-\pi^+$, and $K^-\ell^+\nu$. 
The direct $D$ meson reconstructions used two decay 
modes: $D^+\ra K^-\pi^+\pi^+$ and $D^0\ra K^-\pi^+$. 
CRID kaon identification was used for the $D^0\ra K^-\pi^+$ mode. 
The sample of all $D^{(\ast)}$ candidates had the following composition:
$c\ra D$/$b\ra D$/RCBG=71\%/7\%/22\%.       

The inclusive $D^{\ast +}\ra D^0\pi_s^+$ analysis exploited the 
fact that a high-momentum leading $D^\ast$ in 
a \ccbar\ jet would 
follow the jet axis very closely. Because of the low $Q^2$ of the 
$D^{\ast +}\ra D^0\pi_s^+$ decay, the $\pi_s$ would also travel 
very close to the jet axis. The $\pi_s$ candidates having
momentum transverse to the jet axis
$p_T^2 < 0.01$ (\gevc)$^2$ were selected with a 
signal-to-background ratio of 1:2 and an efficiency three times 
higher than the exclusively reconstructed sample.       
For the combined result in Table~\ref{tab:abacsys}, the  
overlapping candidates between the two analyses were removed 
from the inclusive analysis for the combined \Ac\ result.  

\noindent
%
%
{\bf \Ab\ and \Ac\ with Vertex Charge and Kaon Tag~~~~}
The most precise \Ab\ and \Ac\ measurements at SLD were based
on two novel quark charge assignment techniques: vertex 
charge and identified kaon charge.

The analysis selected $b$ and $c$ hemispheres mainly
using the neural-net $c-b$ separation variable $S_{cb}$ (see 
Section~\ref{sec:analysis-bctag}). The hemisphere $b$  tag 
required $S_{cb}>0.9$ and $M_{\rm corr}<7$\gevcc, and the 
$c$ tag required $S_{cb}<0.4$ and a momentum sum of all
secondary tracks $>5$ \gevc. 

\begin{figure}[tbp]
\begin{center}
  \epsfxsize16cm
  \epsfbox{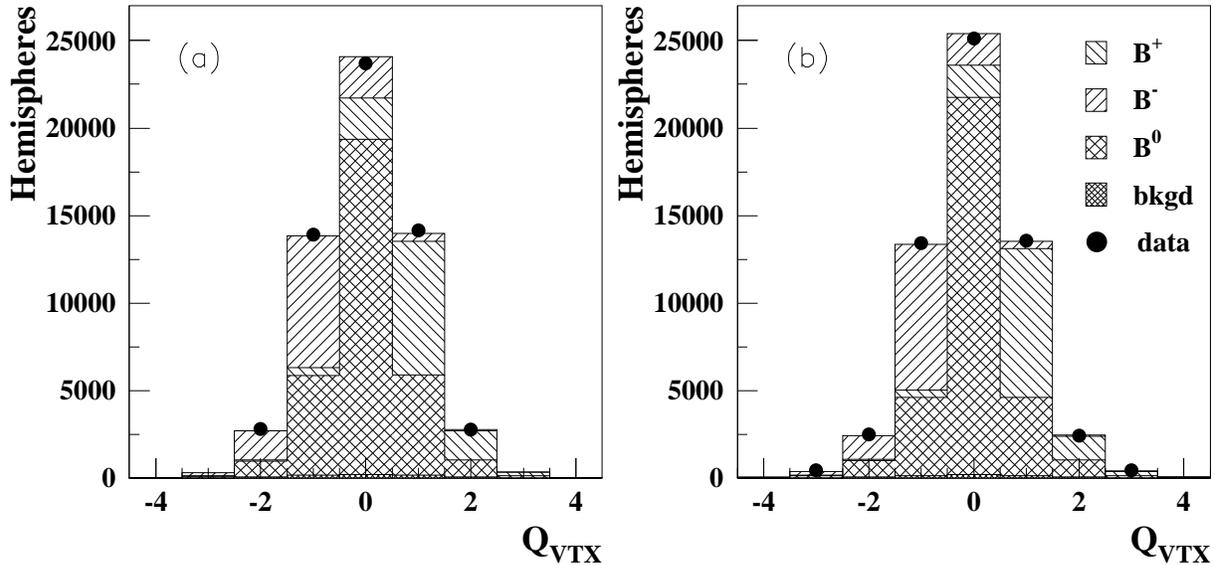}
\end{center}
\vspace{-0.5cm} 
\caption{Vertex charge distributions for the $b$-tagged 
 hemispheres, ($a$) with fully fitted tracks and
 ($b$) including VXD track segments.
 The ``$B^0$'' category shown in the plot includes
 all neutral $b$ hadrons.}
\label{fig:vtxchrg}
\end{figure}

A clean reconstruction of the secondary vertex charge 
tagged the heavy-quark charge. This is especially 
beneficial in the case of $b$ hadrons because both lepton and 
jet-charge techniques suffer from dilution from neutral 
$B$-meson mixing. This method required clean and efficient
separation of primary and secondary tracks, fortunately
achievable with the precision IP and tracking at SLD. The 
secondary track selection procedure can be found in 
Section~\ref{sec:analysis-vertexing}.
Figure~\ref{fig:vtxchrg} shows the vertex charge 
distributions for the $b$-tagged hemispheres,
demonstrating good agreement between
data and Monte Carlo and a clear $+$/$-$ charge separation. Note that the
inclusion of the VXD track segments in the charge 
reconstruction noticeably improved the correct charge 
assignment fraction. For the $b$- ($c$-) tagged hemispheres, the 
fraction of hemispheres with the secondary decay identified
as charged was 56\% (45\%), and the correct $b$- ($c$-) quark charge
identification probability was 81\% (91\%). 

Another quark charge assignment technique was to use the 
dominant $b\ra c\ra s \ra K^-$ and $c\ra s \ra K^-$ decay chains, 
in which the $K^\pm$ were identified by the CRID. 
However, in the case of $b$ hemispheres,
the additional contribution from the $K^\pm$ tag was
found to be small.
This was mostly due to the fact that the hemispheres
with no vertex-charge tag
contained an enhanced fraction of neutral $B$ mesons,
which had inherently poorer $K^\pm$ tag performance
due to $B$ mixing.
On the other hand,
a $K^\pm$ tag fraction of 30\%
and a correct $c$-quark charge identification 
probability of 86\% for the $c$-tagged hemispheres,
provided a very effective additional charge assignment
contribution.
Therefore, the \Ab\ analysis used the vertex-charge tag
only, while the \Ac\ analysis used both vertex-charge
and kaon-charge tags.

The vertex-charge and kaon-charge tags were used to 
make a joint quark charge assignment in the \Ac\ analysis. 
In cases of conflict between a 
vertex-charge and kaon-charge assignment
in the same hemisphere, 
no charge assignment was made. 
Thanks to the high efficiency and the good performance of the 
quark charge assignment, we were able to determine 
the event flavor composition and quark charge assignment 
probability simultaneously from data using a hemisphere
double-tag technique.  
This procedure used the small $uds$ efficiencies and 
hemisphere correlations from the Monte Carlo
simulation and assumed the world 
average values of \Rb\ and \Rc. 

\begin{figure}[tbp]
\begin{center}
  \epsfxsize16cm
  \epsfbox{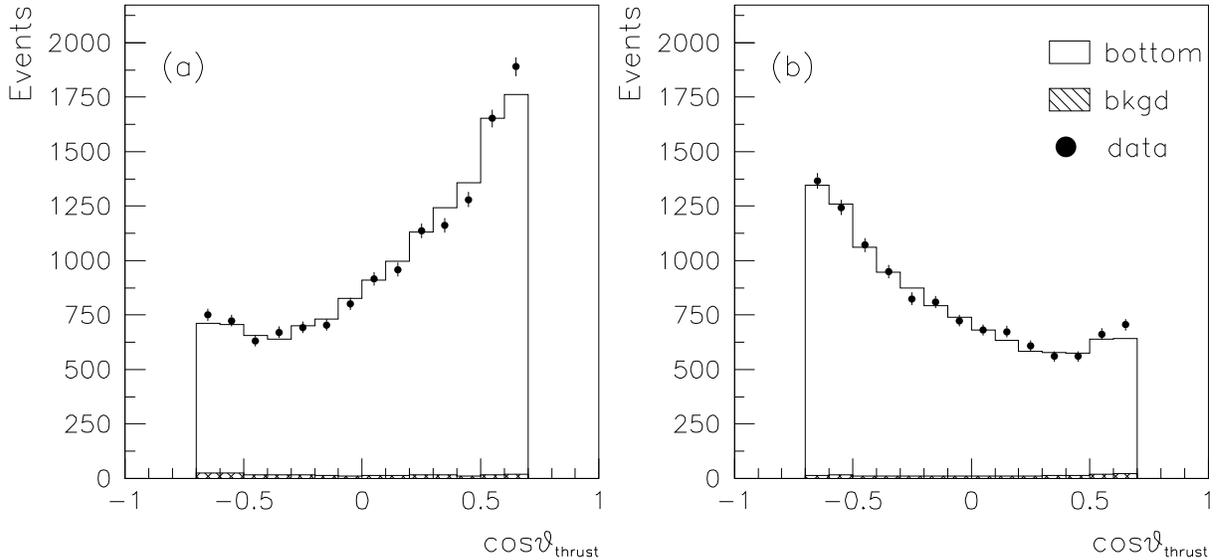}
\end{center}
\vspace{-0.5cm} 
\caption{Event thrust $\cos\theta$ distributions for ($a$) left- and 
($b$) right-handed beams in the vertex-charge \Ab\ analysis, for
data (points) and Monte Carlo simulation (histograms).}
\label{fig:vtxqasym}
\end{figure}

For the \Ab\ and \Ac\ fits, events with either 
hemisphere having a $b$ tag were classified as \bbbar\ events, 
\revise{while}{whereas} events with either hemisphere having a $c$ tag and no 
hemisphere having a $b$ tag were classified as \ccbar\ events.         
Events with two hemispheres having the same charge were discarded,
\revise{while}{and} events with two hemispheres having the opposite charge 
were weighted by the joint correct charge assignment probability. 
The \bbbar\ sample had a $b$ purity of $97.5\pm0.5\%$ and 
a correct $b$-quark charge fraction of $81.5\pm0.5\%$. 
The $b$-quark $\cos\theta$ distributions of $b$-tagged 
events with vertex charge are shown in Figure~\ref{fig:vtxqasym}.  
The most significant systematic uncertainty was due to 
charge assignment calibration statistics.

\noindent
%
%
{\bf \Ab\ and \Ac\ Summary~~~~}
Table~\ref{tab:abacsys} summarizes results for the various 
\Ab\ and \Ac\ measurements. 
\begin{table}[tbp]
\caption{Summary of \Ab\ and \Ac\ results, detailing statistical and 
systematic uncertainties. All measurements are 
preliminary except for the \Ac\ measurement with $D^{(\ast)}$.}
\vspace{0.4 cm}
\label{tab:abacsys}
\begin{center} 
\begin{tabular}{@{}|l|rrr|rrr|@{}}%
\toprule
  Observable                  & \multicolumn{3}{|c|}{$\Ab$}  
                              & \multicolumn{3}{|c|}{$\Ac$} \\
\colrule 
  Measurement type            & Lepton & Jet-Q      & Vtx-Q 
                              & Lepton &$D^{(\ast)}$& Vtx+K \\
  Data sample                 & 1993--98  & 1993--98      & 1996--98 
                              & 1993--98  & 1993--98      & 1996--98 \\
  References                  & \cite{ref:AbAcLept968,ref:AbAcLept935}
                              & \cite{ref:Moriond01,ref:AbJetQ935} 
                              & \cite{ref:Moriond01,ref:AbAcVtxK}   
                              & \cite{ref:AbAcLept968,ref:AbAcLept935}
                              & \cite{ref:AcDsD}   
                              & \cite{ref:Moriond01,ref:AbAcVtxK} \\  
\colrule 
  Measurement value           & 0.924  &   0.907    & 0.921
                              & 0.589  &   0.688    & 0.673 \\
  Statistical error           & 0.030  &   0.020    & 0.018
                              & 0.055  &   0.035    & 0.029 \\
\colrule
  Monte Carlo statistics      & 0.005  &   $-$      & 0.003    
                              & 0.017  &   $-$      & 0.004  \\
  Calibration statistics      &  $-$   &   0.014    & 0.013    
                              &  $-$   &   $-$      & 0.021  \\
  Calibration systematic      &  $-$   &   0.016    & 0.007    
                              &  $-$   &    $-$     & 0.009  \\
  Detector effects            & 0.010  &   0.008    & 0.003    
                              & 0.020  &    $-$     & 0.003  \\
  $b$-tag purity              & 0.009  &   0.005    & 0.007    
                              & 0.036  &    $-$     & 0.003  \\
  Br(semi-$\ell$) \& $B$ mixing & 0.012&   0.000    & 0.000    
                              & 0.022  &   0.010    & 0.000  \\
  $b,c$ decay/prod. model     & 0.011  &   0.000    & 0.003    
                              & 0.018  &   0.018    & 0.004  \\
  $g\ra\ccbar, g\ra\bbbar$    & 0.002  &   0.001    &$<$0.001
                              & 0.001  & $<$0.001   & 0.002  \\ 
  Beam polarization           & 0.005  &   0.005    & 0.005    
                              & 0.004  &   0.003    & 0.004  \\
  QCD correction              & 0.005  &   0.003    & 0.003    
                              & 0.005  &   0.003    & 0.001  \\
  \Ac\ ($\pm 0.035$)          & 0.003  &   0.001    & 0.001          
                              & $-$    &   $-$      &  $-$   \\
  \Ab\ ($\pm 0.035$)          & $-$    &   $-$      &  $-$       
                              & 0.008  & $-$0.003   &$-$0.001  \\
\colrule 
  Total systematic error      & 0.023  &   0.024    & 0.018
                              & 0.053  &   0.021    & 0.024  \\
\botrule
\end{tabular}
\end{center} 
\end{table}
The raw measured asymmetry parameters were 
corrected for QCD and QED radiative effects. 
The QED corrections for \Ab\ and \Ac\ were rather small:  
$-0.2$\% and $+0.2$\%, respectively. These were already taken into
account for results in Table~\ref{tab:abacsys}. 
The size of these corrections can be compared to 
the analogous QED corrections of 2.5\% 
and 9\% applied to unpolarized asymmetries measured at LEP.
   
Whereas the QED corrections were generally calculated reliably 
with small uncertainties, the QCD corrections had significant 
theoretical uncertainties and \revise{}{were} sensitive to analysis details. 
The total integrated QCD correction was at the $\sim3.5\%$ 
level for an ideal measurement of either $A^Q_{FB}$ or $A_Q$. 
However, the analysis event selection and final-state particle 
reconstructions typically tended to suppress the events with hard 
gluon radiation and therefore only needed reduced QCD corrections. 
This can be understood because $b$ tags were typically more efficient 
for high-momentum $b$ hadrons, and high-$x$ $D^*$ reconstruction and 
high-momentum lepton selections also favored high-momentum $c$ and 
$b$ quarks in two-jet events. The QCD corrections were scaled down
by analysis-dependent factors ranging between 25\% and 75\%. 
The SLD \Ab\ and \Ac\ analyses used a first-order QCD correction 
calculated by Stav \& Olsen \cite{ref:StavOlsen}, 
which included heavy-quark mass effects
with explicit $\cos\theta$ 
dependence. The second-order QCD corrections and their 
uncertainties were based on the prescription found in Reference
\cite{ref:LEPEWWG-QCDnote}.  

The individual \Ab\ and \Ac\ measurements were combined, 
taking into account systematic correlations.  
The various \Ab\ measurements had significant event sample 
overlaps. A statistical correlation matrix was built, taking 
into account the different weight of each event used in the
analysis. The effective correlations obtained from this 
procedure were as follows: 
\begin{center} 
\begin{tabular}{lcll}
 Lepton &versus& Jet-Q: & 22\%, \\
 Lepton &versus& Vtx-Q: & 15\%, \\ 
 Jet-Q  &versus& Vtx-Q: & 32\%. 
\end{tabular}\\
\end{center} 
The combined result also included an older measurement 
of \Ab\ using the $K$ tag alone from the 1994--1995 data 
\cite{ref:AbKtag945}.
Given the much smaller samples for the various \Ac\ analyses,
their statistical correlations were deemed to be small. 
The combined preliminary SLD \Ab\ and \Ac\ results are
\begin{eqnarray}
 \Ab & = & 0.916 \pm 0.021, \nonumber\\ 
 \Ac & = & 0.670 \pm 0.027. 
\end{eqnarray}

\begin{figure}[tbp]
\begin{center}
  \epsfxsize16cm
  \epsfbox{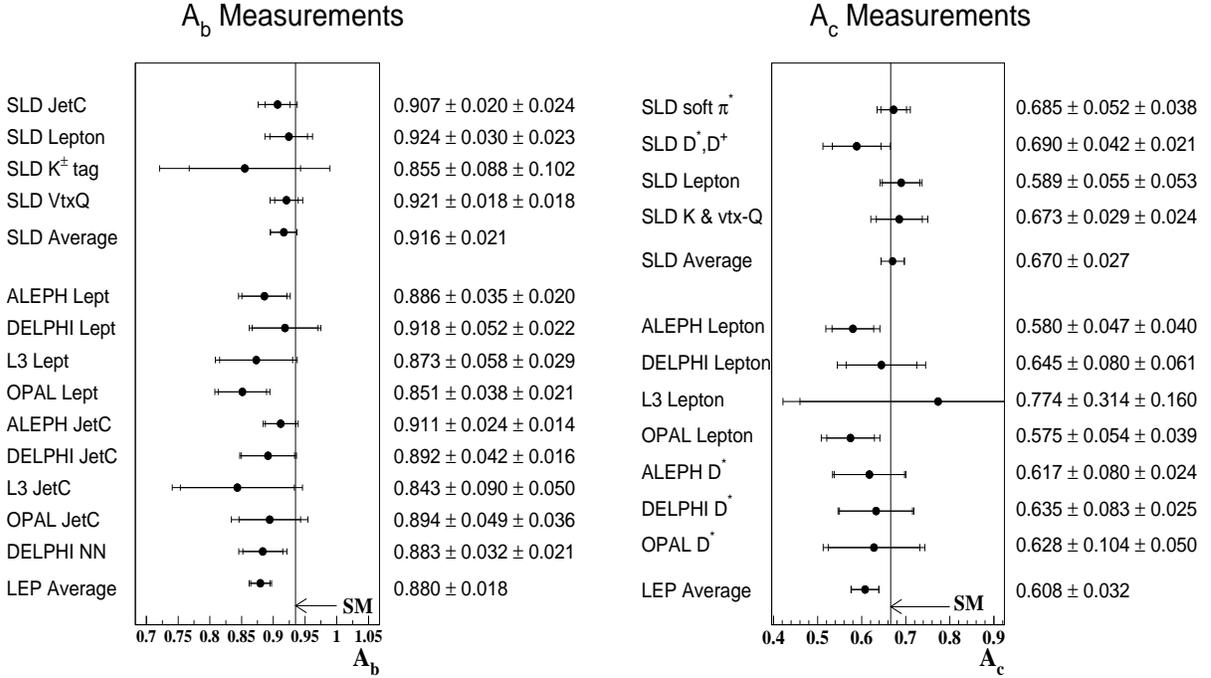}
\end{center}
\vspace{-0.5cm} 
\caption{Comparison between the \Ab\ and \Ac\ measurements by SLD 
and the \Ab\ and \Ac\ values derived from the indirect \Afbb\ and \Afbc\ 
measurements at LEP.}
\label{fig:abacmeas}
\end{figure}

Figure~\ref{fig:abacmeas} lists 
the individual \Ab\ and \Ac\ measurements, as well as the 
SLD average, 
along with the indirect measurements derived from the LEP $A_{FB}$ 
measurements, assuming a measured \Ae\ from the SLD 
and LEP combined $A_{\rm lepton}$ result of 
$\Ae=0.1501\pm0.0016$. 

\input sec5_as.tex

\vspace{0.6 cm}


%% file: sec5_as.tex
\subsubsection{$A_s$ Measurement}

  An important test of the standard model is to verify
the universality of coupling strengths of the \Z\ boson
to fermions with the same charge and weak isospin.
As described in the previous section, a number of precise measurements
of heavy-quark asymmetries have been performed, but
few measurements of light-quark asymmetries have been carried out,
owing to the lack of clear experimental signatures
to separate \Ztouds\ decays from one another.
Nevertheless, it is possible, albeit with low efficiency,
to select a fairly pure
sample of \Ztoss\ decays by requiring each event hemisphere
to contain high-momentum kaons, as outlined below.

  In the \As\ analysis~\cite{ref:As},
\Ztobb\ and \Ztocc\ decays were suppressed by requiring
the events to contain no
more than one well-measured track with normalized impact
parameter in the transverse plane $d / \sigma_d > 2.5$.
Charged kaons were identified with the CRID and required
to have $p > 9$\gevc\ to provide a sample 91.5\% pure in \Kpm.
Neutral kaons were identified via their decay into $\pi^+ \pi^-$
final states with a purity of 90.7\%.
An event was tagged as \Ztoss\ if one hemisphere contained
a \Kpm\ candidate and the other contained either an oppositely
charged \Kpm\ or a \KS\ candidate.
The respective \ssbar\  purities for \Kp \Km\ and \Kpm \KS\ events
were estimated to be 73\% and 60\% from the Monte Carlo simulation,
with an overall \ssbar\ event selection efficiency of 2.6\%.
Identification of the \s-quark hemisphere relied on the
\Kpm\ charge and was correctly assigned with a probability of
97.5\% for \Kp \Km\ events and
85.0\% for \Kpm \KS\ events.

\begin{figure}[tbp]
 \begin{center}
  \epsfxsize13cm
  \epsfbox{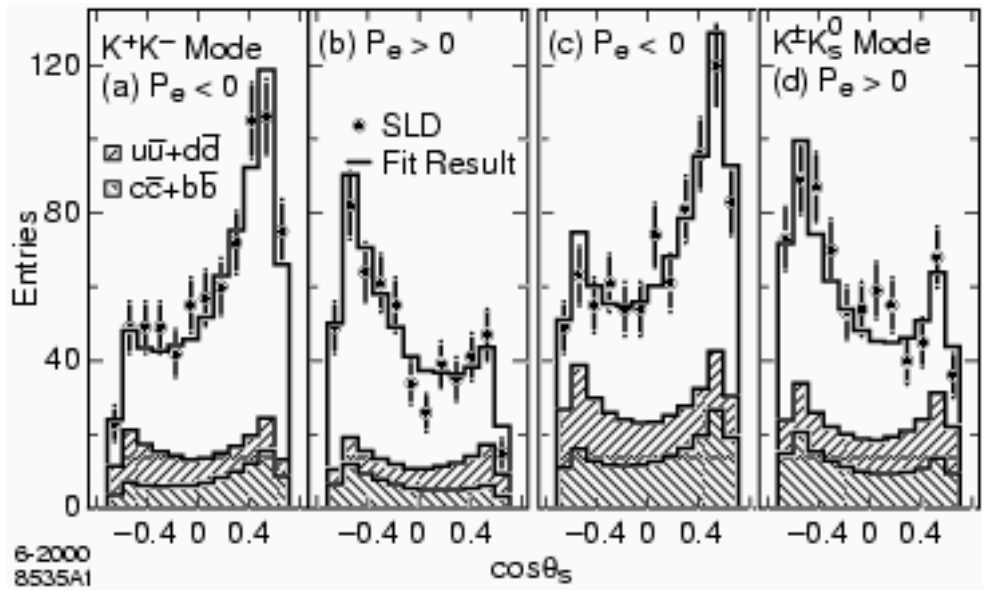}
 \end{center}
 \vspace{-0.5 cm}
\caption{Distributions of the \s-quark \ctheta\ for data (points)
in ($a$) \Kp\ \Km\ events with $\Pe < 0$, ($b$) \Kp\ \Km\ events with $\Pe > 0$,
($c$) \Kpm\ \KS\ events with $\Pe < 0$, ($d$) \Kpm\ \KS\ events with $\Pe > 0$.
The histograms represent the result of the fit.}
\label{fig:KKasym}
\end{figure}

  A maximum likelihood method was used to extract \As\
from the \s-quark $\cos\theta$ distributions for left-
and right-handed electrons in both samples (see Figure~\ref{fig:KKasym}).
From the figure, it is apparent that \bbbar\ and \ccbar\ backgrounds
have asymmetries with the same sign as the signal asymmetry,
but that \uubar\ and \ddbar\ events contribute with the opposite sign.
Therefore, uncertainties in the modeling of light quarks
have the greatest impact
on the value of \As.
Using the 1993--1998 data, the analysis determined
$\As = 0.895 \pm 0.066 {\rm ~(stat.)} \pm 0.062 {\rm ~(syst.)}$,
where the systematic error is dominated by
uncertainties in the \uubar\ and \ddbar\ contributions to
the \ctheta\ distributions.
To reduce modeling uncertainties,
these contributions were constrained using the data, as was
the \s-quark correct charge assignment probability.

%% file: sec6_bphysics.tex
\section{$\bsmix$ MIXING}

\subsection{Introduction}

  The special features of the SLC and SLD environment were particularly
well-suited for studies of $B$ decays.
Measurements of the average $b$-hadron
lifetime~\cite{avgblife} and of the \Bu\ and \Bd\ lifetimes~\cite{blifeprl}
provided the initial impetus for the development of the
inclusive topological vertexing described above.
Recent updates of the \Bu\ and \Bd\ lifetime measurements~\cite{blifelp99}
are among the most precise available to date and
have helped establish a small but
significant difference between the lifetimes of the \Bu\ and \Bd\ mesons,
in agreement with predictions based on the heavy-quark expansion
technique~\cite{Bigi,Neubert}.

  Investigations of the properties of $B$ meson decays at SLD have included the
search for charmless hadronic decays into two-body and quasi--two-body final
states~\cite{charmless}, the search for flavor-changing neutral-current
$b \to s$~gluon decays~\cite{bsg}, and the measurement of the charm hadron
yield~\cite{ncharm}.

  A topic of particular interest is the study of the time dependence of
\bdmix\ and \bsmix\ mixing. Experimental studies of mixing
exploit all the special features of the SLC/SLD environment, i.e.,
beam polarization, small beam spot, high-resolution vertexing,
and good particle identification.
Below, we focus on \bsmix\ mixing studies
in some detail.
A recent study of \bdmix\ mixing with a kaon tag is described
elsewhere~\cite{ThomEPS}.


\subsection{Theory}

  The neutral $B$ meson system consists of \Bq\ and \Bqb\ flavor eigenstates,
which are superpositions of heavy and light mass eigenstates
($q= d$ and $s$ for \Bd\ and \Bs\ mesons, respectively).
The mass eigenstates evolve differently
as a function of time, resulting in time-dependent \bqmix\ oscillations
with a frequency equal to the mass difference \dmq\ between the heavy and light
eigenstates. As a consequence,
an initially pure $|\Bq\rangle$ state may be found to decay as $|\Bq\rangle$
or $|\Bqb\rangle$ at a later time $t$ with a probability
equal to
\begin{eqnarray}
 P(\Bq \to \Bq) & = &
 \frac{\Gamma}{2} e^{-\Gamma t} (1+\cos\dmq\, t) {\rm ~~~~or}
 \label{eq_punmixed} \\
 P(\Bq \to \Bqb) & = &
 \frac{\Gamma}{2} e^{-\Gamma t} (1-\cos\dmq\, t).
 \label{eq_pmixed}
\end{eqnarray}
(Here we neglected the lifetime difference between mass eigenstates.)

  The oscillation frequency can be computed to be~\cite{Buras84}
\begin{equation}
  \dmq = \frac{G_F^2}{6\pi^2} m_{B_q} m_t^2 F(m_t^2 / M_W^2) f_{B_q}^2 B_{B_q}
         \eta_{\rm QCD} \left| V_{tb}^\ast V_{tq} \right|^2,
  \label{eq_dmq}
\end{equation}
where $G_F$ is the Fermi constant, $m_{B_q}$ is the $B^0_q$ hadron mass,
$m_t$ is the top quark mass, $M_W$ is the $W$ boson mass,
$F$ is a function defined in Reference~\cite{Inami81},
and $\eta_{\rm QCD}$ is a perturbative QCD parameter.
The parameter $B_{B_q}$ and the decay constant $f_{B_q}$ parameterize hadronic
matrix elements.
For further details, see the review by Gay~\cite{Gay00}.
Much of the interest in $B$ mixing stems from the fact that
a measurement of the
\Bd\ (\Bs) oscillation frequency allows the magnitude of the
poorly known CKM matrix element
$V_{td}$ ($V_{ts}$) to be determined (see Equation~\ref{eq_dmq}).
However,
the extraction of $|V_{td}|$ from
the fairly precisely measured value
$\dmd = 0.472 \pm 0.017$ ps$^{-1}$~\cite{PDG2000}
is affected by a theoretical uncertainty of about 20\% in the product
$f_{B_q} \sqrt{B_{B_q}}$~\cite{Bernard00}.
Uncertainties are reduced for the ratio
\begin{equation}
  \frac{\dms}{\dmd} = \frac{m_{B_s} f_{B_s}^2 B_{B_s}}{m_{B_d} f_{B_d}^2 B_{B_d}}
                      \left|\frac{V_{ts}}{V_{td}}\right|^2
                    = \frac{m_{B_s}}{m_{B_d}} (1.16 \pm 0.05)^2 \left|\frac{V_{ts}}{V_{td}}\right|^2~,
\end{equation}
which indicates that the ratio $|V_{ts}/V_{td}|$ can be determined
with an uncertainty as small as 5\%~\cite{Bernard00}.

  The CKM matrix describes weak quark mixing in the standard model
and contains an irreducible phase that provides the mechanism
for $CP$ violation in weak decays.
This matrix can be written in terms of four fundamental parameters:
$\lambda = \sin\theta_{\rm Cabibbo}$, $A$, $\rho$, and $\eta$
[as defined in the Wolfenstein parameterization \cite{Wolfenstein}].
The value of $\eta$ is directly related to the amount of $CP$ violation.
The parameters $\lambda$ and $A$ are well-measured,
$\lambda = 0.2237 \pm 0.0033$ and $A = 0.819 \pm 0.040$~\cite{Ciuchini00},
but $\rho$ and $\eta$ are among the least-known parameters of
the standard model.
In terms of these parameters, we have
$\dmd \propto |V_{td}|^2 \simeq A^2 \lambda^6 [(1-\bar{\rho})^2+\bar{\eta}^2]$
and $\dms \propto |V_{ts}|^2 \simeq A^2 \lambda^4$,
where $\bar{\rho} = \rho (1-\frac{\lambda^2}{2})$
and $\bar{\eta} = \eta (1-\frac{\lambda^2}{2})$.
As a result, the combination of \Bd\ and \Bs\ mixing provides one of the
strongest constraints on $\rho$ and $\eta$, and therefore on $CP$ violation,
in the standard model.

\subsection{Experimental Ingredients}

\begin{figure}[tbp]
\vspace{-0.8 cm}
\begin{center}
  \epsfxsize13cm
  \epsfbox{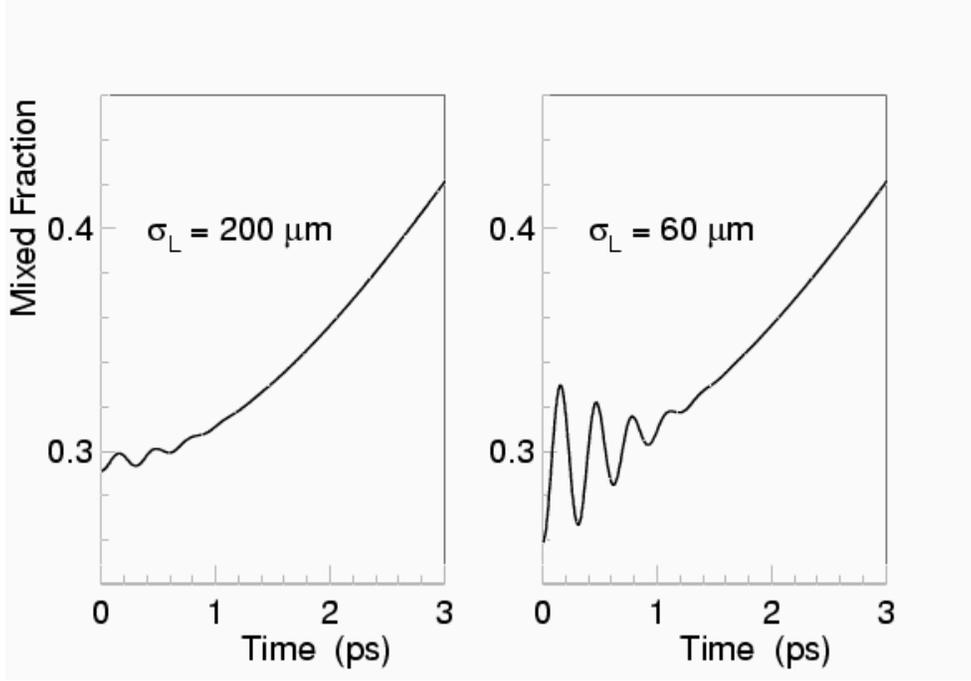}
\end{center}
\vspace{-0.5 cm}
\caption{Fraction of $b$-hadron decays tagged ``mixed'' as a function of
proper time for two different decay length resolutions,
$\sigma_L = 200\:\mu$m and $60\:\mu$m.
Other relevant parameters are
$\dms = 20 \invps$, $f_s = 18\%$, $w = 0.25$,
and $\sigma_p / p = 10\%$.}
\label{fig:fmixillus}
\end{figure}

  Experimental studies of \bsmix\ mixing
require two main ingredients:
($a$) the measurement of the \Bs\ decay proper time, and
($b$) the determination of the \Bs\ or \Bsb\ flavor at both production and
decay to classify the decay as either ``mixed'' (if the tags disagree)
or ``unmixed'' (otherwise).
The significance for a \Bs\ oscillation signal can be approximated
by~\cite{Moser97}
\begin{equation}
  S = \sqrt{\frac{N}{2}}\: f_s\: \left[1 - 2\, w\right]\:
      e^{-\frac{1}{2} (\dms \sigma_t)^2} ,
  \label{eq_signif}
\end{equation}
where $N$ is the total number of decays selected,
$f_s$ is the fraction of \Bs\ mesons in the selected sample,
$w$ is the probability of incorrectly tagging
a decay as mixed or unmixed
(i.e., the mistag rate),
and $\sigma_t$ is the proper time resolution.
The proper time resolution depends on both the decay length resolution
$\sigma_L$ and the momentum resolution $\sigma_p$ according to
$\sigma_t^2 = (\sigma_L / \gamma\beta c)^2 + (t\, \sigma_p/p)^2$.
Based on the Wolfenstein parameterization, we see that
$\dms / \dmd \simeq 1 / \lambda^2$, which is of the order of 20.
Therefore, \Bs\ oscillations are expected to be much more rapid
than \Bd\ oscillations.
The ability to resolve such rapid oscillations thus requires excellent
decay length resolution and benefits from having
a low mistag rate and a high \Bs\ purity.
The importance of decay length resolution is illustrated in Figure~\ref{fig:fmixillus},
which shows the fraction of $b$-hadron decays tagged ``mixed'' as
a function of proper time for $\dms = 20 \invps$, \Bs\ purity of 18\%,
mistag rate $w = 0.25$,
relative momentum resolution $\sigma_p / p = 10\%$, and
either $\sigma_L = 200\:\mu$m or $\sigma_L = 60\:\mu$m,
the former being typical for LEP experiments and the latter typical
for SLD.
It is clear from Figure~\ref{fig:fmixillus} that the decay
length resolution is the most critical ingredient for sensitivity
to the large \Bs\ oscillation frequencies that are expected.

\subsection{Analysis Methods}

  SLD performed three separate analyses using the 1996--1998 data: 
``$D_s$$+$tracks,'' ``lepton$+$$D$,'' and ``charge dipole.''
These analyses differ in the way the \Bs\ decay
candidates were reconstructed and the flavor at decay was tagged (see below).

\begin{figure}[tbp]
\vspace{-0.8 cm}
  \begin{center}
    \epsfxsize11cm
    \epsfbox{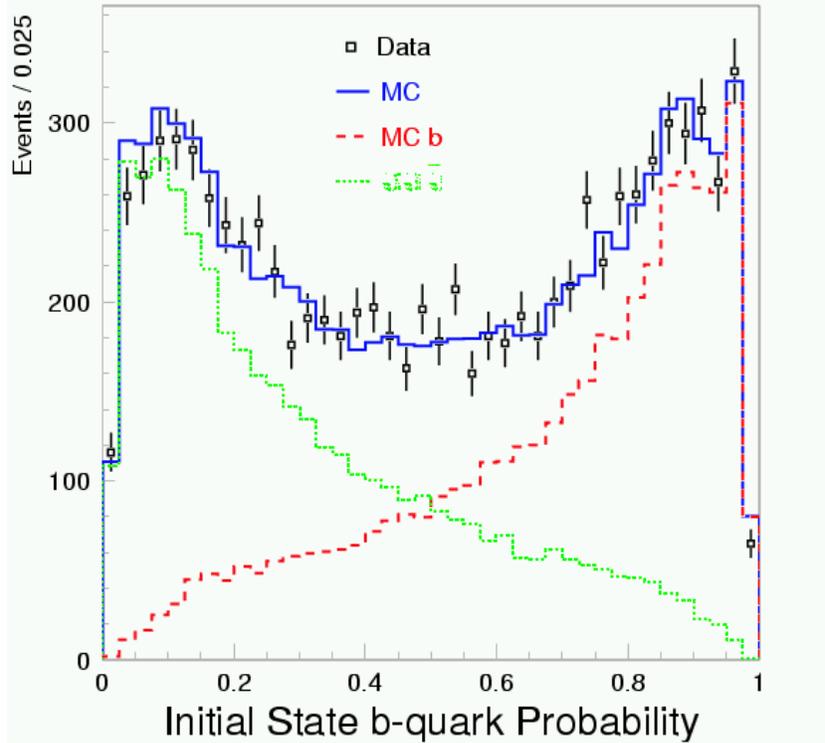}
  \end{center}
  \vspace{-0.5 cm}
  \caption{Distribution of the computed initial state $b$-quark probability for
  data (points) and Monte Carlo (histograms),
  for the events selected in the charge dipole analysis.}
  \label{fig_initag}
\end{figure}

Common to all analyses was the determination of the production
flavor (also known as the initial-state tag).
This was achieved by combining several tagging techniques.
The most powerful technique relied on the large
polarized forward-backward asymmetry in \Ztobb\ decays
and was unique to SLD.
In this case, a left- (right-) handed
incident electron tagged the forward hemisphere quark as a
$b$ ($\overline{b}$) quark.
The probability of correctly tagging a $b$ quark at production
is expressed as
\begin{equation}
P_A(\cos\theta) = \frac{1}{2} + \Ab~{{\Ae - \Pe}\over{1 - \Ae \Pe}}
        ~{{\cos\theta}\over{1+\cos^2\theta}}~.
\end{equation}
The average mistag rate was 28\% for an
average electron beam polarization of 0.73.
Other tags were also used. They relied on charge information from
the hemisphere opposite that of the \Bs\ decay candidate (i.e., the
hemisphere expected to contain the other $b$ hadron in the event):
($a$) momentum-weighted jet charge,
($b$) secondary vertex charge,
($c$) charge of kaon from the dominant decay transition $b \to c \to s$,
($d$) charge of lepton from the direct transition $b \to \ellm$, and
($e$) charge dipole.
The first four of these tags were used to measure \Ab\ and/or \Ac,
as described in Section~\ref{sec:phys-ewquark}, and the charge dipole tag
is described below.
These tags were estimated to have efficiencies (mistag rates)
of 100\% (66\%), 43\% (75\%), 16\% (74\%),
9\% (74\%), and 17\% (70\%), respectively.
The five tags were combined to yield an efficiency of 100\%
and a mistag rate of 28\%.
Overall, the average mistag rate was estimated to be 22\%--25\%, depending
on the $\cos\theta$ acceptance of each particular analysis.
Figure~\ref{fig_initag} shows the combined $b$-quark probability distributions
for data and Monte Carlo in the charge dipole analysis
and indicates a clear separation between
$b$ and $\overline{b}$ quarks.

  Also common to all analyses was the method to estimate
the $b$-hadron momentum.
It utilized the following information:
the momentum of all secondary tracks,
the energy deposited in the electromagnetic section of the calorimeter,
and kinematical constraints stemming from the known $b$-hadron mass
and the \epem\ collision center-of-mass energy.

\subsubsection{$D_s$$+$Tracks Analysis}

  The $D_s$$+$tracks analysis~\cite{bsmixds} aimed to partially reconstruct the
decay $\Bs \to D_s^- X$ with full reconstruction of the
cascade decay $D_s^- \to \phi \pi^-$ or $\Kstarz K^-$. 
As a result, the analysis achieved a high \Bs\ purity and excellent
decay length resolution but at the price of fairly low efficiency.
$D_s$ candidates were selected with a neural-network algorithm
combining kinematical quantities and particle identification information
to yield a sample of 280 $D_s^- \to \phi \pi^-$ and 81 $D_s^- \to \Kstarz K^-$ 
candidates.
These $D_s$ candidates were vertexed with one or more charged tracks to form
the \Bs\ decay.
A $D_s^-$ ($D_s^+$) tagged the decay flavor as \Bs\ (\Bsb).
Table~\ref{tab:bsmixanal} summarizes the performance of this analysis.
Increased sensitivity to \Bs\ decays was obtained by separately parameterizing
decays with a reconstructed track charge sum equal to or different from zero.
For example, \Bs\ purities above 50\% were obtained for the neutral sample.
Furthermore, $\Bs \to D_s^- \ellp \nu X$ candidates were identified
and assigned both higher purity and lower decay flavor mistag rate
than the average.

\subsubsection{Lepton$+$$D$ Analysis}

  The lepton$+$$D$ analysis~\cite{bsmixincl} attempted to
reconstruct $\Bs \to D_s^- \ellp \nu$ decays,
where the $D_s^-$ decay was not reconstructed exclusively but rather
with the inclusive topological vertexing technique described
in earlier sections.
The \Bs\ vertex was reconstructed by intersecting the lepton trajectory
with the $D$ trajectory, which is estimated from
the inclusively reconstructed $D$ vertex and its 
net momentum vector.
Direct leptons from $b \to \ell$ transitions were selected with
a neural-network algorithm that relied on the
following variables:
the transverse momentum of the lepton with respect to the
$B$ vertex direction (vector stretching from the IP to the $B$ vertex),
the $B$ decay length,
the transverse momentum of the lepton with respect
to the $D$ vertex direction (vector stretching from the $B$ vertex to the $D$ vertex),
the mass of the charged tracks associated with the $B$ decay,
and the distance of closest approach of the lepton to the $B$ vertex.
This algorithm effectively suppressed leptons from $b \to c \to \ell$
transitions, which
produce leptons with sign opposite that of direct leptons and
thus dilute the decay flavor tag.
To enhance the \Bs\ purity, the sum of lepton and $D$ vertex track charges was
required to be zero.
The purity was further enhanced from 16\% to 39\% in the subsample containing an
opposite-sign lepton-kaon pair.
Table~\ref{tab:bsmixanal} gives performance parameters.

\subsubsection{Charge Dipole Analysis}

\begin{figure}[tbp]
  \begin{center}
    \epsfxsize=11cm
    \epsfbox{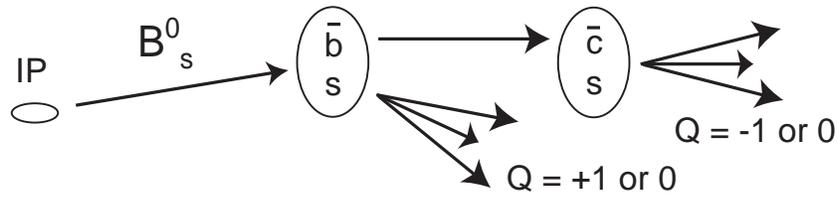}
  \end{center}
  \vspace{-0.5 cm}
  \caption{Sketch illustrating the decay cascade structure for
  $\Bs \to D_s^- X$ decays.}
  \label{fig:dipole_sketch}
\end{figure}

  The charge dipole analysis~\cite{bsmixincl}
was based on a novel method to inclusively reconstruct decays of the type
$\Bs \to D_s^- X$.
For each candidate hemisphere, events were selected in which both
a secondary ($B$) and a tertiary ($D$) vertex could be resolved.
A ``charge dipole'' $\delta Q$ was then defined as the distance between
secondary and tertiary vertices signed by the charge
difference between them such that
$\delta Q < 0$ ($\delta Q > 0$) tagged \Bz\ (\Bzb) decays
(see Figure~\ref{fig:dipole_sketch}).
This analysis, first developed by SLD, relied heavily
on the ability to recognize secondary tracks (from the $B$-decay point)
versus tertiary tracks (from the $D$-decay point).
It also benefited from the fact that \Bs\ decays yield
mostly charged secondary and tertiary vertices, whereas
\Bd\ mesons tend to produce neutral vertices since they
decay mostly into neutral $D$ mesons.

  As in the lepton$+$$D$ analysis, the total charge of all secondary
and tertiary tracks was required to be zero, in order to enhance
the \Bs\ purity from 10\% to 15\%.
\begin{figure}[tbp]
\vspace{-0.6 cm}
  \begin{center}
    \epsfxsize=11cm
    \epsfbox{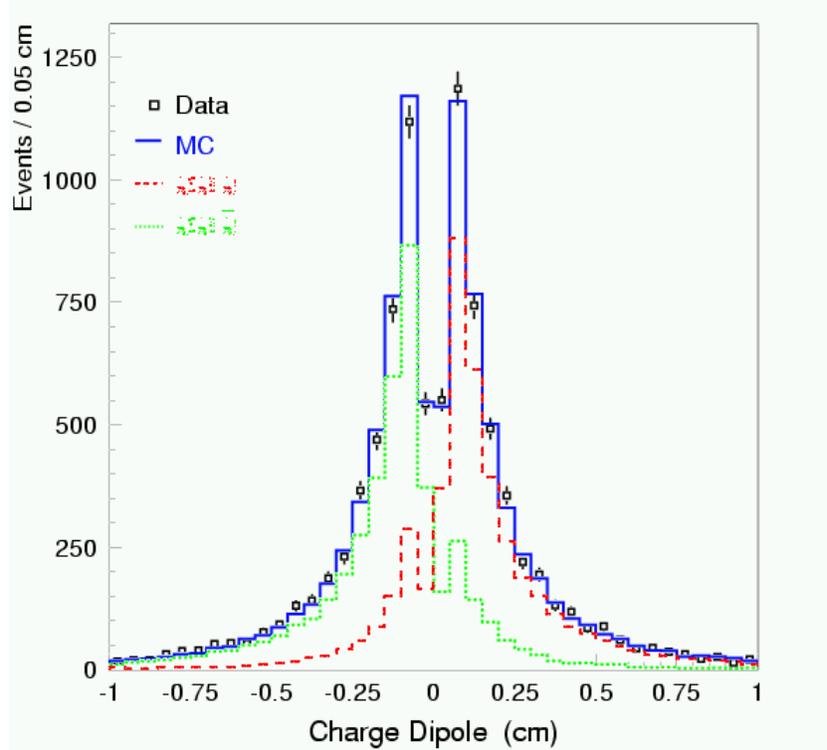}
  \end{center}
  \vspace{-0.5 cm}
  \caption{Distribution of the charge dipole for
  data (points) and Monte Carlo (solid histogram).
  Also shown are the contributions from $b$ hadrons containing
  a $b$ quark (dashed histogram) or a $\bar{b}$ quark (dotted histogram).}
  \label{fig_dipo_distr}
\end{figure}
Figure~\ref{fig_dipo_distr} shows the $\delta Q$ distribution,
illustrating the clear separation between $B$ and $\overline{B}$ decays.
The average decay flavor mistag rate was estimated to be 24\%,
driven mostly by decays of the type $\Bs \to D \overline{D} X$
(representing approximately 20\% of the selected sample)
for which the method assigned the decay flavor randomly.
In contrast, the mistag rate was estimated to be 12\% for
decays of the type $\Bs \to D_s^- X$, where $X$ does not include any charm hadron.

In all the analyses, the sensitivity to \Bs\ oscillations was
enhanced by evaluating the production and decay flavor mistag rates, 
and the decay length and momentum resolutions, 
an event-by-event basis.

\begin{table}[tbp]
\caption{Summary of relevant parameters for the \Bs\ mixing analyses}
\vspace{0.3 cm}
\label{tab:bsmixanal}
\begin{center}
\begin{tabular}{lccc}
\hline
                             & $D_s$$+$Tracks & Lepton$+$$D$ & Charge Dipole \\
\hline
No. decays                        &         361 &       2087 &       8556    \\
\Bs\ purity                       &        38\% &       16\% &       15\%    \\
Production flavor mistag rate     &        25\% &       25\% &       22\%    \\
Decay flavor mistag rate          &        10\% &        4\% &       24\%    \\
$\sigma_L$ (60\% core Gaussian)   &   48 $\mu$m &  55 $\mu$m &  76 $\mu$m    \\
$\sigma_L$ (40\% tail Gaussian)   &  152 $\mu$m & 217 $\mu$m & 311 $\mu$m    \\
$\sigma_p/p$ (60\% core Gaussian) &        0.08 &       0.06 &       0.07    \\
$\sigma_p/p$ (40\% tail Gaussian) &        0.19 &       0.18 &       0.21    \\
Sensitivity                       &  1.4 \invps & 6.3 \invps & 6.9 \invps \\
$\sigma_A$ at \dms = 15 \invps    &        1.60 &       1.14 &       1.11 \\
\hline
\end{tabular}
\end{center}
\end{table}

\subsection{Results}

  The study of the time dependence of \bsmix\ mixing
utilized the amplitude method~\cite{Moser97}.
Instead of fitting for $\dms$ directly,
the analysis was done
at fixed values of $\dms$ and a fit to the amplitude $A$ of the
oscillation was performed, i.e., in Equations~\ref{eq_punmixed}
and \ref{eq_pmixed}, for the unmixed and mixed
probabilities, the term
$\left[ 1 \pm \cos(\Delta m_s t) \right]$ was replaced
with $\left[ 1 \pm A \cos(\Delta m_s t) \right]$.
This method is similar to Fourier transform analysis and
has the advantage of facilitating the combination of results
from different analyses and different experiments.
One expects the amplitude to be consistent with $A=0$ for
values of the frequency sufficiently far from the true oscillation
frequency, whereas the amplitude should peak at $A=1$ at that
frequency.

\begin{figure}[tbp]
  \begin{center}
    \epsfxsize=11cm
    \epsfbox{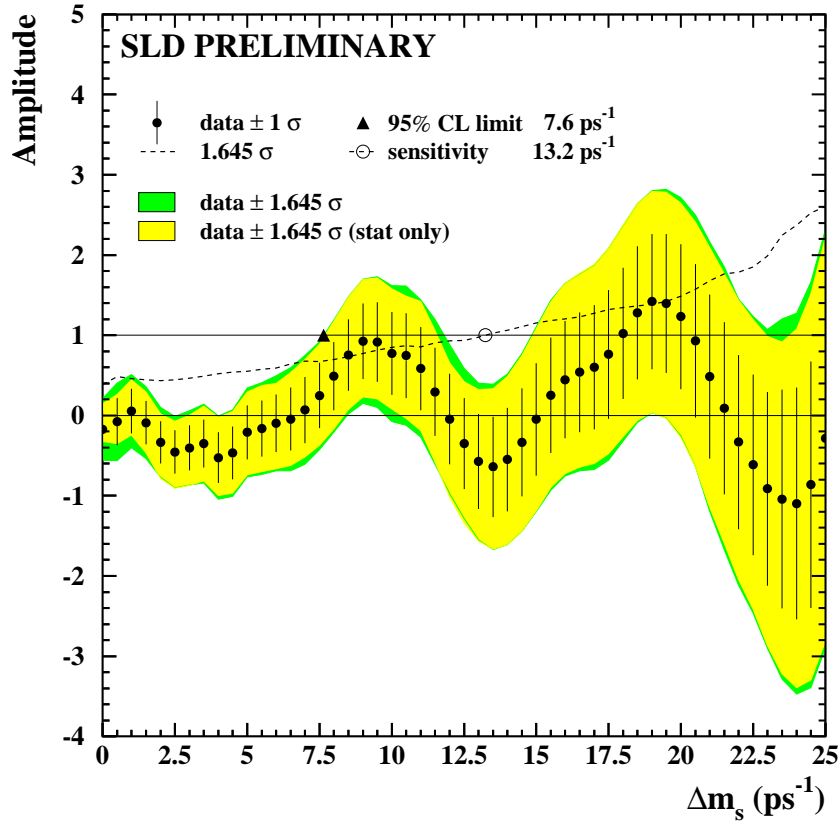}
  \end{center}
  \vspace{-0.7 cm}
  \caption{Measured amplitude as a function of \dms\ for the
  $D_s$$+$tracks, lepton$+$$D$, and charge dipole analyses combined.}
  \label{fig_sld_afit}
\end{figure}

  The $D_s$$+$tracks, lepton$+$$D$, and charge dipole analyses were
combined, taking into account correlated systematic errors.
Events shared by two or more analyses were only kept in the analysis
with the highest sensitivity, thereby removing any statistical correlation
between the analyses.
The dominant contributions to the systematic error were found to be due
to the uncertainties in the \Bs\ production rate, and the decay length
and momentum resolutions, as well as the decay flavor tag in the case
of the charge dipole analysis.
Figure~\ref{fig_sld_afit} shows the measured amplitude as a function
of \dms\ for the combination.
The measured values are consistent with $A = 0$
for the whole range of $\dms$ up to 25 ps$^{-1}$,
and no significant evidence is found for a preferred value
of the oscillation frequency.
The following ranges of \Bs\ oscillation frequencies
are excluded at 95\% CL:
$\dms < 7.6  \invps$ and
$11.8 < \dms < 14.8$ \invps, i.e.,
the condition $A + 1.645\:\sigma_A < 1$ is satisfied for those values.
The combined sensitivity to set a 95\% CL lower limit is
found to be at a \dms\ value of 13.0 \invps,
which corresponds to the frequency that satisfies $1.645\:\sigma_A = 1$.
All results are preliminary.

\begin{figure}[tbp]
  \begin{center}
    \epsfxsize=11cm
    \epsfbox{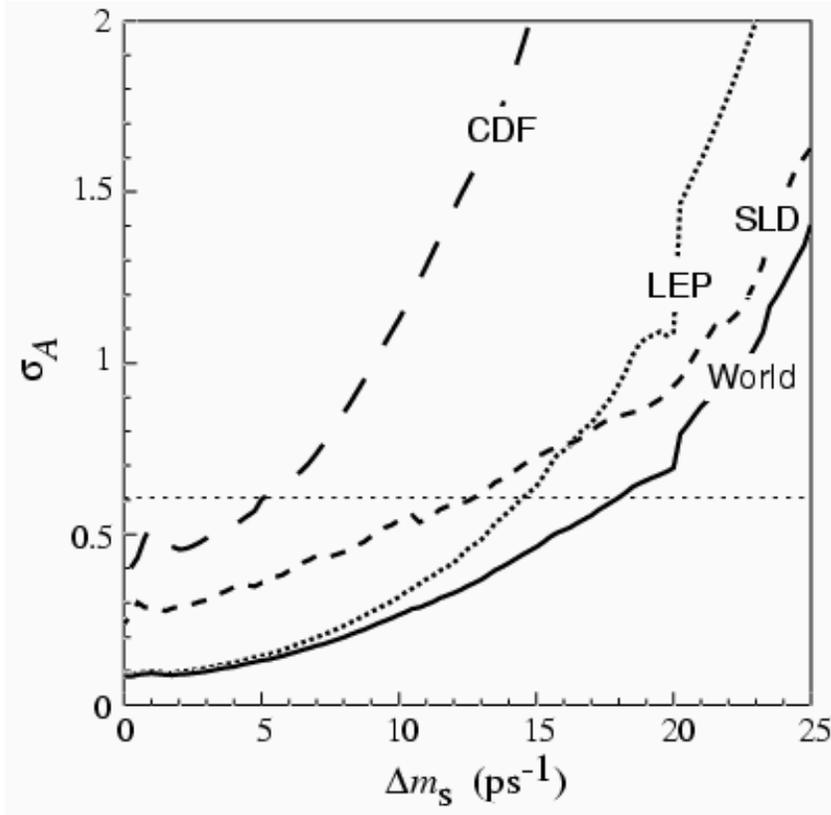}
  \end{center}
  \vspace{-0.7 cm}
  \caption{Amplitude uncertainty as a function of \dms\ for CDF, LEP, SLD,
  and the world average. The dashed horizontal line corresponds to the
  condition $1.645\:\sigma_A = 1$ used to define the 95\% CL sensitivity.}
  \label{fig_sigma_amp}
\end{figure}

The impact of the SLD analyses on the search for \Bs\ oscillations
is apparent in Figure~\ref{fig_sigma_amp}, 
which shows the uncertainty
in the measured oscillation amplitude as a function of \dms\ for
CDF, LEP, SLD, and the world average. The growth of $\sigma_A$ with increasing
\dms\ is due to the finite proper time resolution
and clearly illustrates that the excellent proper
time resolution for SLD yields a smaller increase in uncertainty than in the
case of LEP, for example.
The SLD sensitivity of 13.0 \invps\ can be compared to the
combined LEP sensitivity of 14.5 \invps, obtained with
a data sample 30 times larger than that used by SLD.

%% file: sec7_interp.tex
\section{INTERPRETATION OF RESULTS}

\subsection{\bsmix\ Mixing}
\label{sec:interp-bmixing}
\input sec7_bmixinginterp.tex


\subsection{Electroweak -- Quark Couplings}
\label{sec:interp-quark}
\input sec7_ewinterpquark.tex

\subsection{Electroweak -- Lepton Couplings}
\label{sec:interp-lepton}
\input sec7_ewinterplepton.tex

%% file: sec7_bmixinginterp.tex

\begin{figure}[tbp]
  \begin{center}
    \epsfxsize=16cm
    \epsfbox{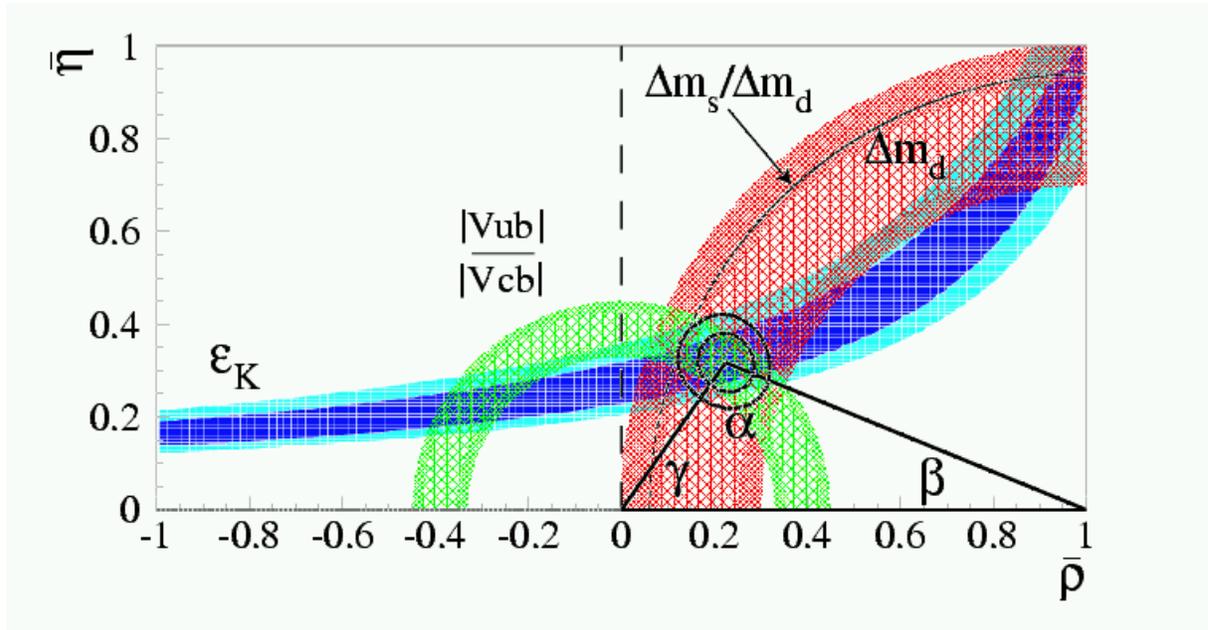}
  \end{center}
  \vspace{-0.9 cm}
  \caption{Constraints on the parameters $\bar{\rho}$ and $\bar{\eta}$ from
   the measurements of $\epsilon_K$, $|V_{ub}/V_{cb}|$, \dmd, and \dms.
   Figure extracted from Ciuchini et al~\cite{Ciuchini00}.}
  \label{fig_ut}
\end{figure}

  As mentioned above, the study of \Bs\ oscillations is
motivated by the fact that the oscillation frequencies \dmd\ and \dms\
provide very powerful constraints on the weak quark mixing sector of
the standard model in general, and on $CP$ violation in particular.
These constraints can be represented as circular bands
centered around the point $(1,0)$ in the $\bar{\rho}$-$\bar{\eta}$ plane
(see Figure~\ref{fig_ut}).
Also shown are the constraints from measurements 
of the $CP$-violating parameter $\epsilon_K$ in the 
$K^0$--$\overline{K}^0$ system
and the ratio of CKM elements $|V_{ub}/V_{cb}|$ measured in charmless
$B$ decays.
The combination of these constraints provides a measurement of
the parameters $\rho$ and $\eta$.
These parameters correspond to the apex of the Unitarity Triangle
(shown in Figure~\ref{fig_ut}).
The triangle represents one of
the unitarity conditions that the CKM matrix must satisfy.
A nonvanishing value of $\eta$ implies the existence of $CP$ violation,
whereas the angles of the triangle are related to the single $CP$-violating
phase of the CKM matrix.

A global fit within the context of the standard model yields
$\bar{\rho} = 0.224 \pm 0.038$ and
$\bar{\eta} = 0.317 \pm 0.040$~\cite{Ciuchini00}.
The fit uses the entire
\Bs\ oscillation amplitude spectrum rather than just
the lower limit on \dms\ to incorporate information
about the exclusion significance as a function of \dms.
It should be noted that the treatment of errors in such a fit
remains controversial because theoretical uncertainties dominate.
Of special interest are the predictions for the $CP$-violating angles
of the Unitarity Triangle $\alpha$, $\beta$, and $\gamma$,
which are currently the subject of intense experimental activity. 
The fit yields
$\sin 2\beta = 0.698 \pm 0.066$,
$\sin 2\alpha = -0.42 \pm 0.23$, and
$\gamma = (54.8 \pm 6.2)^\circ$~\cite{Ciuchini00}.
CP violation in $B$ decays has now been unequivocally 
observed by the BaBar and Belle collaborations, with a world 
average value for $\sin 2\beta$ of $0.79 \pm 0.10$~\cite{sin2b}, 
in good agreement with the
above standard-model prediction.
In the next few years, it is likely that we will learn
the most from the combination
of $\sin 2\beta$ and \dms\ measurements
at the $B$ factories and the Tevatron, 
since they provide
orthogonal constraints on the Unitarity Triangle and are fairly
clean theoretically and experimentally.

%% file: sec7_ewinterpquark.tex

SLD measurements of \Rb, \Rc, \Ab, \Ac, and \As\ are all in
good agreement with the standard model. The SLD \Rb\ measurement precision 
of $\sim$0.5\% is in the interesting regime for testing
the $\sim$1\% level effect due to physics beyond the standard model
through radiative corrections. The \As\ measurement, 
showing good agreement with the \Ab\ measurement, 
confirms quark coupling universality at the $\pm$10\% 
level. These measurements are also mostly in agreement with 
the measurements from LEP. In the case of
\Ab\ and \Ac, the indirect measurements from LEP derived from 
\Afbb\ and \Afbc\ are both lower than the
SLD average at the $1.3\sigma$ and $1.5\sigma$ levels, respectively. 
However, the derived LEP \Ab, using
the LEP and SLD combined $A_\ell$ for the initial state 
\Ae, is by itself $3.1\sigma$ away from the standard model. This is 
essentially the same effect as the case of \Afbb\ and \Afbc\ 
measurements giving higher $\sinsqth$ values than
other measurements when assuming universality in the fermions
(see next subsection).

\begin{figure}[tbp]
\begin{center}
  \epsfxsize13cm
  \epsfbox{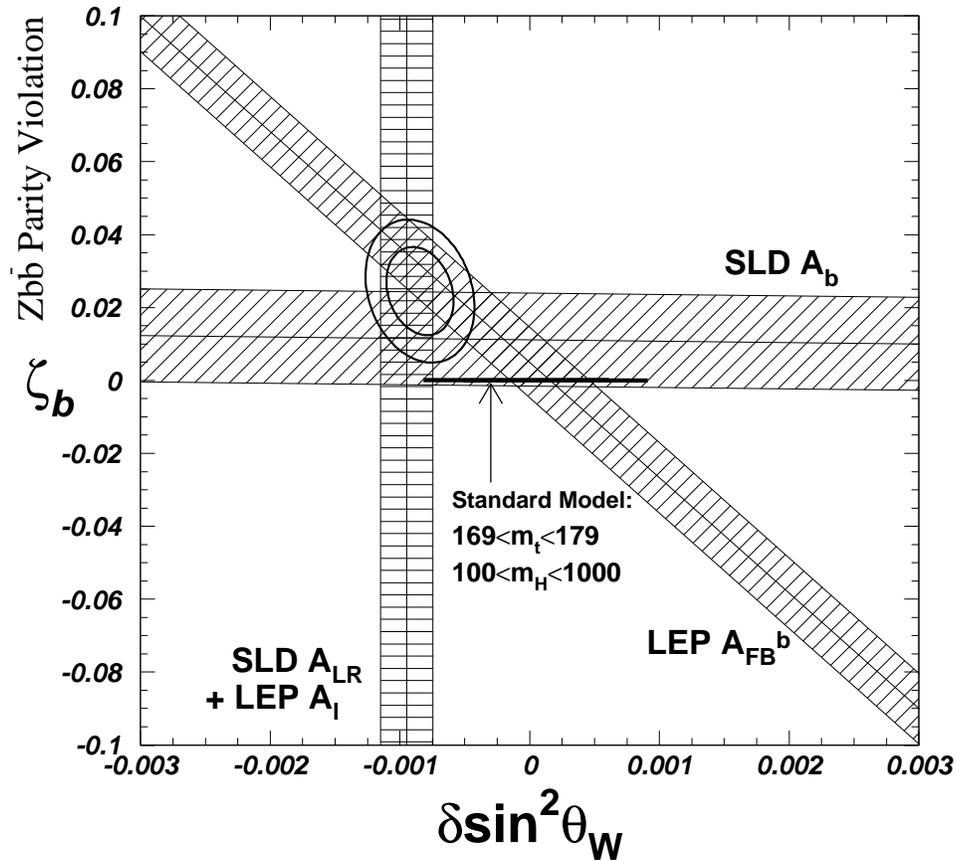}
\end{center}
\vspace{-0.7cm} 
\caption{The \Zbb\ coupling analysis result following 
 Takeuchi~et al. The standard-model point at (0,0) is defined by 
 $m_t$=174\gev, $m_H$=300\gev, $\alpha_s$=0.119, and 
 $\alpha$=1/128.905. The thin horizontal band around (0,0) 
 corresponds to the standard model $m_t$ and $m_H$ variations
 indicated in the plot. 
 The error ellipses represent 68\% and 95\% CL contours for 
 the fit.}
\label{fig:tgr}
\end{figure}

An interesting perspective on the interconnection between
the measurements of $A_\ell$, \Ab, and \Afbb\ is the \Zbb\  
vertex coupling analysis scheme from Takeuchi et al~\cite{ref:TGR}. 
Deviations ($\delta g_L^b,\delta g_R^b$) from the standard
model \Zbb\ coupling can be  
transformed into a pair of variables ($\xi_b,\zeta_b$) such 
that \Rb\ deviations depend only on $\xi_b$ and \Ab\ deviations
\revise{only depend}{depend only} on $\zeta_b$.
Particularly interesting is the relation between
the parity violation type variable $\zeta_b$ and
$\delta\sinsqth$, as shown in Figure~\ref{fig:tgr}. 
The consistency between the various measurements and the standard model
is only at the 1.0\% CL. 
The projection of the different bands on the horizontal line
corresponding to $\zeta_b=0$, i.e., assuming the standard-model \Zbb\
coupling, illustrates the 
discrepancy in $\sinsqth$ between the values derived 
from $A_\ell$ and from \Afbb. The projection along the 
vertical line representing the measured $\sinsqth$ from $A_\ell$ illustrates
the difference in \Ab, comparing the direct measurement from 
SLD and the indirect measurement from \Afbb. 

\begin{figure}[tbp]
\begin{center}
  \epsfxsize16cm
  \epsfbox{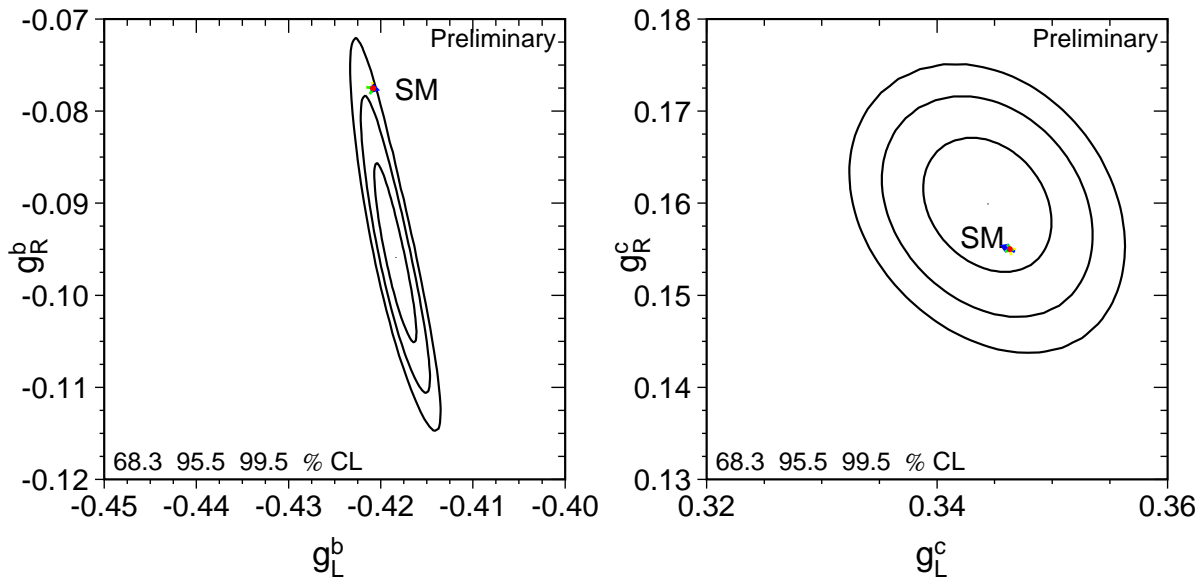}
\end{center}
\vspace{-0.5cm} 
\caption{Left- and right-handed \Zbb\ and \Zcc\ couplings
combining measurements from SLD and LEP.
The standard model (SM) predictions are also shown with small
arrows indicating the effect of the uncertainty
in the top quark mass.}
\label{fig:glgrbc}
\end{figure}

A general fit for the left- and 
right-handed \Zbb\ and \Zcc\ couplings was performed~\cite{Grunewald}
in the context of the standard model (with \sinsqth\ extracted from
the world average $A_\ell$)
using the \Rb, \Rc, \Ab, \Ac, \Afbb, and \Afbc\ measurements.
Table~\ref{tab:glgrbc} summarizes the results of the fit
and the corresponding standard model predictions.
These results are also illustrated in Figure~\ref{fig:glgrbc}.
\begin{table}[tbp]
\caption{Left- and right-handed \Zbb\ and \Zcc\ couplings determined
 from a fit to SLD and LEP measurements;
 the standard model (SM) prediction is also provided}
\vspace{0.4 cm}
\label{tab:glgrbc}
\begin{center}
\begin{tabular}{lrr}
  \hline
  Coupling constant & Fit result           & SM prediction \\ \hline
  $g_L^b$           & $-0.4183 \pm 0.0015$ & $-0.4211$ \\
  $g_R^b$           & $-0.0962 \pm 0.0064$ & $-0.0774$ \\
  $g_L^c$           & $ 0.3443 \pm 0.0037$ & $ 0.3467$ \\
  $g_R^c$           & $ 0.1600 \pm 0.0048$ & $ 0.1548$ \\  \hline
\end{tabular}
\end{center}
\end{table}
There is generally good 
agreement between the measurements and the standard model for
the \Zcc\ coupling at the current precision level.
The apparent departure of the \Zbb\
coupling from the standard model, driven primarily by the 
LEP \Afbb\ measurements, is mainly affecting the right-handed 
coupling value. This is particularly 
difficult to accommodate because no known model can produce 
a deviation from the standard model at this level. The direct measurement
of \Ab\ from SLD provides an important constraint on a possible \Zbb\ 
coupling anomaly, which has relevance to
the validity of setting constraints on the Higgs mass from
measurements of \Afbb,
as is discussed next.   

%% file: sec7_ewinterplepton.tex
The SLD measurement of \sinsqth,
precise and statistics-dominated,
represents a benchmark
for determinations of the weak mixing angle.  What do 
we learn from this result?   \revise{The first step in answering
this question is to}{We must first} 
examine its consistency with the standard model.
We have discussed the role of radiative effects and noted that
the measured top mass agrees with theoretical constraints at the level
of about $7\pm10$ GeV (a relative $4\pm6\%$).  Now 
only one standard-model parameter remains undetermined (although
direct experimental constraints exist)---the Higgs
boson mass $m_H$. Because of radiative effects on the gauge boson
propagators, corrections due to the Higgs boson are expected to
contribute logarithmically in $m_H$, and hence sensitivity 
to this parameter is marginal.  

\subsubsection{Higgs Mass Constraints}

\begin{figure}[tbp]
\vspace{-1.0 cm}
\begin{center}
  \epsfxsize17cm
  \epsfbox{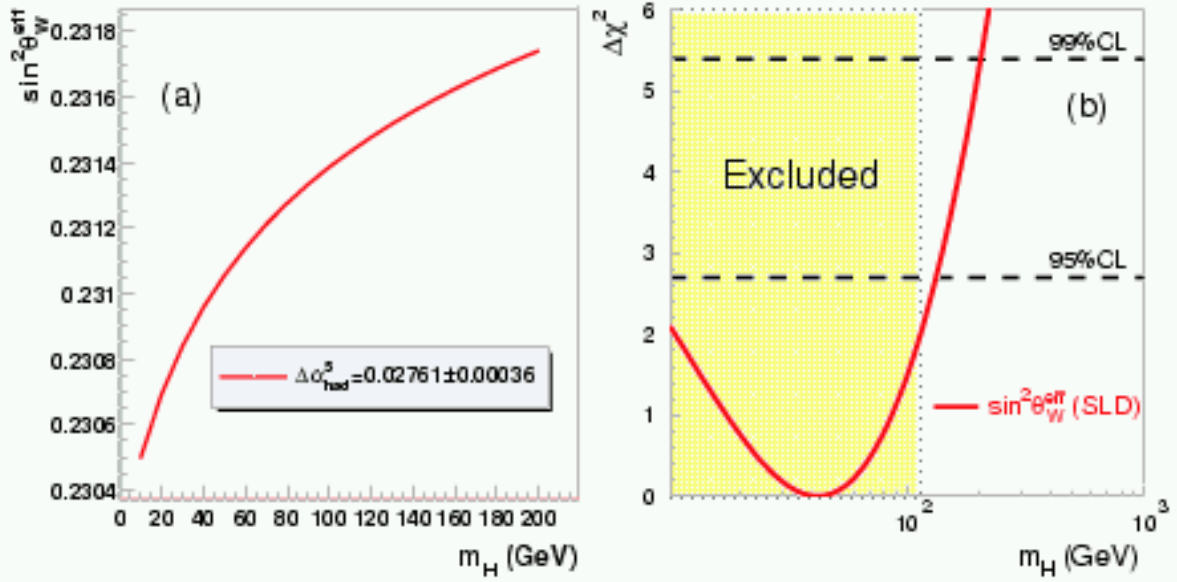}
 \vspace{-1.5 cm}
\caption{($a$) Leading-order radiative effects on \sinsqth\
 as a function of the Higgs boson mass. ($b$) Change in $\chi^2$ curve
 for a fit to the Higgs mass using the SLD \sinsqth\ result.}
\label{fig:mhiggs}
\end{center}
\end{figure}

Nevertheless, the SLD measurement
error is small enough that a meaningful constraint is available
from this measurement alone.  The sensitivity of an observable
$\Omega$ to $m_H$ can be quantified by the 
normalized partial derivative
${1\over{\delta {\Omega}}}{{\partial {\Omega}} \over{\partial{log_{10}(m_H/\rm{GeV})}}}$, 
where $\delta {\Omega}$ is the total uncertainty in $\Omega$.
The value of this
metric for the SLD \sinsqth\ measurement, at 4.4, is larger than
any other presently available electroweak result
[closest are the forward-backward
$b$ quark asymmetry results from LEP (3.9) and the $W$ boson mass result from 
LEP-II (3.7)].
Figure~\ref{fig:mhiggs}$a$ illustrates
the dependence of \sinsqth\ on the Higgs mass. From
this figure it is apparent that the SLD \sinsqth\ result of 0.23097
prefers a low Higgs mass, and that
mass constraints set in this region benefit from the steeper slope
of the curve. The largest uncertainty in these standard-model calculations
is due to the present top quark mass error of 5.1 GeV,
equivalent to $\delta \sinsqth \approx \pm0.00016$.
In addition, an ambiguity in the determination 
of the standard-model Higgs boson mass arises
from uncertainties in the evolution of the fine structure constant
from low $Q^2$ to its value at the \Z\ scale,
$\alpha(M_Z^2)$. This uncertainty stems from the need to use
low-energy $\ee$ annihilation 
cross-section data in the calculation of hadronic loop
effects.  Recent data from the Beijing Electron Synchrotron (BES) 
$\ee$ experiment~\cite{ref:BES} \revise{has}{have} led to
improved errors, which presently correspond to an uncertainty in 
\sinsqth\ of about $\pm0.0001$~\cite{ref:Pietrzyk}.  
However, a number of evaluations of
$\alpha(M_Z^2)$ are in common use, 
leading to variations in \sinsqth\ predictions
over a range of $\sim 0.00015$, and an associated range in quoted Higgs boson
mass limits. The choice of $\alpha(M_Z^2)$ used throughout this
review is due to
Burkhardt and Pietrzyk~\cite{ref:Pietrzyk}.

We perform a $\chi^2$ fit to the 
standard model by including at least one electroweak
observable, along with the fit parameters:
$m_H$, $M_Z$ measured at LEP, $m_t$ measured at Fermilab,  
$\alpha(M_Z^2)$, and the strong coupling constant $\alpha_s$
(the Fermi constant $G_F$ is held fixed).
With the SLD \sinsqth\ result as the sole electroweak observable
included in the fit, 
Figure~\ref{fig:mhiggs}$b$ gives the results in the 
form of a Higgs mass $\chi^2$. 
The resulting one-sided confidence upper limits from the fit
are $m_{H} < 133$ GeV (95\% CL) and 
$ < 205$ GeV (99\% CL).  Also shown is the
current direct 
search lower limit from LEP-II of 114.1 GeV. 
These direct and indirect Higgs bounds are in modest agreement: 
the value of the one-sided confidence limit at 114.1 GeV 
corresponds to a probability of 7\%. 

\begin{figure}[tbp]
\begin{center}
  \epsfxsize9.5cm
  \epsfbox{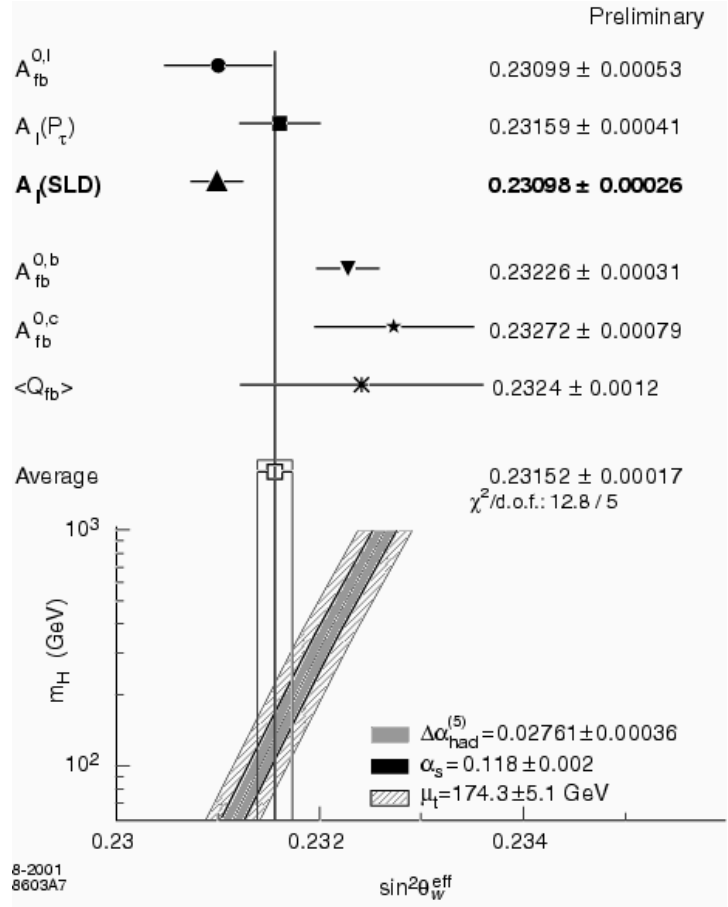}
 \vspace{-0.4 cm}
\caption{SLD and LEP \sinsqth\ results.  The first three results are
based on lepton asymmetries including the SLD result,
the last three results on $b$-quark, $c$-quark, and hadronic-event asymmetries
are based on measurements at LEP.}
\label{fig:LEPEWWGsin}
\end{center}
\end{figure}

As we have seen, the SLD data are 
consistent with lepton universality.  
Measurements at LEP of $\Afbl$,
and of $\tau^{\pm}$ polarization 
(which separately provides 
$A_e$ and $A_\tau$), confirm lepton universality as well.
All available $\Al$ data, expressed in terms of \sinsqth,
are consistent (see Figure~\ref{fig:LEPEWWGsin}). 
Also shown are the LEP \sinsqth\ results
deriving from quark asymmetries, which, taken together, provide
an average ($0.23230\pm0.00029$) that is $3.3\sigma$ 
different from the lepton asymmetries average
($0.23113\pm0.00021$).  Unresolved issues in the  
$\Zbb$ coupling data \revise{have already been}{were} discussed in the 
previous section.   
The $\chi^2$ for the overall \sinsqth\ combination has a probability of 2.5\%. 
The quark asymmetry average is dominated by the $\Afbb$ measurement, the
most precise \sinsqth\ determination at LEP.
 
\begin{figure}[tbp]
\begin{center}
  \epsfxsize12cm
  \epsfbox{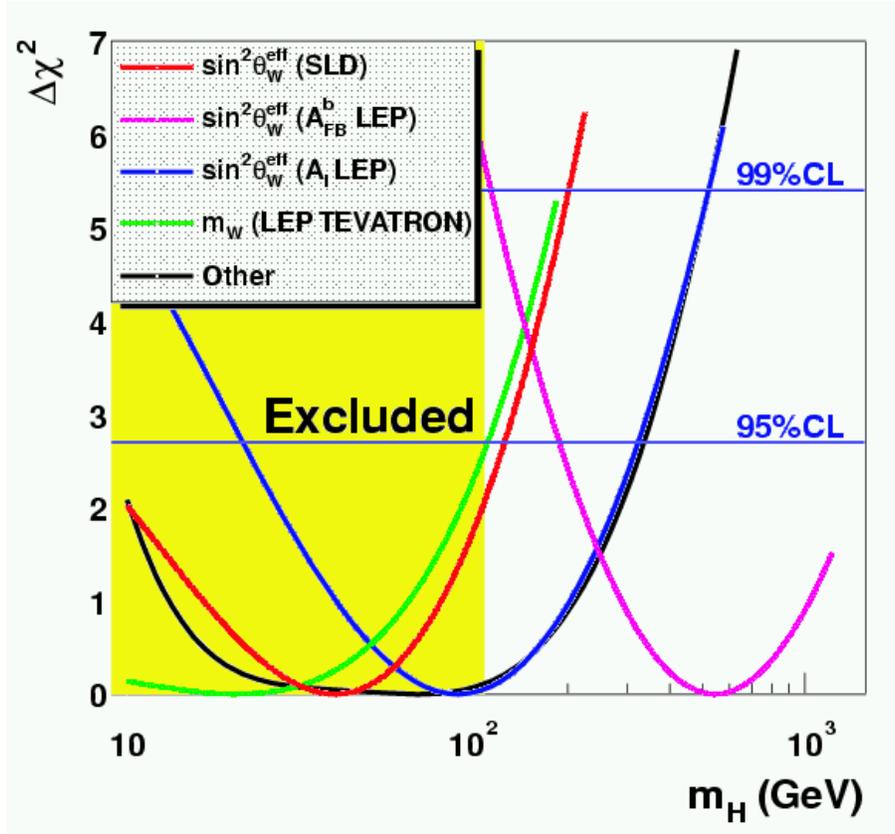}
 \vspace{-1.0 cm}
\caption{Change in $\chi^2$ as a function of the Higgs mass
for various measurements separately. ``Other'' refers here to
all the other electroweak measurements included in our global fit.}
\label{fig:multihiggs}
\end{center}
\end{figure}

Fits are made to worldwide electroweak results~\cite{ref:LEPEWWG2001},
including
all SLD and LEP measurements, LEP-II and Tevatron $M_W$ results,
deep inelastic neutrino scattering data, and 
measurements of atomic parity violation 
in cesium and thallium~\cite{ref:APVissues}. 
Among these, the most sensitive to 
$m_H$ are, in order of decreasing sensitivity, the asymmetries,
$M_W$, and the \Z\ total width $\Gamma_Z$.  With the exception of
the \Afbb\ data mentioned above, all of these measurements 
agree satisfactorily and favor a light Higgs boson.  
Figure~\ref{fig:multihiggs} shows the individual
contributions to the overall fit
broken down into SLD/LEP leptonic and LEP hadronic contributions
to \sinsqth, the world $M_W$ average, and all other electroweak results
(dominated by $\Gamma_Z$).
Including all data results in limits of $m_{H} < 195$ GeV (95\% CL) and 
$ < 260$ GeV (99\% CL), in agreement with the direct lower Higgs mass 
limit at 114.1 GeV. 
The $\chi^2$ per degree of freedom for this 
fit is $23.5/16$, corresponding to a probability of 10\%. 
Excluding the $\Afbb$ result from the fit increases the $\chi^2$ probability
to 47\% and substantially tightens the Higgs mass limits to $m_{H} < 133$ 
GeV (95\% CL) and $ < 185$ GeV (99\% CL). 
The confidence level of this restricted fit at the value of the direct lower
Higgs mass limit is 9\%.  Within the context of the standard model, 
the electroweak
data are suggesting that the Higgs boson is relatively 
\revise{light, and}{light and is} in the region
expected by supersymmetric extensions of the standard model~\cite{ref:MSSM}.  
\revise{While}{Although}
these constraints are not model-independent, they do significantly restrict
possible new physics, even if the Higgs mass limits are
evaded~\cite{ref:Peskinnewphys}.

\subsubsection{Extensions of the Standard Model: The S, T, and U Parameters}

It is desirable to consider the electroweak data in a more general context
than the standard model.
The framework of our analysis will follow 
Lynn et al.~\cite{ref:LPS},
who made the following assumptions:

\begin{itemize}

\item The gauge group of electroweak physics is
$SU(2)_L \times U(1)$ ($Z,Z^\prime$ mixing, for example, is not incorporated).
 
\item The symmetry-breaking sector has a global $SU(2)$
``custodial'' symmetry that insures
$ \rho = M_W^2 / M_Z^2 \cos^2\theta_W = 1 + {\cal O}(\alpha).$
 
\item Vacuum polarization effects dominate compared to
vertex and box corrections (the oblique hypothesis).
 
\item Any new particles are heavy, so that
Taylor expansions are appropriate descriptions of the vacuum
polarization corrections.

\end{itemize}

With these assumptions, the
total number of independent gauge boson correction terms is six.
Using the three precision
constants ($G_F$, $M_Z$, and $\alpha$), this leaves three undetermined
parameters,
which are denoted S, T, and U in the scheme of
Peskin \& Takeuchi~\cite{ref:STU}.
The S parameter is a measure of the weak isospin-conserving
new heavy sector.  The T parameter is equal to $\alpha^{-1} \Delta
\rho$ and is a measure of weak isospin breaking in
the heavy sector (for example, the top quark to bottom quark
mass splitting).
The U parameter introduces small effects, is only
relevant for the observable $M_W$, and is ignored here.
A heavy fermion doublet changes 
the T parameter, which is quadratic in the mass splitting.  The
standard-model Higgs boson contributes offsets that are logarithmic
in the Higgs mass for both S and T, though these effects are
opposite in sign. 

\begin{figure}[tbp]
\begin{center}
  \epsfxsize15cm
  \epsfbox{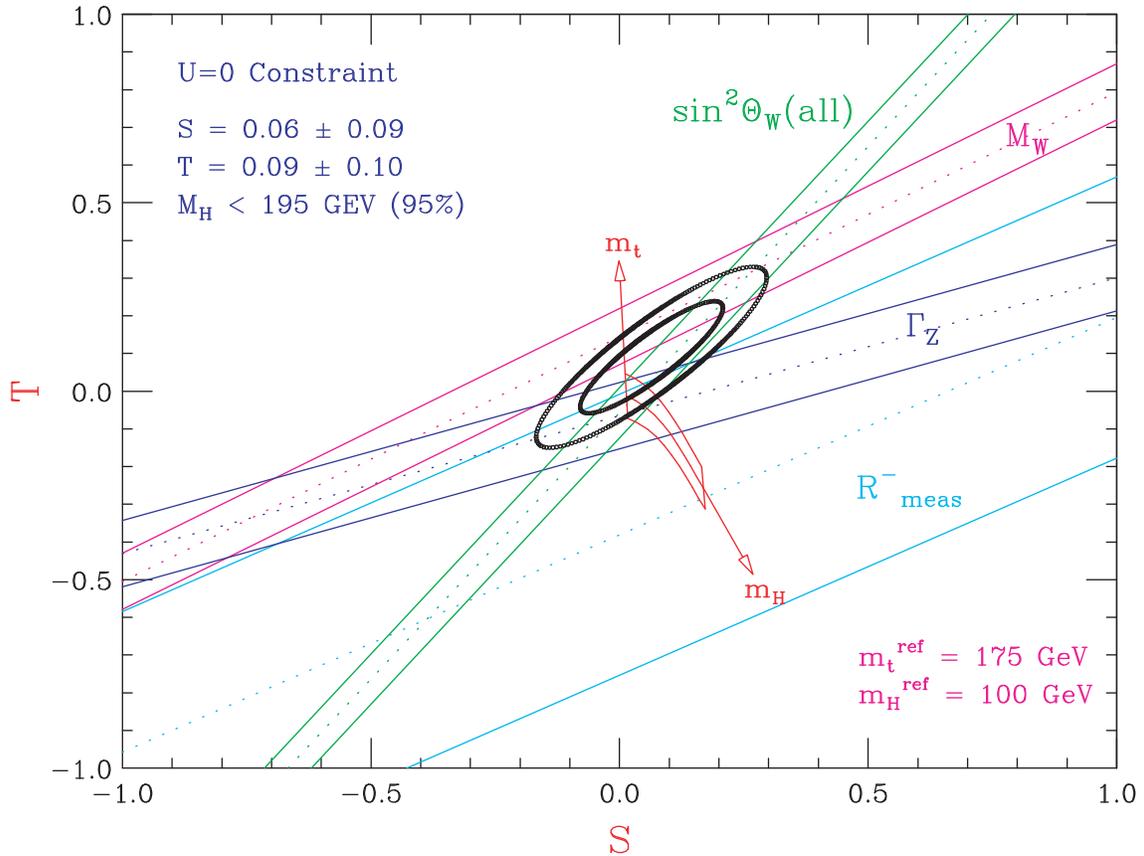}
 \vspace{-0.3 cm}
\caption{Global electroweak fit in the S and T plane.
The ellipses correspond to
the 68\% and 90\% CL contours of the global fit.  Four of the contributors
to the fit are shown as $\pm1 \sigma$ bands.}
\label{fig:STfit}
\end{center}
\end{figure}

We performed a fit to the electroweak data, where offsets from
S and T $=0$ were relative
to a reference value of the top quark and Higgs masses, which
we chose to be 175 GeV and 100 GeV, respectively.
Figure~\ref{fig:STfit} gives the result. 

The fit ellipse (68\% CL and 90\% CL contours are shown) 
is consistent with the 
banana-shaped region of the S-T plane
allowed by the standard model, so long as the Higgs mass is small.
The fit result
is presently consistent with S and T~$=0$.
Also shown are the 
$\pm 1 \sigma$ constraints for four representative electroweak observables:
\sinsqth, $M_W$, $\Gamma_Z$, and $\rm{R}^-$
(from neutrino deep-inelastic scattering),
which appear as bands with different slopes.   
At the present level of precision in the global electroweak
data, the S-T analysis has excluded a subclass of the models that
propose strong symmetry breaking mechanisms 
[conventional ``technicolor'' models~\cite{ref:technicolor}]
that predict significant deviations from S~$=0$.
The S-T analysis is
beginning to probe hypothetical extensions of the standard model, 
such as supersymmetry. For example, 
the region of the S-T plane allowed by the minimal supersymmetric
model (MSSM) overlaps the light Higgs end of the standard-model
zone but also includes
larger deviations from T~$=0$. Some authors have already explored how 
the precision electroweak data constrain the free parameters of
the MSSM~\cite{ref:ErhlerPierce}.

%% file: sec8_epilogue.tex
\section{CONCLUSIONS}

  We have reviewed highlights of the \Z-pole physics program carried out
by the SLD collaboration at the SLC.
The combination of beam polarization, small beam sizes,
and a CCD pixel vertex detector
has led to a series of measurements that complement
those performed at Fermilab and CERN
and that have probed the standard model and its extensions.
Many of these measurements provide unique insight into
the physics of fundamental interactions.
In addition, the novel techniques developed for these measurements
have applicability to future colliding beam experiments.

An important result of the precision electroweak program at SLD is that a 
standard-model Higgs boson, if it exists, should be relatively light.
This conclusion
is also supported by a fit to the complete set of worldwide
precision electroweak data.
In particular, these fits indicate that the Higgs boson
may be within reach of the Tevatron experiments at Fermilab
in the next several years.
If the Higgs boson eludes the Tevatron experiments,
a much greater reach is expected once the CERN LHC program begins in
about 2006.
Also important are the SLD electroweak measurements performed
with individual lepton and quark flavors.
This diverse set of measurements clearly illustrates
the versatility of the powerful flavor-separation methods
available to an experiment at a linear collider.
Finally, studies of \bsmix\ mixing at SLD have
contributed significantly
to the worldwide effort to constrain the quark mixing matrix and $CP$ violation
in the standard model. These played an important role,
complementary to the $CP$ violation measurements currently under way.

Looking a bit further ahead, high-energy ($\sim$500 GeV and up) 
linear collider projects are presently under consideration in the
United States, Germany, and Japan.
These new collider projects would build on the experience
gained from the construction and operation of the SLC and SLD,
as well as from the physics results and techniques developed there.

\section*{ACKNOWLEDGMENTS}

  We would like to thank our SLD colleagues for their contributions
and for making the collaboration an enjoyable and stimulating
environment to work in.
We also wish to thank the SLC accelerator physicists and operations personnel
for their efforts.
\clearpage